\documentstyle [11pt] {article}
{\catcode `\@=11 \global\let\AddToReset=\@addtoreset}
\AddToReset{equation}{section} 

\renewcommand{\pt}{\partial}
\newcommand{\locally}{}
\newcommand{\stationary}{stationary--rotating}

\newcommand{\hS}{\hat\Sigma}
\newcommand{\tS}{\widetilde \Sigma}
\newcommand{\be}{\begin{equation}}

\newcommand{\U}{{\cal U}}
\newcommand{\ds}{{\cal D}(\Sigma)}
\newcommand{\apu}{\stackrel{\approx}{u}}
\newcommand{\apS}{\stackrel{\approx}{\Sigma}}
\newcommand{\ods}{\stackrel{o}{\cal D}\!\!(\Sigma)}

\newcommand{\og}{\stackrel{o}g}

\newcommand{\ee}{\end{equation}}
\newcommand{\bea}{\begin{eqnarray}}
\newcommand{\eea}{\end{eqnarray}}
\newcommand{\beaa}{\begin{eqnarray*}}
\newcommand{\eeaa}{\end{eqnarray*}}

\newtheorem{Definition}{Definition}
\newtheorem{Lemma}{Lemma}
\newtheorem{Proposition}{Proposition}

\newtheorem{Corollary}{Corollary}

\newtheorem{Theorem}{Theorem}
\AddToReset{Conjecture}{section}
\AddToReset{Theorem}{section} 
\AddToReset{Proposition}{section} 
\AddToReset{Corollary}{section} 
\AddToReset{Lemma}{section} 
\AddToReset{Definition}{section} 
\AddToReset{figure}{section} 

\newcommand{\ra}{\rightarrow}
\newcommand{\rai}{\rightarrow\infty}

\newcommand{\N}{I\!\! N}
\newcommand{\R}{I\!\! R}
\newcommand{\Tau}{{\cal T}}

\begin{document}
\title{Maximal Hypersurfaces in Asymptotically Stationary Spacetimes}
\author{Piotr T.\ Chru\'sciel\thanks{Alexander von Humboldt fellow.
On leave of absence from the
Institute of Mathematics,  Polish Academy of Sciences,
Warsaw.   {\em E--mail}: piotr@ibm-1.mpa.ipp-garching.mpg.de.
 Supported in part
by KBN grant \#2 1047 9101.} \\Max Planck Institut f\"ur Astrophysik\\
Karl Schwarzschild Strasse 1\\ D 8046 Garching bei M\"unchen
\\ \\Robert M.\ Wald\thanks{Supported in part by NSF grant PHY--8918388.
{\em E--mail}: rmwa@midway.uchicago.edu.}\\ Enrico Fermi
Institute and Department of Physics\\University of Chicago\\
5640 S. Ellis Ave., Chicago, IL 60637.}
\date{}
\maketitle

\begin{abstract}

Existence of maximal hypersurfaces
and of foliations by maximal hypersurfaces
is proven in two classes of asymptotically  flat spacetimes which possess a
one parameter group of isometries  whose orbits are timelike
 ``near infinity''.
The first class
consists of strongly causal asymptotically flat spacetimes
which contain no ``black hole or white hole" (but may contain
``ergoregions" where the Killing orbits fail to be
timelike).
The second class
of spacetimes possess a black hole and a white hole,
with the black and white hole horizons
intersecting in a compact 2-surface $S$.
\end{abstract}

\newpage

\section{Introduction}
\label{introduction}

The question of the existence of maximal slices ({\em i.e.}, slices with
vanishing trace, $K = {K^a}_a$ of extrinsic curvature, $K_{ab}$)
in asymptotically flat spacetimes has arisen frequently in the analysis of
many issues in general relativity.  The main reason for this is that the
``kinetic term"
\[
K^{ab}K_{ab}-K^2
\]
in the Hamiltonian constraint equation is non--negative when $K = 0$,
thereby simplifying many arguments.  Another reason is that the
momentum constraint becomes conformally invariant when $K = 0$.

The issue of the existence of maximal hypersurfaces arose again recently
in  an analysis of solutions to the Einstein--Yang--Mills equation by
Sudarsky and Wald \cite{SW}.
In theorems 3.3, 3.4, and the discussion following theorem 3.4 of
\cite{SW} two results were proven, which may be roughly summarized as
follows ({\em cf.} \cite{SW} for precise statements of the asymptotic
conditions assumed):

\begin{enumerate}
\item
\label{(l)}  Consider an asymptotically flat (with a single ``end") solution
to the Einstein--Yang--Mills equations (on a trivial $SU(2)$ principal
bundle) with a Killing vector field $X$, timelike at infinity, which has
vanishing electric charge or asymptotically vanishing electrostatic
potential.  Suppose there exists an asymptotically flat maximal slice
$\Sigma$ with compact interior ({\em i.e.},
a two--sphere in the asymptotic region
bounds a compact subset of $\Sigma$), such that $X$ is asymptotically
orthogonal to $\Sigma$ ({\em i.e.}, $X \longrightarrow_{r \rightarrow
\infty}{\pt/\pt t}$ in the asymptotically Minkowskian coordinates in
which $\Sigma\subset\{t=0\}$).   Then the solution is {\em static}
({\em i.e.}, $X$ is orthogonal to $\Sigma$) and has {\em vanishing}
Yang--Mills electric field on the static hypersurfaces.)

\item\label{(2)}  Consider an asymptotically flat solution to the
Einstein--Yang--Mills equations (on a trivial $SU(2)$ principal
bundle) which possesses a Killing vector field $X$ which is timelike
at infinity
and possesses a black and white hole, whose horizons comprise a bifurcate
Killing horizon with bifurcation surface $S$.  Suppose that  $X$  is
normal to the horizon ({\em i.e.}, suppose that the angular velocity of the
horizon vanishes) and suppose that the solution has vanishing electric
charge or asymptotically vanishing electrostatic potential.  Suppose
further that there exists an asymptotically flat maximal hypersurface
$\Sigma$ with boundary $S$, and compact interior such that $X$ is
asymptotically orthogonal to $\Sigma$.  Then the solution is {\em static}
``in the exterior world'' ({\em i.e.}, $X$ is orthogonal to $\Sigma$), with
$X$   strictly timelike outside of the black and white holes, and has {\em
vanishing}  electric field on the static hypersurfaces.  (In the
Einstein--Maxwell (on a trivial $U(1)$ principal bundle) case, the
hypothesis that the electric charge or the asymptotic electrostatic potential
vanishes may be dropped, and it then can be shown that the solution is
static in the exterior world and has vanishing magnetic field on the static
hypersurfaces.)
\end{enumerate}

Clearly, to obtain a more satisfactory picture one has to understand how
restrictive are the conditions of existence of maximal hypersurfaces made
above. In the past decade,
some progress has been made in proving existence  of maximal slices in
asymptotically flat spacetimes \cite{Ba1}, \cite{BCOM}.  In particular, it
has
been shown that asymptotically flat maximal slices always exist in {\em
strictly} stationary, asymptotically flat spacetimes \cite{BCOM}.
Unfortunately, this theorem requires that the
Killing vector field be timelike everywhere, not just ``near infinity".
Furthermore, this theorem does not encompass the situation where one
wishes the maximal hypersurface to pass through a given
2--surface
({\em e.g.} the
bifurcation 2--surface, $S$, of the second result above).

The purpose of
this paper is to extend the previous results on existence of maximal
hypersurfaces to encompass the situations considered by
Sudarsky and Wald, and, in fact, some more general situations as well.
We will prove that in asymptotically stationary spacetimes, there exist
maximal hypersurfaces of the type needed in their results \ref{(l)} and
\ref{(2)} above.  Indeed, our analysis will be more general in that
\begin{enumerate}
\item
Einstein's equation will not be used ({\em i.e.}, no energy conditions will
be imposed), and
\item
we will assume weaker asymptotic conditions on our spacetimes than
assumed in \cite{SW}.
\end{enumerate}
In addition, in an appendix we shall generalize our results to
``asymptotically {\em \stationary}\ spacetimes''.  Thus, under
the conditions of this paper, the hypotheses of the existence of a maximal
hypersurface can be removed from theorems 3.3 and 3.4 of \cite{SW} (as
well as from the discussion in \cite{SW} following theorem 3.4).

We refer the reader to section \ref{preliminaries} below for a precise
specification of the classes of spacetimes considered in our paper and we
refer the reader to section \ref{existence} for the precise statement (and
proof) of our existence theorems on maximal slices.  However, it is worth
elucidating here by some informal discussion and examples the nature of
the spacetimes to which our results apply --- as well as those which are
not treated by our analysis.

To begin, we emphasize that we restrict attention
to strongly causal spacetimes $(M, g_{ab})$ for which there exists a
smooth, acausal
hypersurface $\Sigma \subset M$ (possibly possessing a compact boundary $S$)
with a finite number of
asymptotically flat ``ends", $\Sigma_i$.
Furthermore, we require $(M, g_{ab})$ to possess a
one-parameter group of isometries which has timelike orbits at each
``end''.  (As mentioned above, a generalization to the case where $X$ is
``stationary-rotating" (rather than timelike)
at each end is given in the appendix.)
We shall denote the Killing field which generates these isometries
as $X$ and shall denote the isometry corresponding to parameter $t$
by $\phi[X]_t$. The orbit of $\Sigma_i$ under $\phi[X]_t$ will be denoted
as $M_i$. We
emphasize that we do not impose any conditions on the Killing orbits in
the ``interior portion" of the spacetime.  In particular, ``ergoregions"
where $X$ is null or spacelike are
permitted.

The general form of the spacetime metric in the spacetimes we consider
can be described in concrete terms as follows. In section \ref{existence}
we shall show that in the class of spacetimes which we consider,
there always exists a smooth, spacelike hypersurface $\Sigma'$
such that $X$
is transverse to $\Sigma'$, and the orbits of $X$ passing
through $\Sigma'$ are diffeomorphic to $\R$. We define the set
$M_{\Sigma'}
= \{p \in M : \exists\; t \in \R$ such that $\phi[X]_t(p) \in
\Sigma'\}$. Then $M_{\Sigma'}$ is an open submanifold of $M$
diffeomorphic to $\Sigma' \times \R$. Furthermore, the
hypersurface $\Sigma'$
is given by the equation $t = 0$, where $t$ is the function on
$M_{\Sigma'}$ appearing in the definition of $M_{\Sigma'}$.
It follows that on
$M_{\Sigma'}$ the metric can be written in the form
\begin{eqnarray}
\label{example}
&&{g_{\mu\nu}dx^{\mu}dx^{\nu}} = (|\beta|_\gamma^2 -
\alpha) dt^2 + 2\beta dt + \gamma\ ,\label{(D*)}\\
&&X^\mu \pt_\mu = \pt_t , \quad \frac{\pt\alpha}{\pt t} =
\frac{\pt\beta}{\pt t} = \frac{\pt  \gamma}{\pt t} = 0
\,,\nonumber
\end{eqnarray}
where $\alpha > 0$ is a function on $\Sigma'$, $\beta$ is a one form on
$\Sigma'$, and $\gamma$ is
a Riemannian metric on $\Sigma'$.
Furthermore, we have,
\[
g_{ab} X^a X^b * 0 \quad \Longleftrightarrow \quad
-\alpha + |\beta|_\gamma^2 * 0 ,
\]
where $*$ stands for $=$, $>$ or $<$. It should be stressed that
(\ref{(D*)}) defines a Lorentzian metric regardless of the
sign of $g_{ab} X^a X^b$ 
provided only that $\alpha > 0$ and that $\gamma$
is a Riemannian metric. When
$X$
is spacelike,
then all the coordinate vectors $\frac{\pt}{\pt x^\mu}$, $\mu =
0,\ldots,3$ are spacelike and one loses the {\em a priori} control of the
slopes of the light cones, which is at the origin of
the following difficulty: From the point of view of PDE theory, when
considering existence
of maximal hypersurfaces a natural hypothesis is that of compactness
of domains of dependence of compact sets. Given a metric of the form
(\ref{example}) in which $g_{ab} X^a X^b$ is allowed to change sign, it
seems far from being a trivial exercise to determine whether or
not compactness of domains of dependence of compact sets holds.
Nevertheless, in
Section \ref{structure} we shall show that
this compactness property holds for the spacetimes we consider.

Since the spacetimes we consider possess a hypersurface $\Sigma$
which extends to one or more asymptotically
flat ``ends", $\Sigma_i$ (with orbits $M_i$ under $\phi[X]_t$),
the notions of a ``black hole" and a ``white hole" can
be introduced. In this paper, we define a the {\em black hole}
region of the spacetime to consist
of the events which do not lie in the past of the {\em union} of the
asymptotically flat regions, $M_i$. This notion of black hole
should be sharply contrasted with the set of events which merely
fail to lie in the past
of a particular region $M_i$. (This set of events will be
referred to as the {\em black
hole with respect to
$M_i$}).  The notions of a {\em white hole} and a {\em white
hole with respect to
$M_i$}
are defined similarly, with
"past" replaced by ``future". Note that it is possible for a
spacetime to possess a black and/or
white hole with respect to one or more individual, $M_i$,
and yet fail to have
a black hole and/or white hole; indeed, the spacetime composed
of the union of blocks $B$,
$D$, and $E$ of Figure \ref{F1} provides an example of
such a spacetime. Note also that if more than one asymptotic region is
present in the spacetime, the definition of the black and
white hole regions may depend upon the choice of $\Sigma$, since only the
asymptotic regions, $M_i$, generated by ends, $\Sigma_i$, of $\Sigma$
enter the definition. Thus, for example, in the spacetime of
Figure \ref{F1} below, the black hole region for slice
$\Sigma$ is block $F$, but the black hole
region for slice $\Sigma'$ is the union of blocks $D$, $E$, and $F$,
and the black hole region for hypersurface $\Sigma''$ is the union of
blocks $C$, $D$, $E$, and $F$.

The spacetimes to which our theorems
apply divide into the following two classes\footnote{The reader
should note that the classes of space--times (1) and (2) discussed here
{correspond to}
the classes of space--times (a) and (b)
defined in the main body of the paper {with additional conditions
imposed upon the black and white hole regions of the spacetime}.}:
(1)\ $\Sigma$ is a slice ({\em i.e.}, a
closed hypersurface without boundary)
whose ``interior portion" is compact, and no black hole or white hole is
present. In this case theorem \ref{TM.1} establishes existence
of a maximal slice which is complete,
asymptotically flat at each end and is asymptotically orthogonal to $X$.
 (2)\
$\Sigma$ is a hypersurface with compact boundary $S$
(which need not be connected) and the ``interior portion" of $\Sigma$ is
compact. The spacetime
contains both a black hole and a white hole and
their horizons intersect at $S$.
In this case, Theorem \ref{TM.2} establishes existence of a maximal
hypersurface with boundary $S$, which is complete,
asymptotically flat at the
ends and is asymptotically orthogonal to $X$.

It should be emphasized, that our results on existence of maximal
surfaces are sharp, in the following sense: There exist asymptotically
flat space--times which contain a black hole region, and a white hole
region, and in which no complete, asymptotically flat, maximal surfaces which
are
asymptotically orthogonal to the Killing vector exist. An
example\footnote{This example was
suggested to us by Dieter Brill.  We wish to thank him for useful
discussions
concerning the question of sharpness of Theorem \ref{TM.1}.} of
such a space--time is presented at the beginning of Section \ref{existence}.

We now give some examples of spacetimes which lie in these
classes, as well as some examples which do not. First, class (1)
includes all asymptotically flat spacetimes possessing an
asymptotically flat slice, $\Sigma$, with compact interior, such that
$X$ is globally timelike. (Namely, if $X$ is globally timelike,
it is easily seen
that no black hole or white hole can be present: The event horizon
of a black or white hole is a null surface which is mapped into
itself by the isometries, so if a black or white hole exists,
$X$ must be
null or spacelike on its horizon.) Numerous examples of such spacetimes
are easily constructed using (\ref{(D*)}). Existence of maximal slices when
$X$ is globally timelike was previously proven in \cite{BCOM}. Thus,
Theorem \ref{TM.1} may be viewed as a generalization of this result
to the case where an ``ergoregion" (but no black or white hole) is
present.

An explicit example of a spacetime in class (1) which contains
an ergoregion may be constructed
as follows. Let $g_{ab}$ be a stationary,
axisymmetric metric on $\R^4$ of the form,
\begin{equation}
\label{(I.1)}
ds^2 = -V dt^2 + 2W dt d\varphi + r^2 \sin^2 \theta d\varphi^2 +
r^2 d\theta^2 + dr^2\,.
\end{equation}
Choose $r_1$, $r_2$, $\theta_1$, $\theta_2 \in \R$ with $0 < r_1 < r_2$
and $0 < \theta_1 < \theta_2 < \pi$.  Choose $V = V(r, \theta)$, $W =
W(r, \theta)$ to be smooth functions of their arguments such that (a)\ $V
= 1$ and $W = 0$ whenever $r \not\in (r_1, r_2)$ or $\theta \not\in
(\theta_1, \theta_2)$, (b)\ $V(r_0, \theta_0) < 0$ for some $r_0 \in (r_1,
r_2) $
{and $\theta_0$} $ \in (\theta_1, \theta_2)$, and
(c)\ $W^2 > -V r^2 \sin^2 \theta$
everywhere. Clearly, the orbit of the
asymptotically stationary Killing field $\pt/\pt t$ at $(r_0, \theta_0)$ is
spacelike.  However, no black hole or white hole is present in this
spacetime,
since at every point there exists
some linear combination of the Killing fields $\pt/\pt t$ and $\pt/\pt
\varphi$ which is timelike. (As already noted above, if a black
hole were present, its event
horizon would be null and every Killing field would have to
be tangent to it.
Hence, no Killing field could be timelike on the horizon.)
Spacetimes with this character ({\em i.e.}, possessing an ``ergoregion" but
no black or white hole) which are solutions to the Einstein--perfect--fluid
equations have been numerically constructed by Butterworth and Ipser
\cite{BI}.  Theorem \ref{TM.1} is applicable to such spacetimes.

Another example of a spacetime containing an ergoregion
for which our result for
case (1) is applicable is provided in Figure \ref{F1}.
For the slice $\Sigma$ shown there, there is both a black hole
(block $F$) and a white
hole (block $A$), so this spacetime is {\em not} of class (1).
However, the
union of the blocks $B$,
$D$, and $E$ -- viewed as a spacetime
in its own right -- {\em is} a spacetime of class (1) with respect to
$\Sigma$, despite the fact that the Killing orbits are spacelike
in block $D$.
Thus, Theorem \ref{TM.1}
guarantees existence of a maximal slice extending from $M_1$
to $M_2$.
\begin{figure}

\vspace{10cm}
\caption{
\protect{\label{F1} }
The Penrose diagram of a spacetime which
illustrates some applications of our theorems.
Each point on this Figure represents a two
sphere. The arrows represent the direction and character (spacelike
or timelike) of Killing
orbits. A spherically symmetric,
asymptotically stationary
spacetime, not necessarily satisfying any reasonable
field equations, with the
global structure displayed in this Figure can be easily constructed using the
methods of \protect{\cite{Walker}}.}
\end{figure}
On the other hand, it is worth noting that for the slice $\Sigma'$
of Figure \ref{F1}, the white hole region is block $A$, and the black
hole region includes block $D$. Thus, no subset of this spacetime is
of class (1) for the slice $\Sigma'$ (or for any other slice spanning the
asymptotic regions $M_1$ and $M_3$). Thus, Theorem \ref{TM.1}
cannot be invoked to infer existence of a
maximal slice extending from $M_1$ to $M_3$.

Examples of spacetimes of class (2) include the standard
``non-extremal" stationary black hole solutions, such as the
charged Kerr
solutions with $e^2 + a^2 < m^2$ or the recently discovered
``colored black hole" solutions to the Einstein-Yang-Mills equations.
(Here $\Sigma$ can be taken to be any Cauchy surface for the domain
of outer communications of one asymptotically flat region.)
Note that the charged Kerr solutions with $a \ne
0$ contain an ergoregion
exterior to the black and white holes.
Note also that in the above examples, the intersection surface, $S$,
of the black and white
hole horizons has the topology of a sphere. However, using the
methods of \cite{GH} one can
construct asymptotically stationary spacetimes (not necessarily
satisfying any reasonable field equations
or energy inequalities) such that $S$ has an arbitrary number of
connected components, each of them having arbitrarily specifiable
topology\footnote{The construction of ``toroidal black holes'' given
on p. 682 of \cite{GH} is easily generalized to other topologies. By
construction the Killing orbits in ``opposite'' ends have opposite
time orientation, and the existence of black hole regions follows from
Lemma \ref{Q3a} proved below.}.

Another example of a spacetime of class (2) is provided by
Figure \ref{F1}:
The spacetime shown there is of class (2) with respect to the,
hypersurface $\Sigma''$. Thus,
Theorem \ref{TM.2} guarantees that there
exists a maximal hypersurface with end $M_1$ and boundary $S$, which is
asymptotically flat at
$M_1$ and is asymptotically orthogonal to $X$.

We turn, now, to giving some examples of spacetimes for which our
theorems do {\em not} apply. A simple class of examples of asymptotically
flat, spacetimes with complete Killing field, $X$, timelike
at each end, which contain neither a black
hole nor white hole but fail to be of class (1) can be constructed by
starting with any spacetime of class (1) and removing a Killing orbit
which passes through $\Sigma$. (The compactness requirement on the
``interior portion" of $\Sigma$ then will not be satisfied.) Inextendible
spacetimes with similar properties also may easily be constructed. The
Schwarzschild solution with negative mass provides a good example of
such a spacetime.

A spacetime containing a black hole with respect to one or
more asymptotic regions can fail to be in class
(2) for a variety of reasons.
The following are some
examples of spacetimes which fail to be in class (2)
for any choice of $\Sigma$: (i) Any
spacetime which contains a black hole with respect to one or more
asymptotic regions, but not a white hole. The extended
Schwarzschild spacetime with its white hole region removed provides
a simple example of such a spacetime. (ii) Any spacetime which contains
a black hole and white hole with respect to one or more ends, but the
black and white hole horizons do not intersect. The ``extreme" charged
Kerr solutions (satisfying $e^2 + a^2 = m^2$) and the
Majumdar--Papapetrou black holes ({\em cf.\ e.g.} \cite{HH}) provide examples
of
such spacetimes. Additional examples can be constructed in which
the black hole horizon is a Killing horizon with non-constant surface
gravity. In that case,
R\'acz and Wald \cite{RW} have recently shown that
some of
the generators of the
horizon terminate in a parallelly propagated curvature singularity, and
thus cannot intersect the generators of the horizon of any white hole
that may be present. (iii) Numerous further examples can be constructed
by starting with a spacetime of class (2) and removing suitable Killing
orbits.

Although the above examples make clear
that the assumptions made in case (2)
are rather restrictive
from the mathematical point of view, the following considerations
indicate that these assumptions may not be very
restrictive from the physical point of view. First, if Einstein's
equation holds with matter satisfying
suitable hyperbolic equations and energy conditions, and if
both the matter fields and
the spacetime
are
analytic, then Hawking has argued ({\em cf.\ }Propositions 9.3.5 and 9.3.6 of
\cite{HE}) that each connected component of the event horizon of a black
hole must be a
Killing horizon.  (This implies that either the asymptotically stationary
Killing field  must be normal to the horizon, or there must exist an
additional Killing field in the spacetime.)\ \ Furthermore, if Einstein's
equation with matter satisfying the dominant energy condition holds, then
the surface gravity, $\kappa$, must be constant over each connected
component of the horizon \cite{BCH}: 
However, R\'acz and Wald \cite{RW} have recently shown that if a
spacetime possesses a one-parameter group of isometries with a
Killing horizon such that $\kappa$ is constant and $\kappa \ne
0$, then a neighborhood of the horizon can be smoothly
extended (if necessary) so that the
Killing horizon is a bifurcate horizon. Thus,
these combined results lend plausiblity to the idea
that the only physically relevant stationary black holes
are ones whose event horizon has $\kappa = 0$ ({\em i.e.}, the
``extremal case")
and ones
whose event horizon comprises a portion of a bifurcate Killing horizon.
In the latter case, a white hole must also be present, and, under additional,
presumably mild,
assumptions, the spacetime should be of class (2).

In section \ref{preliminaries} we define precisely the class of spacetimes
we consider, and give some preliminary results.  Some theorems
concerning the structure of these spacetimes are presented in section
\ref{structure}; in particular, relationships are obtained between ``having
no black holes and no white holes" and compactness
of
domains of dependence of compact sets.
Finally, section \ref{existence} contains our theorems on the existence of
maximal hypersurfaces.
In Appendix \ref{rotating} we point out the
existence
of a generalization of our results to space--times which are ``\stationary''.
In Appendix \ref{gluing} we give an alternative proof of one of our
theorems under the additional hypothesis that the ``timelike
convergence condition'' holds.

{\bf Acknowledgements:}  P.T.C.\ acknowledges useful discussions with
R.~Bartnik.
Most of the work on this paper was done when  P.T.C.\ was
visiting the Centre for Mathematics and its Applications of the
Australian National University, Canberra; he wishes to thank
N.\ Trudinger
and the CMA for friendly hospitality.

\section{Preliminary Definitions and Results on Asymptotically 
Stationary Spacetimes}
\label{preliminaries}

Throughout this paper, we will assume that all manifolds are smooth,
connected, Hausdorff, and paracompact, and all spacetimes are time
oriented. For simplicity, unless otherwise specified, only smooth metrics
will be considered, although all the results presented below would hold
under appropropriate finite differentiability conditions.  Unless specified
otherwise, our notation and conventions\footnote{In particular, the
indices $a, b, \ldots\/$ etc.\ are abstract tensor indices, while  latin
indices
$i,j,\ldots\/$ etc.\ are component indices and run from 1 to 3, and greek
indices are component indices and run from 0 to 3.} will follow
\cite{Wa}, with one exception: We shall define the domain of dependence
$\ds$ of an 
achronal set $\Sigma$ as the set of all points $p$ such that every
inextendible timelike (as opposed to causal) curve through $p$ intersects
$\Sigma$. (This agrees with the definition
given in \cite{HE} and
corresponds {when $\Sigma$ is closed}
to the closure of the domain of dependence as defined in
\cite{Wa}.) We define ${\stackrel{o}{\cal D}\!\!(\Sigma)} = {\rm int}
\ds$, where ${\rm int} \Omega$
stands for the interior of a set $\Omega$.

We begin by defining the notion of asymptotic stationarity.  First,
recall that a
{\em hypersurface} is an embedded submanifold (without boundary) of
codimension 1; a {\em hypersurface with boundary} is an embedded
submanifold of codimension 1 with boundary; a {{\em slice} is a}
closed, embedded
submanifold without boundary. 

\begin{Definition}
\label{D1}
A spacetime $(M, g_{ab})$ possessing
an acausal hypersurface $\Sigma$ (possibly with boundary)
expressed in the form {of a disjoint
union} $\Sigma = \Sigma_1 \cup \Sigma'$
will be called
$(k, \alpha)$--asymptotically stationary with respect to the ``end"
$\Sigma_1$ if the following three conditions hold:
\begin{enumerate}
\item
\label{D1.1}
$\Sigma_1$ is diffeomorphic to $\R^3\setminus B(R_1)$, where
$B(R_1)$ is a closed ball of radius $R_1$. There exists a coordinate
system defined on some neighbourhood ${\cal O}_1$ of $\Sigma_1$ in
which the coordinate components, $g_{\mu\nu}$, of the spacetime metric
satisfy, for some $0 < \alpha \leq 1$, $0 < \lambda \le 1$, $k \geq
2$,
\bea
&&g_{00} < -C^{-1}, \quad |g_{\mu\nu} - \eta_{\mu\nu}| \leq C r^{-
\alpha}\label{(D.1)}
\\
&&\forall Z^i\in \R^3\quad g_{ij}Z^iZ^j\ge
C^{-1}\sum_{i=1}^3(Z^i)^2\label{(D.1.0)} \\
&&|\pt_{\sigma_1} \ldots \pt_{\sigma_i} g_{\mu\nu}| \leq C
r^{-\alpha-i} \qquad 1 \leq i \leq k \label{(D.2)}\\
&&|(\pt_{\sigma_1} \ldots \pt_{\sigma_k} g_{\mu\nu})(t,x) -
(\pt_{\sigma_1} \ldots \pt_{\sigma_k} g_{\mu\nu}) (\tau, y)|\nonumber\\
&&\qquad \qquad \qquad \qquad \leq C r^{-\alpha-k-\lambda}(|t-
\tau|^\lambda + |x-y|^\lambda)\label{(D.3)}
\eea
for all $(t,x)$, $(\tau,y)\in {\cal O}_1$ such that $|t-\tau| \leq 1$,
$|x-y| \leq r(x)/2$, and for some constant $C$.
(Here $r \equiv (\sum_{i=1}^3 {x_i}^2)^{\frac{1}{2}}$.)

\item
\label{D1.2}
$\pt \overline{\Sigma}_1 $, where $\overline{\Sigma}_1 $ denotes the
closure of ${\Sigma}_1 $ considered as a subset of $\R^3$, is
identified with a connected component of $\pt \Sigma' $ by a
diffeomorphism.

\item
\label{D1.3}
On $M$, there exists a Killing vector field $X$, the orbits of which are
complete and which is {\em uniformly} timelike on $\Sigma_1$
$(g_{ab}X^aX^b<-C^{-1})$.
Furthermore, in the coordinate system of point \ref{D1.1} we have as $r
\rai$,
\be
X=(A + O(r^{-\alpha}))\frac{\pt}{\pt t}+
O(r^{-\alpha}) \frac{\pt}{\pt x^i}
\label{(D.4)}
\ee
for some constant $A (\ne 0)$.
\end{enumerate}
\end{Definition}

{\bf Remarks}:
\begin{enumerate}
\item
It is not too difficult to show from
(\ref{(D.1)})--(\ref{(D.3)}) and from
\be
\label{3.7.0}
\nabla_\mu \nabla_\nu X_\alpha = R_{\lambda\mu\nu\alpha} X^\lambda
\ee
(this last equation being a well known consequence of the Killing
equations) that there exists constants $A\ne 0$, $B^i$ such that
any Killing
vector which is uniformly timelike for $r \geq R_2$,
for some $R_2 \geq R_1$,
necessarily satisfies, in the coordinate system of definition (\ref{D1}),
\begin{eqnarray}
&&X - A \frac{\pt}{\pt t} - B^i \frac{\pt}{\pt x^i} =
O(r^{-\alpha}),\label{(D.5)}\\
1 \leq i \leq k && \pt_{\sigma_1} \ldots \pt_{\sigma_i}  X^\mu =
O(r^{-\alpha-i}) ,
\label{(D.6)}
\end{eqnarray}
with an obvious uniform weighted H\"older condition satisfied by
$\pt_{\sigma_1} \ldots \pt_{\sigma_k}  X^\mu$.  If $|B|<|A|$
 as must
occur
since $X$ is uniformly timelike on $\Sigma$,
the constants $B^i$ can
then
be set to zero by constructing a coordinate system similar to that of
Proposition \ref{PD.2} below, and subsequently performing an
appropriate Lorentz transformation (``boost").  Thus,
eq.\ (\ref{(D.4)}) involves no loss of generality if $X$ is uniformly
timelike\footnote{If $X$ is merely timelike on $\Sigma_1$, then $X$ could
asymptotically
approach a null vector (so that $|B|=|A|$), in which case it could
not be put in the form eq.\ (\ref{(D.4)}). An example of such a
spacetime is
constructed in the Remark following the proof of Proposition
\ref{APD.2}, Appendix \ref{rotating}. We do not expect there to exist
any metrics with a timelike Killing vector
asymptotically approaching a null vector, and satisfying some
reasonable field equations and/or energy inequalities; however,
it might be of some interest to analyze this question
for vacuum spacetimes.} on $\Sigma_1$.
\item
If (\ref{(D.1)}) and (\ref{(D.2)}) hold with $1 \leq i \leq k+1$, then
(\ref{(D.3)}) will hold as well (with any $0 \leq \lambda \leq 1$).
\item
A generalization of this definition to asymptotically ``\stationary"
spacetimes is given in Appendix \ref{rotating}.
\end{enumerate}

Let $(M, g_{ab})$ with $\Sigma = \Sigma_1 \cup \Sigma'$ be
asymptotically stationary with respect to $\Sigma_1$. We define $M_1$
by
\be
M_1 = \bigcup_{t\in(-\infty, \infty)} \phi[X]_t (\Sigma_1) \label{(D.6a)}
\ee
where throughout this paper $\phi[W]_t$  denotes the
one--parameter group
of isometries generated by a Killing vector field $W$.
The following proposition, which gives
some insight into the structure of asymptotically stationary spacetimes,
will be used throughout (the coordinates the existence of which is claimed
below will be referred to as {\em Killing coordinates based on
$\Sigma\cap M_1
$}):

\begin{Proposition}[Killing time based on $\Sigma\cap M_1
$]
\label{PD.2}
Let $(M, g_{ab})$ with $\Sigma = \Sigma_1 \cup \Sigma'$ be
asymptotically stationary with respect to $\Sigma_1$. Then there exists a
global coordinate system on $M_1$ such that $M_1 \approx \R \times
(\R^3 \setminus B(R_1))$ with $\Sigma_1=\{x^0=0\}$, and
\be
g_{00} < -C^{-1} , \qquad \frac{\pt g_{\mu\nu}}{\pt t}
= 0
\label{(D.7)}
\ee
\begin{eqnarray}
\forall Z^i\in \R^3\quad g_{ij}Z^iZ^j&\ge&
C^{-1}\sum_{i=1}^3(Z^i)^2\label{(D.7.0)} \\
|\pt_{\sigma_1} \ldots \pt_{\sigma_i} (g_{\mu\nu}- \eta_{\mu\nu})| &\leq &C
r^{-\alpha-i} \qquad 0 \leq i \leq k \label{(D.9)}
\end{eqnarray}
for some constant $C$; 
moreover, $\pt_{\sigma_1} \ldots \pt_{\sigma_k}
g_{\mu\nu}$ satisfy an obvious weighted H\"older condition.
\end{Proposition}

A generalization of Proposition \ref{PD.2} (Proposition \ref{APD.2}) is
stated and proven in
Appendix \ref{Killing}.
It should be noted that, as shown in the proof of Proposition
\ref{APD.2} given in Appendix \ref{Killing}, one does {\em not} lose
differentiability and/or uniform
decay bounds of the derivatives of the metric when going to the Killing
coordinates.

We now define the two classes of spacetimes which will be analyzed in
this paper.  These two classes correspond roughly to the two classes of
Einstein--Yang--Mills solutions considered in \cite{SW}.

\begin{Definition}[Spacetimes of class (a)] \label{(a)}
$(M, g_{ab})$ is a strongly causal spacetime which contains a connected,
acausal slice $\Sigma$ expressible as a disjoint union in the form $\Sigma
= \bigcup_{i=1}^I\, \Sigma_i \cup \Sigma_{\rm int}$, where
$\Sigma_{\rm int}$ is compact (with boundary). On $M$ there exists a
Killing vector field $X$ such that $(M, g_{ab})$ is
$(k, \alpha)$--asymptotically stationary with respect to each $\Sigma_i$.
\end{Definition}

Let us emphasize that the hypersurface $\Sigma$ above is closed
and has {\em no boundary} (as opposed to the case (b) defined below in
which
$\Sigma$ has a boundary).
It should be stressed that we assume that $X$ approaches (a multiple
of) $\partial /\partial t$ in {\em all} the asymptotic ends $\Sigma_i$.
Thus, when more then one Killing vector and more than one end are
present, we do not
allow for situations in which
one combination of Killing vectors is asymptotic to $\partial
/\partial t$ in one end, and a different combination of Killing
vectors has  this property
in another end. Note also that, under the conditions of Definition
\ref{(a)}, $\Sigma$ is
necessarily a complete Riemannian manifold with respect to the metric
induced from $(M, g_{ab})$.

\begin{Definition}[Spacetimes of class (b)] \label{(b)}
$(M, g_{ab})$ is a strongly causal spacetime which contains a connected,
closed, acausal hypersurface
$\Sigma$ with boundary $S$, such that $\Sigma$ is
expressible as a disjoint union in the form $\Sigma
= \bigcup_{i=1}^I\, \Sigma_i \cup \Sigma_{\rm int}$, where
$\Sigma_{\rm int}$ is compact (with boundary).
On $M$ there exists a
Killing vector field $X$ such that $(M, g_{ab})$ is
$(k, \alpha)$--asymptotically stationary with respect to each $\Sigma_i$,
and $X$ is tangent to $S$ (so that the isometries, $\phi[X]_t$,
map $S$ into $S$).
\end{Definition}

Note that we do not assume that $S$ is connected. Note also
that under the conditions of Definition \ref{(b)}
$\Sigma$ is necessarily a complete Riemannian manifold with boundary
$S$.

By a standard theorem ({\em cf., e.g.}, \cite{HE,Wa}),
the future Cauchy horizon, $H^+$, of any closed, achronal set is
generated by null geodesics
which either are past inextendible or have past endpoint on the edge
of that set. For spacetimes of class (a), $\Sigma$ is edgeless. However,
for spacetimes of class (b) the edge of $\Sigma$ is $S$. For spacetimes
of class (b), we define
${\cal H}_+$ to consist of the portion of $H^+$
which is generated by null geodesics with
past endpoint on $S$. Similarly,
we define ${\cal H}_-$ to consist of the portion of the of the past Cauchy
horizon, $H^-$, of $\Sigma$ generated by null
geodesics with future endpoint
on $S$. Since $I^+(S)$ and $I^-(S)$ do not lie in $\ds$ and since the
boundary, $\pt \ds$, of $\ds$ is the
entire Cauchy horizon, $\pt \ds = H = H^+ \cup H^-$,
({\em cf., e.g.}, \cite{HE,Wa}),
 it follows that
$$
{\cal H}_+ = J^+(S) \cap \ds,
\qquad {\cal H}_- =
J^-(S) \cap \ds\ .
$$
We define $${\cal H} = {\cal H}_+ \cup {\cal H}_-\ .$$ In our analysis
of spacetimes of class (b) given in the next section, we will focus our
attention on the manifold with boundary, $$M' = {\stackrel{o}{\cal
D}\!\!(\Sigma)} \cup {\cal H}\ .$$
It should
be stressed that ${\cal H}$ may comprise only a portion of the boundary
of $\ds$, as illustrated in Figure \ref{F2}.
\begin{figure}
\vspace{8cm}
\caption{\label{F2} A Penrose diagram of a spacetime in which $\cal H$
is not the entire Cauchy horizon of $\Sigma$. In this spacetime
$\partial \ds$ has two connected components, ${\cal
H}={\cal H}_+\cup{\cal H}_-$
and ${\cal H}^\prime$.}
\end{figure}

For spacetimes of class (a) and (b), we define
\be
M_{\rm ext} = \bigcup_{i=1}^I M_i \ ,
\label{(D.10)}
\ee
where $M_i = \bigcup_{t\in\R} \phi[X]_t(\Sigma_i)$.
and we set
\[
M_{\rm int} =M \setminus M_{\rm ext} \ .
\]
Note that by construction we have $\phi[X]_t [M_{\rm ext}] = M_{\rm ext}$
for all $t$, whence  $\phi[X]_t [M_{\rm int}] = M_{\rm int}$.

In both cases (a) and (b), for an asymptotic end $M_i$ we define ``the
black hole region with respect to $M_i$" by
\be
\label{(D.12.0)}
{\cal B}_i = M\setminus I^-(M_i)\,,
\ee
with the ``white hole region with respect to $M_i$" defined dually,
\be
\label{(D.12.1)}
{\cal W}_i = M \setminus I^+(M_i)\,.
\ee
It is clear that (\ref{(D.12.0)})--(\ref{(D.12.1)}) are equivalent to
the standard
definition\footnote{It is customary to define the black hole,
respectively the white hole, with
respect to the asymptotic end  $M_i$ as $M\setminus J^-({\cal
I}^+;\tilde M)$, respectively $M\setminus J^+({\cal
I}^-;\tilde M)$, where $J^\pm(\Omega;\tilde M)$ denotes the causal
past and future of the set $\Omega$ in the conformally extended
manifold $\tilde M$.} using ${\cal I}$ \cite{Wa} \cite{HE}, whenever
${\cal I}$,
understood as a conformal completion based on a Bondi coordinate
system, exists\footnote{More generally, the equivalence of
(\ref{(D.12.0)})--(\ref{(D.12.1)}) with the standard definition can be
established whenever the manifold $M_i$ admits a ${\cal I}$ satisfying the
conditions of \cite{GHor}, and the Ricci tensor falls off fast
enough (in the sense of the note added in proof (3) of \cite{AX}) in
the asymptotic end in question. The advantage of
(\ref{(D.12.0)})--(\ref{(D.12.1)}) is, that one avoids all the issues
related to the possible low differentiability and/or incompleteness of
${\cal I}$. On the other hand the definition given here does not
generalize in any
obvious way to non--stationary
space--times. It should
be mentioned that if $M_i$ is assumed to be vacuum, then a smooth
${\cal I}$ satisfying the requirements of \cite{GHor} necessarily
exists ({\em cf.\ }\cite{DS}[Appendix], and also
\cite{BeigSimon1,BeigSimon2}).}.

We define the black hole region,
${\cal B}$,
of the spacetime by,
\be
\label{(D.12)}
{\cal B} = M \setminus I^- (M_{\rm ext}) = \bigcap_{i=1}^I {\cal B}_i
\ ,
\ee
and
similarly define the white hole region by,
\be
\label{(D.13)}
{\cal W} = M \setminus I^+ (M_{\rm ext}) = \bigcap_{i=1}^I {\cal
W}_i\,.
\ee
As already mentioned in the Introduction, it should be emphasized
that it is possible for spacetimes of class (a) to have
${\cal B}_i \ne \emptyset$ for some $i$ and yet have ${\cal B} =
\emptyset$. The union of blocks $B$, $D$, and $E$ of Figure
\ref{F1} provides an illustration of such a spacetime. On the other
hand,  in case (b), since $\phi[X]_t[S] =
S$, it follows from Corollary \ref{Q1b}, proven in the next section, that
${\cal B}, {\cal W} \ne \emptyset$, and, indeed, $J^+(S) \subset {\cal
B}$, $J^- (S) \subset {\cal W}$. Finally, it should be mentioned that
Proposition \ref{invariantdefinition} below
shows that ${\cal B}_i$ and
${\cal W}_i$ are independent of that arbitrariness in the definition of
$M_{\rm ext}$, which is related to the
arbitrariness  of separation of $\Sigma$ into $\Sigma_{\rm ext}$ and
$\Sigma_{\rm int}$.

Since $M_{\rm ext}$ is $\phi[X]_s$ invariant, so are $I^{\pm}(M_{\rm
ext})$, we thus have
\be
\label{(D.14)}
\forall \; t \in \R \quad \phi[X]_t({\cal B}) = {\cal B}\,, \quad
\phi[X]_t({\cal W}) = {\cal W}\,.
\ee
It follows immediately
that the boundaries of ${\cal B}$ and ${\cal W}$ -- known as the
{\em event horizons} of ${\cal B}$ and ${\cal W}$, respectively, are
null surfaces which are
invariant under $\phi[X]_t$. As already mentioned in the Introduction,
this implies that if the Killing field $X$ is strictly timelike
in $M$, then ${\cal
B} = {\cal W} = \emptyset$. Furthermore, as shown explicitly in the
Introduction, in case (a), it is possible to have ${\cal B} = {\cal W} =
\emptyset$ even though $X$ becomes spacelike in certain regions (called
ergoregions) of $M$.  Similarly, as explicitly demonstrated by the example
of the Kerr metric, in case (b), $X$ may be spacelike within
${\stackrel{o}{\cal D}\!\!(\Sigma)}$ and yet have
${\stackrel{o}{\cal D}\!\!(\Sigma)}\cap {\cal B} = \emptyset$, $
{\stackrel{o}{\cal D}\!\!(\Sigma)}\cap {\cal W} = \emptyset$.

In this paper, we shall be concerned with
the issue of the existence of asymptotically flat,
maximal hypersurfaces and foliations by maximal hypersurfaces
in spacetimes of class (a) and (b). The notion of
asymptotic flatness of a hypersurface is defined as follows:

\begin{Definition}
\label{D0}
Let $(M, g_{ab})$ be a
space-time in which (\ref{(D.1)})--(\ref{(D.3)}) hold.
A hypersurface $\Sigma \subset M$ will be called
asymptotically flat if $\Sigma\cap{\cal O}_1$ (${\cal O}_1$ -- as in point
\ref{D1.1} of Definition \ref{D1}) is a graph $t = u(x)$ of a function
$u$ satisfying
\[|u| + r |\pt_i u| + \ldots + r^{k+1} |\pt_{i_1} \ldots \pt_{i_{k+1}} u| \leq
\left\{
\begin{array}{ll}
C (1 + r)^{1-\alpha}, & 0 < \alpha < 1\\
C (1 + \ln(1+r)), & \alpha = 1
\end{array}
\right. \]
(together with an appropriate weighted H\"older condition on $\pt_{i_1}
\ldots \pt_{i_{k+1}} u$).
\end{Definition}

By a common abuse of terminology, we shall say that a spacelike
hypersurface $\Sigma$ is maximal if the trace of the extrinsic curvature
of $\Sigma$, perhaps defined distributionally, vanishes.  It follows that a
maximal hypersurface is locally maximal in a variational sense, {\em cf.\
e.g.} \cite{Ba2}.

Let $t$ be a time function on $M$, {\em i.e.}, a continuous function
strictly increasing along any future directed causal curve.  We shall use
the notation
$$\Sigma_s(t) = \{p \in M : t(p) = s\}$$
 for the level sets of $t$.
The sets $\Sigma_s(t)$ are closed, and if $t \in C^k(M)$, $k\ge1$, with
$dt$ non--vanishing on an open subset ${\cal O}$ of $M$, then
$\Sigma_s(t) \cap {\cal O}$ are $C^k$
acausal submanifolds
of $M$.

Let $\Sigma_t$, $t \in I$, where $I \subset \R$, be a family of
hypersurfaces.  We shall say that $\Sigma_t$ foliates ${\cal O}$ if there
exists a function $t$ on ${\cal O}$ such that $\Sigma_s = \Sigma_s(t)$.
Note that this implies in particular that ${\cal O} = \bigcup_{t\in I}
\Sigma_t$, and that the implication $\Big(t \ne t'\Big) \Longrightarrow
\Big(\Sigma_t \cap \Sigma_{t'} = \emptyset\Big)$ holds.   If $\Sigma_t$
foliate ${\cal O}$, then the family $\Sigma_t$ will be called a foliation
(of ${\cal O}$).  The function $t$ will be said associated to the foliation
$\Sigma_t$.  If $dt$ is nowhere vanishing on ${\cal O}$ and $I$ is
connected, with $0 \in I$, then $t$ induces a natural diffeomorphism
$\varphi : {\cal O} \longrightarrow I\times \Sigma_0 $ by setting
$\varphi(p) = (t(p),q(p))$, where $q(p)$ is the intersection of $\Sigma_0$
with the integral curve of $\nabla t$ through $p$.

We conclude this section by introducing some terminology to be used in
the next section ({\em cf.} \cite{Hartman} [Chapter VII]).

\begin{Definition}
\label{D5}
Let $M$ be a manifold, let $\psi_t  :  M \ra M$ denote a one--parameter
group of diffeomorphisms and let $\gamma_p$ denote the orbit of
$\psi_t$ through $p \in M$.  The $\alpha$--limit set $\alpha_p$, of
$\gamma_p$ is defined as the set of all $q \in M$ such that $q$ can be
expressed as $q = \lim_{i\rai} \psi_{t_i}(p)$
for some ${t_i} \ra -\infty$. The $\omega$--limit set, $\omega_\sigma$,
of $\gamma_p$ is defined similarly, except that $\{t_i\}$ is required to
diverge to $+\infty$.
\end{Definition}

\section{Causal Properties of Spacetimes of Classes (a) and (b)}
\label{structure}
In this section we establish some properties of the spacetimes introduced
in the previous section.  Since our results and proofs for case (b) are
closely analogous to those of case (a), in general, we will merely state the
results for case (b) without proof, or will simply indicate the
modifications to the proof for case (a).

Our first result is the following:

\begin{Proposition}
\label{Q1a}
Let $(M, g_{ab})$ be a spacetime of class (a), with slice $\Sigma$.
Then $\ds$ contains complete Killing orbits, {\em i.e.}, if $p \in \ds$
then $\phi[X]_t(p) \in \ds$ for all $t \in \R$.  Moreover $\phi[X]_t(\Sigma)$
are
Cauchy surfaces for ${\stackrel{o}{\cal D}\!\!(\Sigma)}$, for all $t\in \R$.
\end{Proposition}

{\bf Proof}:  We wish to show that $\phi[X]_t [\ds] = \ds$ for all $t \in \R$.
Let
\begin{equation}
\label{(Q.1)}
\Sigma_t = \phi[X]_t[\Sigma]\,.
\end{equation}
Since the $\phi[X]_t$'s are isometries we have $\phi[X]_t[\ds]  = {\cal D}
(\Sigma_t)$, and the proposition will be proven by showing that ${\cal
D}(\Sigma_t) = \ds$.

Let $\gamma\subset{\stackrel{o}{\cal D}\!\!(\Sigma)}$ be an inextendible (in
${\stackrel{o}{\cal D}\!\!(\Sigma)}$), piecewise
differentiable {timelike}
curve, and define
\[
I=\{t\in\R:\phi[X]_t(\Sigma)\cap\gamma\ne\emptyset\}\ .
\]
Since $\gamma$ is a {timelike}
curve and all the $\phi[X]_t(\Sigma)$'s are
spacelike it follows that $I$ is open. From the fact that $\Sigma$ is a
Cauchy surface for ${\stackrel{o}{\cal D}\!\!(\Sigma)}$ we have $0\in I$, thus
$I\ne\emptyset$. Let
$t_i\in I$ be such that $t_i \ra t\in\R$, let $q_i=\phi[X]_{-t_i}
\Big(\gamma\cap\phi[X]_{t_i}(\Sigma)\Big) \in\Sigma$.  Asymptotic flatness
({\em cf.} Proposition \ref{PD.2}) implies that $q_i\in K$ for some
compact subset of $\Sigma$. Passing to a subsequence, still denoted by
$t_i$, there exists $q\in K$ such that $q_i \ra q$. Continuity of
$\phi[X]_s(p)$ with respect to $s$ and $p$ imply that the sequence
$\phi[X]_{t_i}(q_i)=\gamma\cap \phi[X]_{t_i}(\Sigma )$ accumulates at
$\phi[X]_t(q)$, thus $\gamma$ accumulates at $\phi[X]_t(q)$ and
inextendability of $\gamma$ implies that $t\in I$. Thus $I=\R$ and the
result follows. \hfill$\Box$

By similar arguments, we obtain the following Proposition for case (b).
(The only modifications needed to the proof concerns analysis of the
possibility that the accumulation point $x$ might lie on
$S$; in that case, the invariance of $S$ under the isometries
should be used.)

\begin{Proposition}
\label{Q1b}
Let $(M, g_{ab})$ be a spacetime of class (b).  Then for $p \in \ds$,
respectively $p \in {\stackrel{o}{\cal D}\!\!(\Sigma)}$,
$t \in \R$, we have $\phi[X]_t(p) \in \ds$, respectively $\phi[X]_t(p) \in
{\stackrel{o}{\cal D}\!\!(\Sigma)}$.
Moreover the hypersurfaces $\phi[X]_t(\Sigma\setminus S)$ are Cauchy
surfaces for ${\stackrel{o}{\cal D}\!\!(\Sigma)}$, and ${\cal
D}[\phi[X]_t(\Sigma)] = \ds$.
\end{Proposition}

As a corollary of Propositions \ref{Q1a} and \ref{Q1b} we have
the following:

\begin{Corollary}
\label{NEW1}
Let $(M, g_{ab})$ be a spacetime of class (a) or (b).
Then $M \setminus {\stackrel{o}{\cal D}\!\!(\Sigma)} \subset {\cal B} \cup
{\cal W}$.
In particular, we have
 $M_{\rm ext}\subset {\stackrel{o}{\cal D}\!\!(\Sigma)}$.
\end{Corollary}

{\bf Proof}: That $M_{\rm ext}\subset {\stackrel{o}{\cal D}\!\!(\Sigma)}$
follows immediately from
the definition of $M_{\rm ext}$ and Propositions \ref{Q1a} and \ref{Q1b}.
Now, suppose that $p \not\in {\cal B} \cup {\cal W}$ and
$p \not\in {\stackrel{o}{\cal D}\!\!(\Sigma)}$. Then there exist
$q_-, q_+ \in M_{\rm ext}$ such that $p \in I^-(q_+) \cap I^+(q_-)$.
Without loss of generality, we may assume that
$q_- \in I^-[\Sigma]$ and $q_+ \in I^+[\Sigma]$.
Since
$p \not\in {\stackrel{o}{\cal D}\!\!(\Sigma)}$, there exists $p' \in I^-(q_+)
\cap I^+(q_-)$ such that
$p' \not\in \ds$.
Let  ${\lambda'}_-$ be a
future directed timelike curve which connects $q_-$ to $p'$,
and let ${\lambda'}_+$ be a past directed timelike curve which
connects $q_+$ to $p'$. Since $\Sigma$ is achronal, ${\lambda'}_-$
and ${\lambda'}_+$ cannot both intersect $\Sigma$. Suppose
that ${\lambda'}_-$ fails to intersect $\Sigma$. Then since
$p' \not\in \ds$, there exists a future inextendible timelike curve
starting at $p'$ which fails to intersect $\Sigma$. By adjoining this
curve to ${\lambda'}_-$, we obtain a future inextendible timelike
curve from $q_- \in I^-[\Sigma]$ which fails to intersect $\Sigma$.
This contradicts the fact that $q_- \in M_{\rm ext} \subset {\stackrel{o}{\cal
D}\!\!(\Sigma)}$. A
similar argument yields a contradiction if ${\lambda'}_+$ fails to
intersect $\Sigma$.
\hfill$\Box$

For some arguments it will be convenient to have global hyperbolicity
of the objects at hand (but it should be stressed that we are {\em
not}
assuming global hyperbolicity unless indicated otherwise).
The main significance of Proposition \ref{Q1a} is that it will have the
effect of allowing us to restrict consideration in case (a) to globally
hyperbolic spacetimes for which the hypersurface $\Sigma$ is a Cauchy
surface.  Namely, in case (a), since $\ds$ is invariant under the isometries
$\phi[X]_t$, so is $M' = {\stackrel{o}{\cal D}\!\!(\Sigma)}$.  Hence, the
spacetime $(M',
g_{ab})$ also is of class (a) (with the same asymptotic regions) but is
globally hyperbolic with Cauchy surface $\Sigma$.  Existence of a
maximal hypersurface in $(M', g_{ab})$ implies existence of a maximal
hypersurface in $(M, g_{ab})$ and existence of a maximal foliation of
the spacetime $(M', g_{ab})$ implies existence of a maximal foliation of
a subset of $(M, g_{ab})$ which covers the entire asymptotic region.
Similarly, in case (b) Proposition \ref{Q1b} will have the effect of
allowing us to restrict consideration to $M' = {\stackrel{o}{\cal
D}\!\!(\Sigma)} \cup {\cal H}$.

Most of the remainder of this section will be devoted to proving the
equivalence of a ``no black or white hole" condition to a compactness
condition.  In preparation for this, we prove the following lemmas:

\begin{Lemma}\label{Q2}
Let $(M, g_{ab})$ be a spacetime of class (a) or (b) and let $Q \subset
M$ be any isometry invariant subset, {\em i.e.}, $\phi[X]_t[Q] = Q$.  Let
$M_i$ denote any ``end" ({\em cf.} eq.\ {(\ref{(D.6a)}))}.  Then either
$I^+(Q) \supset M_i$  or $I^+(Q) \cap M_i = \emptyset$.  Similarly
either $I^-(Q) \supset M_i$ or $I^-(Q) \cap M_i = \emptyset$.
\end{Lemma}

{\bf Proof:}  Suppose $I^+(Q) \cap M_i \ne \emptyset$.  Let $x \in
I^+(Q) \cap M_i$.  Since $\phi[X]_t[I^+(Q)] = I^+[\phi[X]_t(Q)] = I^+(Q)$ and
$\phi[X]_t(M_i) = M_i$, it follows that $\phi[X]_t(x) \in I^+(Q) \cap M_i$ for
all $t \in \R$, {\em i.e.}, $\gamma_x \subset I^+(Q) \cap M_i$ where
$\gamma_x$ denotes the orbit of $x$ under $\phi[X]_t$.  However, from the
asymptotic form of the metric given by Proposition \ref{PD.2}, it
follows that given any $x,y \in M_i$, there exists a sufficiently large
negative $t$ such that  $y \in I^+[\phi[X]_t(x)]$.  This implies that  $M_i
\subset I^+(\gamma_x) \subset I^+(Q)$.  \hfill $\Box$

We note the following Corollaries ({\em cf.} also \cite{Carter}
[Corollary, p.\ 136]):
\begin{Corollary}
\label{CQ2}
Let $(M, g_{ab})$ be a space-time of class (a), let $Q$ be a compact
set such that $\phi[X]_t(Q) = Q$, $t \in \R$.  Then
we have
${\cal B}\cup {\cal W} \ne \emptyset$.  If moreover $Q \subset
{\stackrel{o}{\cal D}\!\!(\Sigma)}$ then ${\cal B} \cap {\cal W}$ is nonempty,
with $Q \subset {\cal B}
\cap {\cal W}$. In particular if  $p \in {\stackrel{o}{\cal D}\!\!(\Sigma)}$
and
$X|_p=0$, then $p\in{\cal B} \cap {\cal
W}$.
\end{Corollary}

{\bf Proof}:  Suppose that $Q \subset {\stackrel{o}{\cal D}\!\!(\Sigma)}$.
Since $ {\stackrel{o}{\cal D}\!\!(\Sigma)}$ is globally hyperbolic and $Q$ is
also compact as a subset of
${\stackrel{o}{\cal D}\!\!(\Sigma)}$ the inclusions $M_i \subset I^+(Q)$, $M_i
\subset I^-(Q)$
are not possible ({\em cf.\ e.g.} \cite{Geroch}). Thus Lemma \ref{Q2}
gives $M_i \cap I^+(Q) = M_i \cap I^-(Q) = \emptyset$, $i = 1,\ldots,I$,
which implies $Q \subset {\cal B} \cap {\cal W}$.  If on the other hand
$Q \not\subset {\stackrel{o}{\cal D}\!\!(\Sigma)}$, then $M \setminus
{\stackrel{o}{\cal D}\!\!(\Sigma)} \ne
\emptyset$ and {${\cal B}\cup {\cal W} \ne \emptyset$}
follows from {Corollary \ref{NEW1}}.

\begin{Corollary}
\label{CQ3}
Let $(M, g_{ab})$ be a spacetime of class (b).  Then $S \subset {\cal B}
\cap {\cal W}$.
\end{Corollary}

{\bf Proof}:  We have $\phi[X]_t [S] = S$, but $I^+(S) \cap \Sigma_{\rm ext}
= I^-
(S) \cap \Sigma_{\rm ext} = \emptyset$ since $\Sigma$ is achronal. \hfill$\Box$

Before stating our final corollary of lemma \ref{Q2}, we introduce the
following terminology:

\begin{Definition}
\label{definition}
For $p\in {\stackrel{o}{\cal D}\!\!(\Sigma)}$ we shall say that an orbit
$\gamma_p(t)\equiv\phi[X]_t(p)$
is {\em future oriented}, respectively {\em past oriented},
if there exists an increasing
sequence $t_i\in \R$, $i\in{\bf Z}$, with $|t_i|
\longrightarrow_{|i|\rightarrow \infty}\infty$
such that $\phi[X]_{t_{i+1}}(p)\in I^+(\phi[X]_{t_{i}}(p))$, respectively
$\phi[X]_{t_{i+1}}(p)\in I^-(\phi[X]_{t_{i}}(p))$.
An orbit will be said {\em time oriented} if it is future or past oriented.
\end{Definition}

If $X_p$ is timelike future pointing, respectively past pointing, then
$\gamma_p(t)$ is clearly future oriented, respectively past oriented.
It should be noted, that if there
exists $p$ and $\Delta t \in \R$ such that
$\phi[X]_{\Delta t}(p)\in I^+(p)$, then the orbit $\gamma_p$ of
$\phi[X]$ through
$p$ is necessarily time oriented, for then for any $q\in\gamma_p$ we
have $\phi[X]_{\Delta t}(q)\in I^+(q)$, and the desired sequence is
given by $i\Delta t$, $i\in ${\bf Z}.

\begin{Corollary}
\label{W1}
Let $(M, g_{ab})$ be a spacetime of class (a) or (b) such the orbits of $X$
are future oriented at each ``end", $M_i$. Let $p \not\in{\cal W}$.
Then,  there exists $T \in \R$
such that for all $t > T$, we have $\phi[X]_t [p] \in {}
I^+[\Sigma_{\rm ext}]$.
Similarly, if
$p \not\in{\cal B}$, then there exists $T' \in \R$
such that for all $t < T'$, we have $\phi[X]_t [p] \in I^-[\Sigma_{\rm ext}]$.
Furthermore, {if $p \not\in {\cal B} \cup {\cal W}$}, then
the set $\{t \in \R : \phi[X]_t(p) \in \Sigma\}$ is nonempty and compact.
\end{Corollary}

{\bf Proof}:  If $p \not\in{\cal W}$, then by lemma \ref{Q2},
the past of the Killing orbit through $p$ must contain at least one ``end",
$M_i$. Consequently, there exists $T \in \R$
such that $\phi[X]_T [p] \in I^+[\Sigma_{\rm ext}]$. The
first claim of the
corollary then follows immediately by applying $\phi[X]_{t'}$ to this
relation, for any $t' > 0$, using the fact that for $t' > 0$, $\phi[X]_{t'}$
takes $\Sigma_{\rm ext}$ into $I^+[\Sigma_{\rm ext}]$. Similar arguments
establish the second claim. Finally, if $p \not\in {\cal B} \cup {\cal W}$,
then $\{t \in \R : \phi[X]_t(p) \in \Sigma\}$ is a closed subset of
$[T',T]$, and hence is compact. To prove that this set is
nonempty, we note that $\phi[X]_t$ enters both
$I^-[\Sigma]$ and $I^+[\Sigma]$. Furthermore, {by Corollary \ref{NEW1},
$p \in \ods$, and by Propositions \ref{Q1a}
and \ref{Q1b}, $\phi[X]_t \subset {\stackrel{o}{\cal D}\!\!(\Sigma)}$.}
However, ${\stackrel{o}{\cal D}\!\!(\Sigma)}$
is globally hyperbolic, so it can be expressed as the disjoint union
${\stackrel{o}{\cal D}\!\!(\Sigma)} = [I^-[\Sigma] \cap {\stackrel{o}{\cal
D}\!\!(\Sigma)}] \cup [\Sigma \cap {\stackrel{o}{\cal D}\!\!(\Sigma)}] \cup
[I^+[\Sigma] \cap {\stackrel{o}{\cal D}\!\!(\Sigma)}]$. Hence, it follows that
a continuous curve
in ${\stackrel{o}{\cal D}\!\!(\Sigma)}$ which enters both $I^-[\Sigma]$ and
$I^-[\Sigma]$ must
intersect $\Sigma$.
\hfill$\Box$

The following Proposition gives an invariant characterization of
${\cal B}_i$ and ${\cal W}_i$:

\begin{Proposition}
\label{invariantdefinition}
Let ${\cal O}_i$ be the connected component of the set
$\{p\in {\stackrel{o}{\cal D}\!\!(\Sigma)}:$ the orbit
$\gamma_p$ through $p$ is time oriented$\}$ which contains $M_i$. Then
${\cal B}_i=M\setminus I^-({\cal O}_i)$, ${\cal
W}_i=M\setminus I^+({\cal O}_i)$.
\end{Proposition}

{\bf Proof:}  By definition $M_i \subset{\cal O}_i$, thus  $M\setminus
I^-({\cal O}_i)\subset {\cal B}_i$. Now $\partial (M\setminus
I^-(M_i))\cap{\cal O}_i$ is an achronal subset of ${\cal O}_i$ which is
invariant under $\phi[X]_t$ by construction. Because all the orbits of
$X$ in ${\cal O}_i$ are time oriented, the only such
subset of ${\cal O}_i$ is the empty set, and  the reverse inclusion follows.
\hfill $\Box$

\begin{Lemma}
\label{Q3a}
Let $(M, g_{ab})$ be a spacetime of class (a) such that the orbits of
$\phi[X]_t$ fail to have the same (past or future) orientation on all ends,
{\em i.e.} suppose $A_iA_j < 0$  for some $i$ and $j$, where $A_i$ is
the constant $A$ of Definition \ref{D1} ({\em cf.} (\ref{(D.4)})) in the
region $M_i$.  Then ${\cal B} \cup {\cal W} \ne \emptyset$, where
${\cal B}$ and ${\cal W}$ were defined by eqs. (\ref{(D.12)}) and
(\ref{(D.13)}).  Moreover 
$I^+(M_i) \cap M_j = \emptyset$, $I^- (M_i) \cap M_j = \emptyset$.
\end{Lemma}

{\bf Remark}:  Note that the last part of Lemma \ref{Q3a} says, that two
ends with different Killing--time orientations cannot communicate with
each other (in fact, the argument here shows the stronger statement that
if the orbit $\gamma_p$ has  time orientation which is opposite to that of
the orbit $\gamma_q$, then $I^+(p)\cap I^-(q)=I^-(p)\cap I^+(q) =
\emptyset$), while the first part says that, reversing time orientation if
necessary, there exists a region which cannot {send a signal to}
any of
the ends $M_i$.

{\bf Proof}:  Let $M_+ =\bigcup_{A_i>0} M_i$ and let $M_- =
\bigcup_{A_i<0} M_i$.  We show first that $I^+(M_-) \cap M_+ =
\emptyset$.  Namely, suppose that $x \in M_+$  and $y \in M_-$ were
such that $x \in I^+(y)$.  By choosing $t$ sufficiently large {negative},
 we can
ensure that $\phi[X]_t(x) \in {I^-}[\Sigma]$,
$\phi[X]_t(y) \in {I^+}[\Sigma]$.
However, by isometry invariance, we have $\phi[X]_t(x) \in I^+[\phi[X]_t(y)]$,
which contradicts the achronality of $\Sigma$.  Similarly, we have $I^-
(M_-) \cap M_+ = \emptyset$ and the same relations with $M_+$  and
$M_-$ interchanged.

Now, $I^+(M_+)$ and $I^+(M_-)$ are open sets.  If their union failed
to cover $M$, then ${\cal W} \ne \emptyset$.  If their union covers $M$,
connectedness of $M$ implies that they cannot be disjoint.  Let $x \in
I^+(M_+) \cap I^+(M_-)$.  Then $I^+(x) \subset I^+(M_+) \cap I^+(M_-
)$ and hence $I^+(x) \cap M_- = \emptyset$, $I^+(x) \cap M_+ =
\emptyset$.  Thus, $I^+(x) \cap M_{\rm ext} = \emptyset$, {\em i.e.}
$x \in {\cal B}$, so ${\cal B} \ne \emptyset$. \hfill$\Box$

Similarly one proves:

\begin{Lemma}
\label{Q3b}
Let $(M, g_{ab})$ be a spacetime of class (b) such that the orbits of
$\phi[X]_t$ fail to have the same (past or future) orientation on all ends.
Then
\[
({\cal B} \cup {\cal W}) \cap 
{\stackrel{o}{\cal D}\!\!(\Sigma)} \ne \emptyset\,.
\]
\end{Lemma}

\begin{Lemma}
\label{Q4}
Let $(M, g_{ab})$ be a spacetime of class (a) or (b)
with $A_i > 0$ for all $i$.  Let  $p \in {\stackrel{o}{\cal D}\!\!(\Sigma)}$
and let
$\omega_p$, respectively $\alpha_p$, denote the $\omega$--limit set,
respectively the $\alpha$--limit set of the orbit $\phi[X]_t(p)$.  Then
$\omega_p \subset {\cal B}$, $\alpha_p\subset {\cal W}$.
\end{Lemma}

{\bf Proof}:  By duality it is sufficient to prove $\omega_p \subset
{\cal B}$. It follows immediately from definition \ref{D5} that
$\phi[X]_t[\omega_p] = \omega_p$.  Hence, by lemma \ref{Q2} it suffices to
show that $I^+(\omega_p) \not\supset M_{\rm ext}$.  However, since
$A_i > 0$, for all $t > 0$ we have  $I^+(p) \cap M_{\rm ext} \supset
\phi[X]_t[I^+(p) \cap M_{\rm ext}] = I^+[\phi[X]_t(p)] \cap M_{\rm ext}$,
from which it follows that $I^+(p) \cap M_{\rm ext} \supset
I^+(\omega_p) \cap M_{\rm ext}$.  However, since $p \in {\stackrel{o}{\cal
D}\!\!(\Sigma)}$,
$J^+(p)$ intersects $\Sigma$ compactly ({\em cf.\ e.g.} \cite{Geroch}),
so there exists $y \in \Sigma_{\rm ext} \subset M_{\rm ext}$ such that $y
\not\in I^+(p)$.  Thus, $I^+(\omega_p) \not\supset M_{\rm ext}$, as we
desired to show. \hfill$\Box$

Finally, in the course of proving the second half of theorem \ref{Q6a},
we shall appeal to the following lemma:

\begin{Lemma}
\label{Q5}
Let $(M, g_{ab})$ be a spacetime of class (a), and suppose that ${\cal
D}(\Sigma_{\rm int})$ is compact.  Then ${\cal D}(\Sigma_{\rm int})
\subset {\stackrel{o}{\cal D}\!\!(\Sigma)}$.
\end{Lemma}

{\bf Proof}:  Clearly, ${\cal D}(\Sigma_{\rm int}) \subset \ds$, so we
need only prove a contradiction with the possibility that there exists $p\in
{\cal D}(\Sigma_{\rm int})$ with $p \in \pt \ds$.  However, we have $\pt
\ds = H^+(\Sigma) \cup H^-(\Sigma)$ where $H^+$ and $H^-$ denote the
future and past Cauchy horizons.  Without loss of generality,  we may
assume $p \in H^+(\Sigma)$.  Since $\Sigma$  is edgeless, $p$ lies on a
past inextendible null geodesic $\lambda$ which remains in
$H^+(\Sigma)$.  However, since $p \in {\cal D}^+(\Sigma_{\rm int})$
and $\lambda$ does not intersect $\Sigma_{\rm int}$, it follows that
$\lambda \subset {\cal D}^+(\Sigma_{\rm int}) \subset {\cal
D}(\Sigma_{\rm int})$.  Since ${\cal D}(\Sigma_{\rm int})$ is compact,
this violates strong causality ({\em cf.\ e.g.} lemma 8.2.l of \cite{Wa}).
\hfill$\Box$

Similarly, we have

\begin{Lemma}
\label{Q6}
Let $(M, g_{ab})$ be a spacetime of class (b) and suppose that ${\cal
D}(\Sigma_{\rm int})$ is compact.  Then ${\cal D}(\Sigma_{\rm int})
\subset {\stackrel{o}{\cal D}\!\!(\Sigma)} \cup {\cal H}$.
\end{Lemma}

We now prove the main theorems of this section.

\begin{Theorem}
\label{Q6a}
Let $(M, g_{ab})$ be a spacetime of class (a).  Then
\[({\cal B} \cup {\cal W}) \cap {\stackrel{o}{\cal D}\!\!(\Sigma)} =
\emptyset\]
if and only if \begin{quote}
(i) $A_i\ A_j > 0$ for all $i, j$,
where the $A_i{}$'s are the constants $A$ of definition \ref{D1} ({\em
cf.\ }eq.\ (\ref{(D.4)})) in the $i$'th end,
and
\end{quote}\begin{quote}
(ii)
${\cal D}(\Sigma_{\rm int})$ is compact.
\end{quote}
\end{Theorem}

{\bf Remark:} 
It is easily seen that compactness of domains of dependence of
compact sets can hold
even if black hole and white hole regions occur when Killing orbits
having different time orientations exist in $(M,g_{ab})$ --- an
example is given by {\em e.g.\ }{the slice $\Sigma'$ in}
the space--time of Figure \ref{F1}
from which the blocks $E$ and $F$ have been removed. (After these
excisions the metric on
this space--time can actually be taken to be that of the
$|e|<m$ Reissner--Nordstr\"om space--time).

{\bf Proof of $({\cal B} \cup {\cal W}) \cap {\stackrel{o}{\cal D}\!\!(\Sigma)}
= \emptyset
\Longrightarrow$ (i), (ii)}:  That property (i) holds follows from lemma
\ref{Q3a} applied to the spacetime $M' = {\stackrel{o}{\cal D}\!\!(\Sigma)}$.
We shall prove
compactness of ${\cal D}(\Sigma_{\rm int}) \cap {\stackrel{o}{\cal
D}\!\!(\Sigma)}$, from
which it follows
by elementary topology considerations that ${\cal D}(\Sigma_{\rm int})
\subset {\stackrel{o}{\cal D}\!\!(\Sigma)}$
and hence that ${\cal D}(\Sigma_{\rm int})$ is compact.  To prove
compactness of ${\cal D}(\Sigma_{\rm int}) \cap {\stackrel{o}{\cal
D}\!\!(\Sigma)}$,
 we consider a
sequence $\{x_i\}$ in ${\cal D}(\Sigma_{\rm int}) \cap {\stackrel{o}{\cal
D}\!\!(\Sigma)}$
and will show existence of an accumulation point.  By
{corollary \ref{W1}}
the orbit
through $x_i$, $\gamma_i \equiv \bigcup_{t\in(-\infty, \infty)}
\phi[X]_t(x_i)$, must
intersect $\Sigma$ (possibly more than once).  Let
$y_i$  denote an intersection point of $\gamma_i$ with $\Sigma$, and let
$t_i$ be such that $\phi[X]_{t_i}(y_i) = x_i$.
It is easily seen, {\em e.g.} using the coordinates of Proposition
\ref{PD.2} that we have  ${\cal D}(\Sigma_{\rm int}) \cap M_{\rm ext}
= \emptyset$, which implies $x_i \not\in M_{\rm ext}$, and hence
$y_i \in \Sigma_{\rm int}$.
Since $\Sigma_{\rm int}$ is compact, there exists a subsequence
$\{{\tilde y}_i\}$ which converges to $y \in \Sigma_{\rm int}$.  Let
$\{{\tilde t}_i\}$ denote the corresponding sequence of numbers
satisfying $\phi[X]_{{\tilde t}_i}({\tilde y}_i) ={\tilde x}_i$.

Now, {by lemma \ref{Q2} the orbit, $\gamma$, through $y$
satisfies} $I^+(\gamma)
\supset M_j$ for some $j$.  Let $t_- \in \R$ be such that
$I^+[\phi[X]_{t_-}(y)] \cap
\Sigma_{\rm ext} \ne \emptyset$.  Then $I^+[\phi[X]_t(y)] \cap \Sigma_{\rm
ext} \ne \emptyset$ for all $t < t_-$.  However, $I^+[{\cal
D}(\Sigma_{\rm int})] \cap \Sigma_{\rm ext} = \emptyset$, so $\phi[X]_t(y)
\not\in {\cal D}(\Sigma_{\rm int})$ for  $t < t_-$.  Similarly, there
exists $t_+ \in \R$ such that $\phi[X]_t(y) \not\in {\cal D}(\Sigma_{\rm
int})$ for all $t > t_+$.  However, since $\phi[X]_{{\tilde t}_i}({\tilde y}_i)
\in {\cal D}(\Sigma_{\rm int})$ it follows by continuity of $\phi[X]_t(y)$
with respect to\ $t$ and $y$ that $\limsup {\tilde t}_i \leq t_+$,
$\liminf {\tilde
t}_i \geq t_-$.  Thus, $\{{\tilde t}_i\}$ is a bounded sequence and hence
has an accumulation point $t$.  The point $x = \phi[X]_t(y)$ is then an
accumulation point of the original sequence $\{x_i\}$.  We have $x \in
{\cal D}(\Sigma_{\rm int})$ since ${\cal D}(\Sigma_{\rm int})$ is
closed, and we have $x \in {\stackrel{o}{\cal D}\!\!(\Sigma)}$ since all orbits
in $
{\stackrel{o}{\cal D}\!\!(\Sigma)}$ are complete ({\em cf.}\ Proposition
\ref{Q1a} and the discussion
following Proposition \ref{Q1b}).

{\bf Proof of (i) and (ii) $\Longrightarrow ({\cal B} \cup {\cal W})
\cap {\stackrel{o}{\cal D}\!\!(\Sigma)} = \emptyset$}:  Given (i), by changing
time orientation
if necessary we may assume without loss of generality that $A_i > 0$ for
all $i$.  We restrict attention to the globally hyperbolic region $M' =
{\stackrel{o}{\cal D}\!\!(\Sigma)}$ which, by Proposition \ref{Q1a}, also is a
spacetime of class
(a).  By Lemma \ref{Q5} and condition (ii), we have that ${\cal
D}(\Sigma_{\rm int}) \subset M'$, so ${\cal D}(\Sigma_{\rm int})$ is
compact as a subset of $M'$.  Furthermore, ${\cal B} \cap M'$ is
precisely the black hole region of the spacetime $M'$, and similarly for
${\cal W} \cap M'$. This means that, without loss of generality, we may
assume that $M$ is globally hyperbolic with Cauchy surface $\Sigma$,
and we then must prove that ${\cal B} \cup {\cal W} = \emptyset$.

Suppose, now, that ${\cal W} \ne \emptyset$.  We note, first, that it
follows straightforwardly from the definition of ${\cal B}$ and ${\cal
W}$ that ${\cal B} \subset J^+[{\cal D}(\Sigma_{\rm int})]$ and ${\cal
W} \subset J^-[{\cal D}(\Sigma_{\rm int})]$. Furthermore, $J^-[{\cal
W}] \subset {\cal W}$.  Let $x \in {\cal W}$ and consider a past
inextendible causal curve $\lambda$ with future end point $x$.  Clearly,
we have $\lambda \subset {\cal W}$.  Since $M$ is globally hyperbolic,
$\lambda$ must enter and remain in $I^-(\Sigma)$.  However, if
$\lambda$ remained in the compact region ${\cal D}(\Sigma_{\rm int})$,
we would violate strong causality.  This implies that $\lambda$ must enter
$I^-[\Sigma_{\rm ext}]$ so that
there exists a point $p \in {\cal W} \cap I^-[\Sigma_{\rm ext}]$. Since
$I^-[\Sigma_{\rm ext}]\cap {\cal B}=\emptyset$  we have $p\not\in {\cal
B}$.
Choose any $q \in I^-(p)$, and let $\tau_0 > 0$ denote the Lorentz
distance, $\tau(q, p)$, between $q$ and $p$.  Clearly, we also have $q \in
{\cal W} \cap I^-[\Sigma_{\rm ext}]$, and $q \not\in {\cal B}$.

Consider, now, the orbit, $\gamma_q$, of $q$ under the isometries.
Since ${\cal W}$ is isometry invariant, we have $\gamma_q \subset {\cal
W} \subset J^-[{\cal D}(\Sigma_{\rm int})]$, so $\gamma_q \cap
J^+[\Sigma_{\rm ext}] = \emptyset$, {\em i.e.}, for all $t \in \R$ we have
$\phi[X]_t(q) \cap J^+[\Sigma_{\rm ext}] = \emptyset$.  On the other hand
for each $j$ such that $I^+(\gamma_q) \cap M_j \ne \emptyset$, by
lemma \ref{Q2}, we have $M_j \subset J^+[\gamma_q]$.  Equivalently,
for each $j$ such that $I^+(\gamma_q) \cap M_j \ne \emptyset$, $J^+(q)$
intersects every orbit in $M_j$, and in particular, $J^+(q)$ intersects
every orbit on $\pt M_j$.  Since $\pt \Sigma_{\rm ext}$ has only a finite
number of connected components each of which is compact, it follows
that there exists $T \in \R$ such that $\phi[X]_T [J^+(q) \cap \pt M_{\rm
ext}] \subset J^+(\Sigma)$.  This implies that for all $t \geq T$, we have
$\phi[X]_t(q) \cap J^-[\Sigma_{\rm ext}] = \emptyset$.  Combined with the
previous result, this shows that for all $t \geq T$ we have $\phi[X]_t(q) \in
{\cal D}(\Sigma_{\rm int})$.  Similarly, increasing $T$ if necessary, for
all $t \geq T$ we have $\phi[X]_t(p) \in {\cal D}(\Sigma_{\rm int})$.

Now for $n \in \N$, consider the sequences $\{p_n \equiv \phi[X]_n(p)\}$,
$\{q_n \equiv \phi[X]_n(q)\}$.  Since $p, q \not\in {\cal B}$, we have $p_n,
q_n \not\in {\cal B}$ for all  $n$.  Since $\{p_n\}$ enters and remains in
the compact subset ${\cal D}(\Sigma_{\rm int})$, there exists a
subsequence $\{p_{n_i}\}$ which converges to a point $x \in {\cal
D}(\Sigma_{\rm int})$.   Similarly, the corresponding subsequence
$\{q_{n_j}\}$ has a sub--subsequence $\{q_{n_{i_j}}\}$ converging to
$y \in {\cal D}(\Sigma_{\rm int})$.  By isometry invariance, we have
$\tau(q_{n_{i_j}}, p_{n_{i_j}}) = \tau_0$. By continuity of the Lorentz
distance function $\tau$ in a globally hyperbolic spacetime, we have
$\tau(y, x) = \tau_0 > 0$.  In particular this implies that $x \in I^+(y)$.
However, since $x$ is in the $\omega$--limit  set of the orbit of $p$, and
$y$ is in the $\omega$--limit set of the orbit of $q$, by lemma \ref{Q4}
we have $x,y \in {\cal B}$.  Since $q_{n_{i_j}}, p_{n_{i_j}} \not\in
{\cal B}$, we must have $x,y \in \pt {\cal B}$.  However, $\pt {\cal B}$
is an achronal set,
which
contradicts $x \in I^+(y)$.  The hypothesis that ${\cal B} \ne \emptyset$
leads to a contradiction in a similar manner. \hfill$\Box$

Before proving the analogous theorem for spacetimes of class (b), we shall
need the following lemma:

\begin{Lemma}
\label{AQ4}
Let $(M, g_{ab})$ be a spacetime of class (b) such that $({\cal B} \cup {\cal
W}) \cap
{\stackrel{o}{\cal D}\!\!(\Sigma)}= \emptyset$. Then, for
$p\in {\cal H}_+$, the $\alpha$--limit set,
$\alpha_p$, of the orbit $\phi[X]_t(p)$ is a nonempty subset of $S$.
Similarly for $p\in {\cal H}_-$,
$\omega_p$ is a nonempty subset of $S$.
\end{Lemma}

{\bf Remark:} Note that the proof below 
shows that the
null geodesic generators of ${\cal H}_+$ are actually
future complete. [This follows from the fact that
for each $n \in \N$ the
minimum value of affine parameter on the cross-section
$\phi[X]_{nt_+}[S^+]$ is $\beta^n \epsilon$].

{\bf Proof:}
Since by Lemma \ref{Q4}, any $\alpha$-limit set is contained in
${\cal W}$, and since ${\cal H}_+ \setminus S$ is contained in
$I^+[{\stackrel{o}{\cal D}\!\!(\Sigma)}]$, it follows immediately that the
$\alpha$-limit set of any
orbit on ${\cal H}_+$ must be a subset of $S$. Thus, we need only prove
that this subset is nonempty. To do so,
on $S$, define $k^a = n^a + r^a$,
where $n^a$ is the future-directed unit normal to
$\Sigma$ and $r^a$ is the unit normal to $S$ in
$\Sigma$ (pointing ``towards
$\Sigma$"). Then $k^a$ is tangent to the null geodesic generators of
${\cal H}_+$.
[This can be seen as follows:
Clearly, $I^+[S] \cap \ds$ is empty,
so the null geodesic determined by $k^a$ can't lie in
${\stackrel{o}{\cal D}\!\!(\Sigma)}$. On the other
hand, a simple local argument  can be used to show that sufficiently near
$S$, this null geodesic lies in $\ds$.]
Let $\lambda$ denote the affine parameter along the null
generators of ${\cal H}_+$ determined by $k^a$. Consider the map,
$\psi$, defined
perhaps only
on
a sufficiently small neighborhood, ${\cal O} \subset
S\times \R^+$, of $S\times \{0\}$
which takes $(p,\lambda)$ into the
point lying at affine parameter $\lambda$ on the null generator of
${\cal H}_+$ through $p \in S$. By decreasing ${\cal O}$ if necessary, we
may assume that this map is a diffeomorphism of ${\cal O}$ onto
a neighborhood of $S$ in ${\cal H}_+$. Consequently, we may choose
$\epsilon > 0$ such that $\psi [S,\epsilon]$ is a smooth cross-section,
$S^+$, of ${\cal H}_+$. Decreasing $\epsilon$ if necessary, we may
assume that $S^+\subset{\cal D}(\Sigma_{\rm int})$.

Let $p\in S^+$. Then there
exists $q\in{\stackrel{o}{\cal D}\!\!(\Sigma)}$ such that $q\in I^-(p)$. Since
${\cal W} \cap
{\stackrel{o}{\cal D}\!\!(\Sigma)}= \emptyset$, by Corollary \ref{W1} there
exists $t_q$ such that for $s\ge t_q$ we have $\phi[X]_s(q)\in
I^+(\Sigma_{\rm ext})$. Set $t_p=t_q$, so that for all $s\ge t_p$ we
have $\phi[X]_s(p)\in
I^+(\Sigma_{\rm ext})$.
Since $I^+(\Sigma_{\rm ext})$ is open, there exist open neighbourhods
${\cal U}_p$ of $\phi[X]_{2t_p}(p)$ satisfying ${\cal
U}_p\subset I^+(\Sigma_{\rm ext}) $. Set ${\cal
O}_p=\phi[X]_{-2t_p}({\cal
U}_p)\cap S^+$.
It follows (passing to open subsets of ${\cal U}_p$ if
necessary) that  $\{ {\cal O}_p\}_{p\in S^+}$ is a
covering of $S^+$ by open sets such that the implication [$q\in {\cal
O}_p \Rightarrow \forall s\ge 2t_p \ \phi[X]_s(q)\in
I^+(\Sigma_{\rm ext})$] holds. By compactness of $S^+$ a finite
covering $\{ {\cal O}_{p_i}\}_{i=1}^N$ can be chosen. Set $t_+=2\max_i
t_{p_i}$, define
\[
S^+_t\equiv \phi[X]_{t}(S^+)\ .
\]
By construction we have $S^+_{t_+}\subset I^+(\Sigma_{\rm ext})$.
Let $\alpha$ be the minimum value of affine parameter on $S^+_{t_+}$.
Then, clearly $\alpha > \epsilon$.

Now, for all $t$, $\phi[X]_t$
maps affinely parametrized null geodesic generators of ${\cal H}_+$
into
affinely parametrized null geodesic generators of ${\cal H}_+$. In
particular, the above argument shows that
$\phi[X]_{t_+}$ maps each null geodesic segment on ${\cal H}_+$
with past endpoint on $S$ and of affine
length $\epsilon$
into a null geodesic segment on ${\cal H}_+$
with past endpoint on $S$ and
of affine length at
least $\beta \epsilon$, where $\beta = \alpha/\epsilon > 1$. It follows
that $\phi[X]_{t_+}$ maps any null
geodesic segment on ${\cal H}_+$ starting at $S$ and having affine
length $\lambda$ into a similar null geodesic segment of affine length at
least $\beta \lambda$. It then follows immediately that $\phi[X]_{nt_+}$ maps
each such null geodesic segment of affine
length $\lambda$ into a null geodesic segment of affine length at
least $\beta ^n \lambda$. Conversely, $\phi[X]_{-nt_+}$ maps
each such null geodesic segment of affine
length $\lambda$ into a similar null geodesic segment of affine length
no larger than $\beta ^{-n} \lambda$.

Now, let $p\in {\cal H}_+$. Consider the sequence $\phi[X]_{-nt_+}(p)$.
Let $\{p_n\} \in S$ denote the intersection with $S$ of the generator of
${\cal H}_+$ through $\phi[X]_{-nt_+}$. Compactness of $S$ implies
existence of a subsequence which converges to $q \in S$. It then follows
immediately that $q$ is an $\alpha$-limit set of the orbit through
$p$, so the $\alpha$-limit
set of $\phi[X]_t(p)$ is nonempty, as we desired to show.
\hfill $\Box$

The theorem for spacetimes of class (b) analogous to Theorem \ref{Q6a}
is the following:

\begin{Theorem}
\label{Q6b}
Let $(M, g_{ab})$ be a spacetime of class (b).  Then
\[({\cal B} \cup {\cal W}) \cap
{\stackrel{o}{\cal D}\!\!(\Sigma)}= \emptyset\]
if and only if
\begin{quote}
(i) $A_i\ A_j > 0$ for all $i, j$,
where the  $A_i{}$'s are the constants $A$ of definition \ref{D1} ({\em
cf.\ }eq.\ (\ref{(D.4)})) in the $i$'th end,
and
\end{quote}
\begin{quote}
(ii)
${\cal D}(\Sigma_{\rm int})$ is compact.
\end{quote}
\end{Theorem}

{\bf Remark:}
As shown in the first paragraph of the first part of the
proof below, under the hypothesis
that $({\cal B} \cup {\cal W}) \cap {\stackrel{o}{\cal D}\!\!(\Sigma)} =
\emptyset$, we have
${\cal B} \cap M' = {\cal H}_+$ and ${\cal W} \cap M' =
{\cal H}_-$. This implies that ${\cal H}_+$ comprises (a portion of)
the event horizon of the black hole, ${\cal H}_-$ comprises (a portion
of) the white hole horizon, and that the intersection of the black
and white hole horizons is $S$. Thus, spacetimes of class (b) which
satisfy the hypothesis of theorem \ref{Q6b} correspond to the spacetimes
of ``class (2)" discussed in the Introduction.

{\bf Proof of $({\cal B} \cup {\cal W}) \cap {\stackrel{o}{\cal D}\!\!(\Sigma)}
= \emptyset
\Longrightarrow$ (i), (ii)}:  That property (i) holds follows from lemma
\ref{Q3b}. To prove (ii), we first note that by corollary
\ref{CQ3}, we have $S \subset ({\cal B} \cap {\cal W})$.  Since
$J^+({\cal B}) \subset {\cal B}$, we have ${\cal H}_+ \subset {\cal B}$.
However, we must have $({\cal H}_-\setminus S) \cap {\cal B} =
\emptyset$, since otherwise we would have ${\cal B} \cap {\stackrel{o}{\cal
D}\!\!(\Sigma)} \ne
\emptyset$.  This implies that ${\cal B} \cap M' = {\cal H}_+$, where
$M' \equiv {\stackrel{o}{\cal D}\!\!(\Sigma)} \cup {\cal H}$.  Similarly, we
have ${\cal W} \cap M' =
{\cal H}_-$.

We shall prove compactness of ${\cal D}(\Sigma_{\rm int}) \cap M'$
from which it follows by elementary topology that ${\cal
D}(\Sigma_{\rm int}) \subset M'$ and
hence that ${\cal D}(\Sigma_{\rm int})$ is compact.  Let $\{x_i\}$ be an
infinite sequence in ${\cal D}(\Sigma_{\rm int}) \cap M'$.  We wish to
show existence of an accumulation point.  Without loss of generality, we
may assume that infinitely many of the $x_i$ lie in ${\cal
D}^+(\Sigma_{\rm int}) \cap M'$.

Let $X'$
be a vector field on $M$ which is
transverse to $\Sigma$, (strictly) timelike on a neighbourhood of
$\Sigma_{\rm int}$
and is such that $X' = X$ in $M_{\rm ext}$, at least for
$r$ large enough.  Since
$\Sigma_{\rm int}$ is compact, there exists $\epsilon > 0$ such that the
induced action $\phi[X']_t(\Sigma)$ generated by $X'$ for
$t \in (-2\epsilon, 2\epsilon)$
yields a spacelike foliation, such that $M' \cap {\cal
O}$
is a closed subset of the spacetime region ${\cal O}$ covered by this
foliation and such that $X'$ is timelike in
$(I^+(\Sigma_{\rm int})\cup I^-(\Sigma_{\rm int}))\cap{\cal O}$.
We denote the
hypersurface at $t = \epsilon$ by $\tS$.  Clearly $\tS \subset
I^+(\Sigma)$, and the intersection of $\tS$ with ${\cal H}$ yields a
cross--section of ${\cal H}_+\setminus S$.
We denote by ${\cal R}$ the
region covered by the foliation for $t \in [0, \epsilon]$.  We denote by
${\cal R}_{\rm int} \subset {\cal R}$ the portion of ${\cal R}$ connected
to $\Sigma_{\rm int}$ by integral curves of $X'$.  Finally, we write
$\tS_{\rm int} = \tS \cap {\cal R}_{\rm int}$ and note that the
construction ensures that $J^-(\tS_{\rm int}) \cap {\cal
D}^+(\Sigma_{\rm int}) \cap M' \subset {\cal R}_{\rm int}$ and ${\cal
D}(\Sigma_{\rm int}) \cap M' \setminus {\cal R}_{\rm int} \subset
J^+(\tS_{\rm int}) \cap M'$.

Now ${\cal R}_{\rm int}$ is compact, since it has topology $[0, \epsilon]
\times \Sigma_{\rm int}$.  Since both ${\cal D}^+(\Sigma_{\rm int})$
and $M' \cap {\cal R}$ are closed, ${\cal D}^+(\Sigma_{\rm int}) \cap M
\cap {\cal R}_{\rm int}$ is compact.  Hence, $\{x_i\}$ must have an
accumulation point unless all but finitely many elements of this sequence
lie in ${\cal D}^+(\Sigma_{\rm int}) \cap M' \setminus {\cal R}_{\rm
int}$.  Hence, without loss of generality, we may assume that $\{x_i\}$ is
a sequence in ${\cal D}^+(\Sigma_{\rm int}) \cap M' \cap J^+(\tS_{\rm
int})$.  For each $i$, either $x_i \in {\stackrel{o}{\cal D}\!\!(\Sigma)} \cap
J^+(\tS_{\rm int})$ or $x_i
\in {\cal H}_+$.  If $x_i \in {\stackrel{o}{\cal D}\!\!(\Sigma)}$, then by
hypothesis $x_i \not\in {\cal
B}$ and by the same type of argument as in theorem \ref{Q6a}, the half-
orbit $\gamma_i = \cup_{(-\infty, 0]} \phi[X]_t(x_i)$ must intersect $\tS$ in
the compact set $\tS_{\rm int} \cap M'$.
On the other hand, if $x_i \in
{\cal H}_+$, using
lemma \ref{AQ4}, it follows immediately that the similar half-orbit
$\gamma_i$ also intersects $\tS$ in $\tS_{\rm int} \cap M'$.
Since ${\cal
W} \cap (\tS_{\rm int} \cap M') = \emptyset$, existence of an
accumulation point for $\{x_i\}$ follows by an argument similar to that
used in the proof of the corresponding part of theorem \ref{Q6a}.

{\bf Proof of (i) and (ii) $\Longrightarrow ({\cal B} \cup {\cal W})
\cap {\stackrel{o}{\cal D}\!\!(\Sigma)} = \emptyset$}:  By Lemma \ref{Q6} we
have ${\cal D}(\Sigma_{\rm int}) \subset {\stackrel{o}{\cal D}\!\!(\Sigma)}
\cup {\cal H}$.  As in the
proof of theorem \ref{Q6a}, we assume that ${\cal W} \cap {\stackrel{o}{\cal
D}\!\!(\Sigma)} \ne
\emptyset$ and obtain a contradiction.  Let $x \in {\cal W} \cap
{\stackrel{o}{\cal D}\!\!(\Sigma)}$
and let $\lambda$ be a causal curve through $x$ which is past
inextendible in ${\stackrel{o}{\cal D}\!\!(\Sigma)}$.  Then either $\lambda$ is
past inextendible in
${\stackrel{o}{\cal D}\!\!(\Sigma)} \cup {\cal H}$ or $\lambda$ has a past
endpoint on ${\cal H}_-$.
In the latter case, we have $x \in I^+(\gamma)$, where $\gamma$ is a
null geodesic generator of ${\cal H}_-$ on which
$\lambda$
has a past
endpoint.  We then may choose a causal curve $\lambda'$ through $x$
which lies in $I^+(\gamma) \cap I^-(x) \subset {\stackrel{o}{\cal
D}\!\!(\Sigma)}$ and is past
inextendible in $M$ ({\em cf.\ }lemma 8.1.4 of \cite{Wa}).  Thus, in either
case, we obtain a causal curve in ${\stackrel{o}{\cal D}\!\!(\Sigma)}$ which is
past inextendible in
${\stackrel{o}{\cal D}\!\!(\Sigma)} \cup {\cal H}$.  By strong causality, this
curve cannot lie entirely
within ${\cal D}(\Sigma_{\rm int})$, and we thereby obtain a point $p
\in {\cal W} \cap {\stackrel{o}{\cal D}\!\!(\Sigma)}$ with $p \not\in {\cal
B}$.  The remainder of the
proof parallels the proof of the corresponding portion of theorem
\ref{Q6a}.
\hfill $\Box$

We end this Section with some additional results. {First, for any
set $K \subset M$ (where $K$ is not necessarily achronal)
we define the ``effective domain of
dependence" of $K$, denoted $\tilde{\cal D}(K)$, to consist of all $p \in M$
such that every (past and future) inextendible timelike curve through $p$
intersects $K$. We then have the following:}

\begin{Corollary}
\label{compactness1}
Let $(M, g_{ab})$ be a space-time of class (a), suppose that
$$
({\cal B} \cup {\cal W})\, \cap  {\stackrel{o}{\cal D}\!\!(\Sigma)} =
\emptyset\,.
$$
For any compact $K\subset {\stackrel{o}{\cal D}\!\!(\Sigma)}$, {$\tilde{\cal
D}(K)$} is compact.
\end{Corollary}

{\bf Proof}:
Suppose first that
$K\subset\Sigma$.
One can enlarge $K$ or $\Sigma_{\rm int}$ or both
so that $K=\Sigma_{\rm int}$ holds, and
compactness of ${\tilde{\cal D}(K)} = {\cal D}(K)$ follows
by Theorem \ref{Q6a}.

In the general case, let $p\in K$. By {Corollary \ref{W1}}
there
exists $t^\pm_p$ such that
$\phi[X]_{t^\pm_p}(p)\in I^\pm(\Sigma_{\rm ext})$.
Since $I^\pm(\Sigma_{\rm ext})$ is open, there exist open neighbourhods
${\cal V}^\pm_p$ of $\phi[X]_{t_\pm}(p)$ satisfying ${\cal
V}^\pm_p\subset I^\pm(\Sigma_{\rm ext}) $. Set ${\cal
O}_p=\phi[X]_{-t^\pm_p}({\cal
V}^\pm_p)\cap K$.
It follows that $\{{\cal O}_p\}_{p\in K}$ 
is a covering
of $K$ by
open sets satisfying $[q\in{\cal O}_{p}\Rightarrow
\phi[X]_{t^\pm_p}(q)\in I^\pm(\Sigma)]$; by compactness of $K$ a finite
covering $\{{\cal O}_{p_i}\}_{i=1}^N$ can be chosen. 
Let $t_\pm=\max_i\{-
t^\mp_{p_i}\}$; it follows that 
$K\subset I^\mp(\phi[X]_{t_\pm}(\Sigma))$. By
Proposition \ref{Q1a} 
$\phi[X]_{t_\pm}(\Sigma)$ are
Cauchy surfaces, so that by \cite{Geroch} $K^\pm\equiv J^\pm (K)\cap
\phi[X]_{t_\pm}(\Sigma)$ are compact. Now {$\tilde{\cal D}^\pm(K)$}
are closed
subsets of ${\cal D}^\pm(K^\mp)$, which have already been shown to be
compact, and the result follows. 
\hfill $\Box$

\begin{Corollary}
\label{compactness2}
Let $(M, g_{ab})$ be a space-time of class (b), suppose that
$$
({\cal B} \cup {\cal W})\, \cap  {\stackrel{o}{\cal D}\!\!(\Sigma)} =
\emptyset\,.
$$
For any compact $K\subset \, {\stackrel{o}{\cal D}\!\!(\Sigma)}\cup{\cal H}$,
{$\tilde{\cal D}(K)$} is compact.
\end{Corollary}
{\bf Proof:}
It suffices to prove compactness
of $J^+(\Sigma) \cap {\tilde{\cal D}(K)}$.
Enlarging $\Sigma_{\rm int}$ if necessary we may assume $K\cap M_{\rm
ext}=\emptyset$. Let ${\cal R}$, $\tilde \Sigma$, etc., be as in the
proof of Theorem \ref{Q6b}, set $K_0=K\cap{\cal R}$,
$K_+=K\cap J^+(\tilde \Sigma)$. Arguing as in the proof of Corollary
\ref{compactness1} there exists $t_+$ such that $K_+\subset
\cup_{t\in[0,t_+]}
\phi[X]_t(\tilde\Sigma)$. Replacing ${\cal R}$ by ${\cal R}\cup_{t\in[0,t_+]}
\phi[X]_t(\tilde\Sigma)$ and $\tilde \Sigma$ by
$\phi[X]_{t_+}(\tilde\Sigma)$ if necessary we may without loss of
generality assume $t_+=0$. Now the argument of the proof of Theorem
\ref{Q6b} shows
that {$\tilde{\cal D}({\cal R}_{\rm int})$} is compact,
and the result follows
since {$\tilde{\cal D}(K)$} is a closed subset of
{$\tilde{\cal D}({\cal R}_{\rm
int})$}.
\hfill $\Box$

Let us remark that it follows immediately from
the inclusion
$$
M\setminus {\stackrel{o}{\cal D}\!\!(\Sigma)} \subset {\cal B}\cup {\cal W}\ ,
$$
proved in Corollary \ref{NEW1}
that a space--time of class (a) in which ${\cal B}\cup {\cal
W}=\emptyset$
is globally hyperbolic.
We shall end this Section by proving a
{related result:}

\begin{Proposition}
\label{inextendibility}
Let $(M,g_{ab})$ be a space--time of class (a), and suppose that
\be
\label{oneend}
({\cal B}_1\cup{\cal W}_1)\cap\ods = \emptyset\ ,
\ee
where ${\cal B}_1$ and ${\cal W}_1$ are the black hole region and the
white hole region with respect to a given end $M_1$. Then
$$
M = \ods \ .
$$
In other words, $\ods$ is inextendible in the class of space--times
of class (a).
\end{Proposition}

{\bf Remarks:}
\begin{enumerate}
\item
If $M$ has only one end, then clearly (\ref{oneend}) is equivalent to
$({\cal B}\cup{\cal
W})\cap\ods = \emptyset$. It is, however, possible to have
(\ref{oneend}) and more than one end, as is {\em e.g.\ }shown by the metric
$-dt^2+g_{ij}dx^idx^j$, where $g_{ij}dx^idx^j$ is the space--part of
the
``Schwarzschild wormhole'' metric (two asymptotically flat ends
connected
by a throat). Other examples of this kind (with an arbitrary number of
ends) can be
constructed
by appropriately gluing together  along boundaries of
world--tubes space--times of class (a) in which the Killing
vector is timelike everywhere.
\item
The following example shows that the hypothesis of strong causality
of the potential extensions is necessary in general: Consider the
space--time $M'$ consisting of blocks $B$,$D$ and $E$  of
Figure \ref{F1}, let $t$ be a Killing time function on $M$ as given
by Proposition \ref{PT.1a} below, let $\Sigma_0$ be the
zero level set of $t$; use $t$ to identify $M'$ with $\R\times\Sigma_0$.
On $\Sigma_0$ remove two balls $B_a\subset
\Sigma_0\cap M_a$, $a=1,2$, and identify $(t,p_1), p_1\in\partial B_1$
with $(t,p_2), p_2=\phi(p_1)\in\partial B_2$, where $\phi$ is a smooth
diffeomorphism
from $\partial B_1$ to $\partial B_2$. Smoothing out the metric near
$\R\times \partial B_a$, $a=1,2$ one obtains  a space--time $M$ with a
Killing vector field which has complete orbits. Note that $t$ passes
from $M'$
to the quotient space--time $M$
and therefore $M$ is stably causal, thus $M$ is a space--time of class
(a).
Clearly $M$ 
is extendible because $M'$ is, and the identifications which
lead to the construction of $M$ are done only in the interior of $M'$.

\item
Let us point out that neither the  hypothesis of existence of Killing vectors
in the (potential) extended space--time, nor the the hypothesis of stable
causality
of the (potential) extended space--time are needed to show inextendibility
of $M$ when the Killing vector is strictly
timelike.
For suppose that $(M,g_{ab})$ were extendible to a space--time
$(M',g'_{ab})$ with a $C^2$ metric $g'_{ab}$, it follows from equation
(\ref{3.7.0}) that $X$ can be extended to a continuous vector field
defined on the closure $\bar M$ of $M$ in $M'$.
Now $\partial M$ is an 
achronal topological hypersurface, so that every timelike vector is
necessarily transverse to $\partial M$.
Let
$p\in\partial M$,  if  $X(p)$ were timelike the orbits of $X$ could
not
be complete in $M$,
hence $g_{ab}X^aX^b(p_i)$ must tend to zero
on any  sequence $p_i\in M$ such that
$p_i\rightarrow
p$. But the hypothesis that
$X$ is strictly timelike on $M$ together with asymptotic flatness
imply that there exists $\epsilon >0$ such that
$g_{ab}X^aX^b<-\epsilon $ on $M$, and the result follows.

\end{enumerate}
{\bf Proof:} We note first that
$({\cal B}_1\cup{\cal W}_1)\cap\ods = \emptyset$ implies that
for each $i$, we have $I^-(M_i) \supset \ods$, since by
Lemma \ref{Q2} we have $I^-(M_i) \supset M_1$, and, hence,
$I^-(M_i) \supset I^-(M_1) \supset \ods$. Similarly, for each $i$ we have
$I^+(M_i) \supset \ods$.

Suppose, now, that $M \ne \ods$. Then $H \ne \emptyset$, where
$H$ denotes the Cauchy horizon of $\Sigma$. Without loss of generality,
we may assume that $H^+ \ne \emptyset$. Let $p \in H^+$. By theorem
\ref{Q6a} and Lemma \ref{Q5}, we have
$p \not\in {\cal D}^+(\Sigma_{\rm int})$,
and, hence, $p \in I^+(\Sigma_{\rm ext})$. Replacing $\Sigma$ by
$\phi[X]_t(\Sigma)$ in this argument, we see that for all $t \in \R$
we have $p \in I^+(\phi[X]_t[\Sigma_{\rm ext}])$. However, this implies that
$I^-(p) \supset M_i$ for some $i$, and, hence,
$I^-(p) \supset I^-(M_i) \supset \ods$.

Now, let ${\cal O}$ be any open
neighborhood of $p$ which does not intersect $M_{\rm ext}$,
and let ${\cal V}$ be any open neighborhood of $p$ contained in ${\cal O}$.
Since $p$ lies on the
boundary of $\ods$, there exists $q \in \ods \cap {\cal V}$.
Since
$q \not\in {\cal B}$, there exists a future-directed causal curve
$\lambda_1$ from
$q$ to a point $r \in M_{\rm ext}$. Since $I^-(p) \supset \ods$, there exists
a future-directed causal curve $\lambda_2$ from $r$ to $p$. Clearly,
the causal curve $\lambda = \lambda_1 \cup \lambda_2$
has more than one connected component in ${\cal
V}$,
thus showing that strong causality is
violated at $p$. \hfill$\Box$

The above proof caries over immediately to
 spacetimes of class (b), in the following sense:

\begin{Proposition}
\label{inextendibility2}
Let $(M,g_{ab})$ be a space--time of class (b), and suppose that
$$
({\cal B}_1\cup{\cal W}_1)\cap\ods = \emptyset\ ,
$$
where ${\cal B}_1$ and ${\cal W}_1$ are the black hole region and the
white hole region with respect to a given end $M_1$. Then
$H = {\cal H}$, where $H$ denotes the Cauchy horizon of $\Sigma$.
\end{Proposition}

\section{Maximal Slicings}
\label{existence}
The main results of this section, and indeed, of this paper, are the
following:

\begin{Theorem}
\label{TM.1}
Let $(M,g_{ab})$ be a space-time of class (a) and suppose that
\begin{equation}
\label{(M.1)}
({\cal B} \cup {\cal W})\, \cap  {\stackrel{o}{\cal D}\!\!(\Sigma)} =
\emptyset\,.
\end{equation}
Then ${\stackrel{o}{\cal D}\!\!(\Sigma)}$ can be covered by a family of maximal
(spacelike)
asymptotically flat slices $\hat\Sigma_s$, $s\in\R$, which are
Cauchy surfaces for
${\stackrel{o}{\cal D}\!\!(\Sigma)}$, such that
\begin{equation}
\label{(M.2)}
\phi[X]_t(\hat\Sigma_s) = \hat\Sigma_{s+t} , \quad \mbox{and} \quad
\hat\Sigma_s
\approx \Sigma ,
\end{equation}
where $\approx$ means ``diffeomorphic to''.
If moreover  $X$ is timelike on ${\stackrel{o}{\cal D}\!\!(\Sigma)}$ or if
the timelike convergence condition holds,
\begin{equation}
\label{(TCC)}
R_{ab} Z^a Z^b \ge 0 \quad \mbox{for all timelike} \quad Z^a ,
\end{equation}
then $\{\hat\Sigma_t\}_{t \in \R}$ foliates ${\stackrel{o}{\cal
D}\!\!(\Sigma)}$.
\end{Theorem}


\begin{Theorem}
\label{TM.2}
Let $(M, g)$ be a space-time of class (b)
and suppose that (\ref{(M.1)})
holds.  Then ${\stackrel{o}{\cal D}\!\!(\Sigma)}$ can be covered by a family of
maximal (spacelike)
asymptotically flat hypersurfaces $\hat\Sigma_s$, $s\in\R$,  with boundary
$\pt \hat\Sigma_s = \pt \Sigma = S$, such that (\ref{(M.2)}) holds.
Moreover $\hat\Sigma_s\setminus S$ are Cauchy
surfaces for  ${\stackrel{o}{\cal D}\!\!(\Sigma)}$.

If moreover $X$ is timelike on ${\stackrel{o}{\cal D}\!\!(\Sigma)}$ or if the
timelike
convergence condition (\ref{(TCC)}) holds,
then $\{\hat\Sigma_t\setminus S\}_{t  \in \R}$ foliates
${\stackrel{o}{\cal D}\!\!(\Sigma)}$.
\end{Theorem}

{\bf Remarks}.\begin{enumerate}
\item Theorem \ref{TM.1} generalizes a similar Theorem proved in
\cite{BCOM}, where strict stationarity (the Killing vector timelike
everywhere) is assumed.

\item As mentioned in Corollary 4.4 of \cite{BCOM}, the existence of
maximal slices restricts the allowed topologies of $\Sigma$;  thus there
exist no space-times satisfying the hypotheses of Theorem \ref{TM.1}
and the timelike convergence condition
(\ref{(TCC)}) for
which $\Sigma$'s topology
satisfies a genericity condition.

\item It should be noted that, even though asymptotically flat, the maximal
hypersurfaces will in general {\em not\/} be asymptotic to the original $t$
slicing of the exterior region.  Even if $\alpha = 1$,
where $\alpha$ is the fall--off rate of the metric in the asymptotic
region, {\em cf.\ }Definition \ref{D1},
one will only have
$|u| \le C(1 + \ln(1+r))$, where $u$ is the ``height function" of
$\hat\Sigma_0$ in $M_{\rm ext}$.

\item  If the timelike convergence condition holds, one has uniqueness
of $\{\hat\Sigma_t\}_{t  \in \R}$ under some supplementary rather weak
conditions \cite{BCOM} [Theorem 5.5].  Proposition 1 and Theorem 1 of
\cite{CHAH} can be used to remove the ``boost-domain" conditions of
\cite{BCOM} [Theorem 5.5] if, {\em e.g.}, the metric satisfies some
hyperbolic field equations for which the ``boost problem" ({\em cf.}
\cite{CM}) is well posed.
\end{enumerate}

Before passing to the proofs of our results, let us present in some
detail an example which shows that the
hypothesis
$({\cal B}\cup{\cal W})\cap\ods = \emptyset$ of Theorem \ref{TM.1} cannot
be removed without imposing some other conditions: Let $(\hat M,\hat
g_{ab})$ be the maximally analytically extended Schwarzschild space--time,
with ``stationary'' Killing vector $X$.
Let $\hat \Sigma_0$ be the standard ``Einstein--Rosen bridge'' maximal
surface
extending from one end $\hat M_1$ to the other end $\hat M_2$. Let $t$ denote
the Killing time function based on $\hat \Sigma_0$, defined in the
ends
$\hat M_a$, $a=1,2$. Consider any complete, asymptotically flat,
maximal
hypersurface $\hat \Sigma$ in $\hat M$ which  is
asymptotically orthogonal to the Killing vector. With a little work
one shows that there exist $t_1,t_2\in\R$ such that $\hat \Sigma $ is
asymptotic to the level sets $t=t_a$ in $\hat M_a$, $a=1,2$. This
together with the uniqueness results in \cite{BCOM} implies that $\hat
\Sigma$ is necesssarily spherically symmetric.

We let $\Omega$ consist of the union of the
 causal future of the "north pole" $p\in S\equiv\{r\in\hat M|X(r)=0\}$
 with the causal past of the "south pole"  $q\in S$ of the bifurcation sphere
$S$.
Consider the space--time $(M,g_{ab})$
defined as
$M\equiv \hat M\setminus \Omega$, $g_{ab}=\hat g_{ab}|_M$.
As $\Omega$ is invariant under the flow of $X$ in $\hat M$, so is $M$,
hence
all the Killing orbits in $M$ are complete.
``Pushing'' $\hat \Sigma_o$ slightly to the future in a neighbourhood
of $q$ and slightly to the past in a neighbourhood of $p$ one obtains
a complete asymptotically flat hypersurface $\Sigma _0$ in $M$, thus
$M$ is of class (a).
Moreover $M$ has
both a black hole and a white
hole region, thus $M$ does {\em not} satisfy the hypotheses of Theorem
\ref{TM.1}.
Let $\Sigma$ be any complete, asymptotically flat, maximal surface in
$M$ which is asymptotically orthogonal to $X$. 
It follows that such a surface
is also a maximal surface in $\hat M$ which enjoys the same
properties. By what has been said above it follows that
$\Sigma$ must be spherically symmetric. But it is clear from {\em
e.g.\ } the Penrose diagram for the maximally extended Schwarzschild
space--time 
that there are no spherically symmetric
complete hypersurfaces in $M$. Consequently, there are no
complete, asymptotically flat, maximal surfaces in
$M$ which are asymptotically orthogonal to $X$, and sharpness of
Theorem \ref{TM.1} follows.

The proofs of Theorems \ref{TM.1} and \ref{TM.2} will
largely run in
parallel.
The idea is to construct some
appropriate time functions (given in Propositions \ref{PT.1a}
and \ref{PT.1b} below),
and to prove a height estimate by moving the
surfaces by the isometry group.  Higher order {\em a priori} estimates
follow then from Bartnik's work \cite{Ba1}, and existence is obtained by
standard arguments. Note that from Theorems \ref{Q6a} and \ref{Q6b}, it
follows that, without loss of generality
we may assume that for spacetimes of class (a) and (b) the Killing orbits in
$M_{\rm ext}$ are future oriented.

If the Killing field $X$ were timelike throughout $M$, the desired time
functions for cases (a) and (b) could be constructed very simply as the
parameter along the integral curves of $X$ starting at $\Sigma$; {\em
i.e.} by solving ${\cal L}_X t = 1$ with the initial condition $t = 0$ on
$\Sigma$.  However, in case (a), $X$ need not be timelike in $M_{\rm
int}$, so $X$ need not be transverse to $\Sigma$, in which case this
simple construction does work.  The situation is worse in case (b), since
the condition $\phi[X]_t [S] = S$ implies that no time function $t$ can satisfy
${\cal L}_X t = 1$ throughout $\ds$.
 Nevertheless, we shall show that in case (a) an asymptotically
flat slice, $\Sigma'$ (which is a Cauchy surface for ${\stackrel{o}{\cal
D}\!\!(\Sigma)}$)
can be constructed which is everywhere transverse to $X$. Similarly,
in case (b) an asymptotically flat hypersurface, $\Sigma'$
with boundary $S$ can be constructed such that $\Sigma' \setminus S$
is transverse to $X$ and is a Cauchy surface for ${\stackrel{o}{\cal
D}\!\!(\Sigma)}$. These results will
enable us to prove the following propositions:

\begin{Proposition}
\label{PT.1a}
Let $(M, g_{ab})$ be a space-time of class (a), suppose that
\begin{equation}
\label{(empty)}
({\cal B} \cup {\cal W})\, \cap  {\stackrel{o}{\cal D}\!\!(\Sigma)} =
\emptyset\,.
\end{equation}
Then there exists a time function $\tau \in C^\infty({\stackrel{o}{\cal
D}\!\!(\Sigma)})$ with
asymptotically flat level sets
$\Sigma_s(\tau) = \{p : \tau(p) = s\}$ (which are Cauchy surfaces for
${\stackrel{o}{\cal D}\!\!(\Sigma)}$) such that
$\phi[X]_t$ acts by translations in $\tau$:
\begin{equation}
\label{(transl)}
\phi[X]_t(\Sigma_s(\tau)) = \Sigma_{s+t}(\tau) .
\end{equation}
\end{Proposition}

\begin{Proposition}
\label{PT.1b}
Let $(M, g_{ab})$ be a space-time of class (b), suppose that
(\ref{(empty)}) holds.
Then there exists a neighbourhood ${\cal U}$ of ${\stackrel{o}{\cal
D}\!\!(\Sigma)}\cup {\cal H}$
and a time function
${\tilde \tau} \in
C^\infty({\cal U})$, with asymptotically flat level sets
$\Sigma_s({\tilde \tau}) = \{p :
{\tilde \tau}(p) = s\}$ such that 
\[
\phi[X]_t(\Sigma_s(\tilde\tau)\cap \ds) = \Sigma_{s+t}(\tilde\tau)
\cap \ds\ ,
\]
for $s \ge
s_0$, $t \ge 0$ and for $s
\le -s_o$, $t \le 0$, for some $s_0$.  Moreover, for all $t,s \in \R$
\[
\phi[X]_t\big(\Sigma_s(\tilde\tau) \cap M_{\rm ext}\big) =
\Sigma_{s+t}(\tilde\tau)
\cap M_{\rm
ext}\,.
\]
Furthermore, $\Sigma_0({\tilde \tau})\cap{\stackrel{o}{\cal D}\!\!(\Sigma)} $
is a Cauchy
surface for ${\stackrel{o}{\cal D}\!\!(\Sigma)}$  and $\pt\{ \Sigma_0({\tilde
\tau})\cap\ds\}=S$.
\end{Proposition}

{\bf Remark:} If the timelike convergence condition (\ref{(TCC)})
holds, for spacetimes of class (b), a somewhat shorter though
certainly less elegant proof of Proposition
\ref{PT.1b} can be given. The relevant argument is
outlined in the proof
of Theorem \ref{TA.3}, Appendix \ref{gluing}.

Before beginning the proofs of Propositions \ref{PT.1a} and \ref{PT.1b},
we introduce the following notation. For any hypersurface $\hat
\Sigma\subset M$ we set
\begin{eqnarray*}
\hat\Sigma_{0,{\rm ext}} &= &\hat\Sigma_{\rm ext} = \hat\Sigma \cap
M_{\rm ext} ,\\
\hat\Sigma_{t,{\rm ext}} &= &\phi[X]_t(\hat\Sigma_{\rm ext}) ,\\ \hat
\Sigma_{0,\rm int}&= &\hat \Sigma_{\rm int}=\hat
\Sigma\cap M_{\rm int}  , \\
\hat\Sigma_{t,{\rm int}} &= &\phi[X]_t(\hat \Sigma_{\rm int})\ ,
\end{eqnarray*}
and
\[
\delta \le 0 , \quad\hat \Sigma_{0,\delta} =  \{p \in \hat\Sigma_{\rm
int}:d_{\hat\Sigma}(p, \pt
\hat\Sigma_{\rm int}) \ge |\delta|\} 
\]
\begin{equation}
\label{hypdelta}
\delta \ge 0 , \quad\hat \Sigma_{0,\delta} = \hat
\Sigma_{\rm int} \cup_{i=1}^I \{p \in\hat \Sigma_i : R_i \le r(p) \le
R_i + \delta\} ,
\end{equation}
where $d_{\hat
\Sigma}(p, \pt\hat \Sigma_{\rm int})$ is the Riemannian distance
on $\hat\Sigma$ from
$p$ to $\pt\hat \Sigma_{\rm int}$,
$r(p)$ is the coordinate radius function in the end under
consideration, and the $R_i{}$'s are the constants $R_1$ of Definition
\ref{D1} in the $i$'th end.
 There exists $\epsilon_0 > 0$ such that
for $\delta > -
\epsilon_0$ the sets $\hat \Sigma_{0, \delta}$ are manifolds with smooth
boundary.

{\bf Proof of proposition \ref{PT.1a}}: Let $(t, x)$ be coordinates
on $M_{\rm ext}$ as given by
Proposition
\ref{PD.2}, normalized so that $A=1$ in all
the ends, $M_i$, and so that $g_{\mu\nu}$ is asymptotically
conformal (with constant
conformal factor) to the standard
Minkowski metric.
Let $\og$ be any smooth Lorentzian metric on $M$ which,
coincides with $g$ on $M_{\rm int}$, and such that
\[
\og = -dt^2 + \sum_{i=1}^3 {(dx^i)}^2
\]
for all $x$ such that $r(x) \ge R$, for some $R \ge R_i$, $i =
1,\ldots,I$, on all ends
$M_i$.  Let  ${\widetilde
X}$ be any smooth vector field defined on a neighborhood of
${\Sigma}$, which is transverse to  ${\Sigma}$, and
which coincides with $X$ for $r(x) \ge R$. On $M_{\rm ext}$ define
\[\rho = \Box_{ \og} \,t ,
\]
where $\Box_{ \og}$ is the d'Alembertian of the metric $\og$.  Let
$\rho_1 \in
C^\infty(M)$ be any function satisfying $\rho_1 = \rho$ for $r \ge R$.
Define $t_1 \in
C^\infty({\stackrel{o}{\cal D}\!\!(\Sigma)})$ as the unique solution of the
Cauchy problem
\begin{eqnarray}
\Box_{ \og} \, t_1 &= &\rho_1 ,\nonumber\\
t_1 \big|_{\Sigma} &= &0 ,\label{(problem)}\\
({\cal L}_{{\widetilde X}} t_1)\big|_{\Sigma} &= &1 ,\nonumber
\end{eqnarray}
Let
${\widetilde{\cal O}} \subset\,
{\stackrel{o}{\cal D}\!\!(\Sigma)}$ be defined by ${\widetilde{\cal O}} = \{p
\in\, {\stackrel{o}{\cal D}\!\!(\Sigma)} : \nabla
t_1 \; \mbox{is
timelike}\}$. Let ${\cal O} \subset {\widetilde{\cal O}}$ be the
domain of dependence
of $\Sigma$ in the space-time $({\widetilde{\cal O}}, g_{ab})$.  Lie
dragging
along integral curves of $\nabla t_1$, ${\cal O}$ can be identified with
$\{(t_1, p) \in \R \times
\Sigma : t^-(p) < t_1 < t^+(p)\}$ for some functions
$t^\pm(p)\in\R\cup\{\pm\infty\}$.
Compactness of
$\Sigma_{0, 2R}$ implies that there exists $\epsilon > 0$ such that
$(-4\epsilon,
4\epsilon) \times \Sigma_{0, 2R} \subset {\cal O}$.  From uniqueness
of solutions of the
problem (\ref{(problem)}) it follows that for $(t_1, x) \in \Omega = \{r
\ge R, |t| \le r-R\}$
we have $t_1 = t$. Decreasing $\epsilon$ if necessary we thus have
${\cal O}_{4\epsilon}
\subset {\cal O}$, where ${\cal O}_{s} = (-s,s) \times
\Sigma=\{p:|t_1(p)|<s\}$. Further decreasing $\epsilon$ if necessary,
we have $t_1 = t$ at all points in ${\cal O}_{4\epsilon}$ with $r > 2R$.

Next, we claim that for all $p \in {\stackrel{o}{\cal D}\!\!(\Sigma)}$, the set
$\{t\in\R:\phi[X]_t(p)\in {\cal O}_{2\epsilon}\}$ is nonempty and
bounded. {Namely, nonemptiness is an immediate consequence
of Corollary \ref{W1}.}
To prove boundedness, we note that if the orbit of $X$ through $p$
intersects $\Sigma$ at $r > 2R$ in $\Sigma_{\rm ext}$, then boundedness
follows immediately from the fact that $t_1 = t$ at all points in
${\cal O}_{2\epsilon}$ with $r > 2R$. Hence, if
$\phi[X]_{t_i}(p) \in {\cal O}_{2\epsilon}$ for an unbounded sequence
$\{t_i\}$, then infinitely many of these points lie in the compact set
$[-2\epsilon,2\epsilon] \times \Sigma_{0, 2R}$. Passing to a convergent
subsequence, we obtain a nonempty $\alpha$-limit set or $\omega$
-limit set (or both). However, this contradicts Lemma \ref{Q4}.

The desired time function $\tau$ may now be constructed as follows:
Let $t_2$ be the
Lipschitz continuous,
piecewise smooth function defined on $M$ by
\[
t_2(p) = \left\{
\begin{array}{ll}
t_1(p) , & p \in {\cal O}_{2\epsilon} ,\\
2\epsilon , &p \in I^+(\Sigma)\setminus{\cal O}_{2\epsilon} ,\\
-2\epsilon , &p \in I^-(\Sigma)\setminus{\cal O}_{2\epsilon} .
\end{array} \right.\]

Let $\varphi \in C^\infty(\R)$ be any function satisfying
\[
\varphi(x) =
\left\{
\begin{array}{ll}
0 , & x \le -\epsilon\\
1 , & x \ge \epsilon ,
\end{array} \right.
\]
$\varphi'(x)\big|_{(-\epsilon, \epsilon)} > 0$, and
\begin{equation}
\label{(PT.*.0)}
\varphi(x) - \frac{1}{2} = -\left(\varphi(-x) - \frac{1}{2}\right)\,.
\end{equation}
Define
\[
t_3 = \varphi \circ t_2 .
\]
It follows that $t_3 \in C^\infty(M)$ with
\begin{equation}
\label{(PT.1)}
t_3(p) = \left\{
\begin{array}{ll}
0 , & p \in I^-(\Sigma)\setminus{\cal O}_\epsilon\\
1 , & p \in I^+(\Sigma)\setminus{\cal O}_\epsilon ,
\end{array} \right.
\end{equation}
and
we also have
\begin{equation}
\label{(PT.2)}
\nabla t_3(p) = \left\{
\begin{array}{ll}
0 , & p \in M\setminus{\cal O}_\epsilon\\
\mbox{timelike, past pointing} , & p \in {\cal O}_\epsilon .
\end{array} \right.
\end{equation}
For $p\in{\stackrel{o}{\cal D}\!\!(\Sigma)}$ we define
\begin{equation}
\label{(PT.*)}
\tau(p) = \int_{-\infty}^0 t_3(\phi[X]_s(p)) ds + \int_0^\infty (t_3-1)
(\phi[X]_s(p)) ds .
\end{equation}

We now claim:

\quad 1)
\begin{eqnarray}
\label{(PT.3)}
\hfill \qquad \qquad \qquad \qquad &&\mbox{for} \quad p \in {\stackrel{o}{\cal
D}\!\!(\Sigma)} ,
\qquad |\tau(p)| <
\infty .
\end{eqnarray}

Indeed let $p \in\, {\stackrel{o}{\cal D}\!\!(\Sigma)}$. Then, as shown above,
there exists $t_\pm$,
$|t_\pm| < \infty$,
such that for $t \ge t_+$ $t_3 (\phi[X]_t(p)) = 1$, and for $t
\le t_-$ $t_3 (\phi[X]_t(p)) = 0$,
 so that the equation (\ref{(PT.*)})
reads
\be
\label{4.12}
\tau(p) = \int_{t_-}^0 t_3(\phi[X]_s(p)) ds + \int_0^{t_+} (t_3-1)
(\phi[X]_s(p)) ds
\ee
which proves (\ref{(PT.3)}).
Standard results about integrals with
parameters and (\ref{4.12}) imply $\tau\in
C^\infty({\stackrel{o}{\cal D}\!\!(\Sigma)})$.

\quad 2)
\begin{eqnarray}
\label{(PT.4)}
\hfill \qquad \qquad \qquad \qquad &&X(\tau) = 1 .\qquad \qquad
\qquad \qquad \hfill
\end{eqnarray}
Namely, by a change of variables we have
\[
\tau(\phi[X]_t(p)) = \int_{-\infty}^t t_3(\phi[X]_s(p)) ds + \int_t^\infty
(t_3-1) (\phi[X]_s(p)) ds ,
\]
so that
\[
X(\tau)\big|_p = \frac{d}{dt} \tau(\phi[X]_t(p))\big|_{t=0} = 1 .
\]
[Let us note, that (\ref{(PT.4)}) implies that for $p \in {\stackrel{o}{\cal
D}\!\!(\Sigma)}$, we
have
$X_p \ne 0$ ({\em cf.} also Corollary \ref{CQ2})].

\quad 3)
\begin{eqnarray}
\label{(PT.5)}
\hfill \qquad \qquad \qquad &&\nabla \tau \quad \mbox{is timelike
on} \quad {\stackrel{o}{\cal D}\!\!(\Sigma)}
.\qquad \qquad \qquad \hfill
\end{eqnarray}
Namely, we have
\[
\nabla \tau(p) = \int_{-\infty}^\infty ({\phi[X]_{s}}_* \nabla t_3)(p) ds ,
\]
so that (\ref{(PT.5)}) follows from (\ref{(PT.2)}) and from the fact,
that for any
isometry
$\psi$ defined on a connected set the tangent map $\psi_*$ maps
timelike, consistently 
oriented
vectors to timelike, consistently oriented vectors.

\quad 4)
\begin{eqnarray}
\label{(PT.6)}
\qquad \qquad&& \mbox{the level sets of} \quad \tau \quad \mbox{are
asymptotically flat} .\qquad \qquad \hfill
\end{eqnarray}
To prove (\ref{(PT.6)}), consider $p = (t, x) \in
\{r\ge R+\epsilon,|t|\le r-R\}$, with $R$, $\epsilon$ as at the beginning
of the proof of this
Proposition. For
such $p $  we have
$t_3(\phi[X]_s(p)) = \phi[X](s+t)$, so that $\partial\tau/\partial x^i
=0$, thus $\tau=\tau(t)$.
(\ref{(PT.4)}) gives $\tau = t +
\tau_0$ 
for some constant 
$\tau_0$. 
The symmetry condition
(\ref{(PT.*.0)})
gives $\tau_0 = 0$, so that
in  $
\{r\ge R+\epsilon,|t|\le r-R\}$
$\tau$ coincides
with the original Killing
time $t$.

\quad 5)
\begin{eqnarray}
\label{(WT.1)}
\qquad \qquad&& \mbox{the level sets of} \quad \tau \quad \mbox{are
Cauchy surfaces for} \quad {\stackrel{o}{\cal D}\!\!(\Sigma)}.\qquad \qquad
\hfill
\end{eqnarray}
To prove (\ref{(WT.1)}), we note first that (\ref{(PT.5)}) implies that
any level surface of $\tau$ is achronal. Consider the
restriction of $\tau$ to the original
slice $\Sigma$. By the previous argument, we have $\tau|_{\Sigma} = 0$
 for all
$r > R$, so $\tau|_{\Sigma}$
can be nonvanishing only on a compact subset of
$\Sigma$, and, hence, there exists $T \in \R$ such that $|\tau| < T$ on
$\Sigma$. Applying $\phi[X]_T$ to this result (using (\ref{(PT.4)})), we see
that $\tau > 0$ everywhere on $\Sigma_T$. Similarly, we have $\tau < 0$
on $\Sigma_{-T}$. However, by Proposition \ref{Q1a}, $\Sigma_T$ and
$\Sigma_{-T}$ are Cauchy surfaces for ${\stackrel{o}{\cal D}\!\!(\Sigma)}$.
Hence, every inextendible
{timelike}
curve in ${\stackrel{o}{\cal D}\!\!(\Sigma)}$ intersects $\Sigma_T$ and
$\Sigma_{-T}$, and, thus,
passes from positive to negative values of $\tau$. By continuity, every
such inextendible {timelike}
curve intersects the level set $\tau = 0$, thus
implying that this level set is a Cauchy surface for ${\stackrel{o}{\cal
D}\!\!(\Sigma)}$. That all the
other level sets of $\tau$ also are Cauchy surfaces for ${\stackrel{o}{\cal
D}\!\!(\Sigma)}$ then
follows by application of Proposition \ref{Q1a}.

The above results complete the proof of Proposition \ref{PT.1a}.
\hfill$\Box$

In order to prove Proposition \ref{PT.1b}, we first shall need the following
Lemma:

\begin{Lemma}
\label{W.2b}
Let $(M, g_{ab})$ be a space-time of class (b) and suppose that
(\ref{(empty)}) holds.
Then there exists a smooth spacelike
hypersurface $\Sigma_0$ with boundary
$S$ such that $\Sigma_0$ coincides with $\Sigma$ outside of a compact
set, $\Sigma_0\cap{\stackrel{o}{\cal D}\!\!(\Sigma)} $ is a Cauchy
surface for ${\stackrel{o}{\cal D}\!\!(\Sigma)}$, and $X$ is everywhere
transverse to $\Sigma_0
\setminus S$
\end{Lemma}

{\bf Proof:}
The idea of the proof is to construct a time function $\tau$
in ${\stackrel{o}{\cal D}\!\!(\Sigma)}$ in a way somewhat similar to that
in the proof of Proposition \ref{PT.1a}.
Supplementary difficulties arise because we need to ensure
that the hypersurface $\Sigma_0$ defined by $\tau = 0$
smoothly intersects $S$ and remains
spacelike ``up to boundary''. Thus, the proof of the Lemma will consist mainly
of
 constructing a function $t_1$ [{\em cf.\ }eq. (\ref{(problem)})] near $S$ so
that
$\Sigma_0$ will have the desired properties there.

Let $N$ denote the normal bundle of $S$, i.e., a point of $N$
consists of a point $p \in S$ together with a vector $V^a$ at $p$ which
is normal to $S$. As in the proof of Lemma \ref{AQ4}, let $n^a$ denote the
future-directed unit normal field to $\Sigma$ on $S$, and let $r^a$
denote the inward (``towards $\Sigma$") pointing unit normal to $S$ in
$\Sigma$. Then at each $p \in S$, $(n^a, r^a)$ is a basis for normal
vectors to $S$ at $p$, so we can uniquely express each normal
vector, $V^a$, at $p$ as $V^a = Tn^a + Zr^a$. Thus, each $x \in N$ is uniquely
characterized by $(p,T,Z)$ with $p \in S$ and $T,Z \in \R$.

Consider the ``wedge", ${\cal R}$, of $N$ defined by $Z > |T|$.
It is convenient
to introduce ``Rindler coordinates" $(t',z')$ in ${\cal R}$ by,
\begin{eqnarray}
 t' &= &\tanh^{-1}(T/Z) ,\label{Rindlert'}\\
z' &= &\sqrt{Z^2 - T^2} ,\label{Rindlerz'}
\end{eqnarray}
so that $z'$ is just the length of the normal vector to $S$.
Now, since $\phi[X]_t$ maps $S$ into $S$ and maps vectors normal to $S$
into vectors normal to $S$, it induces an action on $N$, which we shall
denote by $\phi'[X]_t$. Since $\phi[X]_t$ preserves the length
of vectors normal to $S$, it
follows immediately that on ${\cal R}$, the action of $\phi'[X]_t$ takes
the form,
\begin{equation}
\label{phi'}
\phi'[X]_t (p,t',z') = (\phi[X]_t(p), t' + f_t(p),z')
\end{equation}
for some function $f_t$ on $S$. Furthermore, it follows
directly from the proof of Lemma \ref{AQ4} that
there exists $a > 0$ such that we have
$f_{t_+}(p) > a$ for all $p \in S$, where $t_+$ was defined in Lemma
\ref{AQ4}. Consequently, for all $n \in \N$ we have
$f_{nt_+}(p) > na$
for all $p \in S$, from which it follows that
$f_{t}(p)\rightarrow \infty$ as $t \rightarrow \infty$ and
$f_{t}(p)\rightarrow - \infty$ as $t \rightarrow - \infty$. In other words,
along every orbit, $\phi'[X]_t$, in ${\cal R}$, we have
$t'\rightarrow \infty$ as $t \rightarrow \infty$ and
$t'\rightarrow - \infty$ as $t \rightarrow - \infty$.

We now repeat
the construction used to obtain
the time function $\tau$ of
Proposition \ref{PT.1a}, replacing ${\stackrel{o}{\cal D}\!\!(\Sigma)}$ by
${\cal R}$,
replacing $\phi[X]_t$ by $\phi'[X]_t$ and replacing
the time function $t_1$ (defined in a neighborhood of $\Sigma$ in
${\stackrel{o}{\cal D}\!\!(\Sigma)}$ by eq. (\ref{(problem)})) by the function
$t'$ (which is globally
well defined in ${\cal R}$, so we choose an arbitrary $\epsilon > 0$
to define the analog, $t'_2$,
of $t_2$).
{}From what has been said at the end of the previous paragraph it
follows that the
resulting function, denoted $\tau'$,
will be well defined smooth
on ${\cal R}$ (although it will be singular
on the boundary of ${\cal R}$). Furthermore, it follows directly from
the construction that throughout ${\cal R}$, we have
$\pt \tau'/\pt t' > 0$ and
$\pt \tau'/\pt z' = 0$, where the partial derivatives are taken holding
$p$ fixed. (The first relation is the analog of property (3) in Proposition
\ref{PT.1a}, while the second relation follows from the fact that
both $t'$ and the action of $\phi'[X]_t$ [{\em cf.\ }eq.\
(\ref{phi'})] are independent
of $z'$.) It follows from the above properties of $\tau'$ that in ${\cal R}$,
the level surface, $\Sigma'_0$, defined by
$\tau' = 0$ is given by an equation of the form
$t' = g(p)$, where g is a smooth function on $S$. However, returning to
the original, globally nonsingular coordinates $(T,Z)$ on $N$, we see
immediately that any hypersurface of this form in ${\cal R}$
can be smoothly extended to a hypersurface $\widetilde{\Sigma}'_0$
through $S$, and the magnitude of the ``slope"
of $\widetilde{\Sigma}'_0$ at $S$ satisfies $|\pt T/\pt Z| < 1$.

Now, let $\psi : N \rightarrow M$ be the map which takes each
$(p, V^a) \in N$ to the point in $M$ lying at unit affine parameter
along the geodesic in $M$ determined by the initial conditions
$(p, V^a)$. Then, by the same arguments as used for the ordinary
exponential map, it follows that $\psi$ is a diffeomorphism from some
open neighborhood, ${\cal U}$, of $S$ in $N$ to an open neighborhood,
${\cal V}$ of $S$ in $M$. In particular, the image, $\widetilde{\Sigma}_0$,
under $\psi$ of, $\widetilde{\Sigma}'_0$ is a smooth hypersurface through
$S$ in $M$. Furthermore, the above ``slope" condition implies that
$\widetilde{\Sigma}_0$ is spacelike at $S$.

The pullback under $\psi^{-1}$ of the function $t'$ on ${\cal R} \cap
{\cal U}$ defines a smooth function, $\hat{t}$, on ${\stackrel{o}{\cal
D}\!\!(\Sigma)} \cap {\cal V}$.
In ${\stackrel{o}{\cal D}\!\!(\Sigma)} \cap {\cal V}$, we define the function
$\hat{\rho}$ by
\[\hat{\rho} = \nabla^a \nabla_a \hat{t}
\]
Note that from the definition of $t'$, it follows
immediately that in a neighborhood
of $S$ on $\Sigma$, we have
{$\hat{\lambda} \equiv n^a \nabla_a \hat{t} > 0$},
where $n^a$ here denotes the unit normal to $\Sigma$;
indeed, we have
{$\hat{\lambda} \rightarrow \infty$} in the limit as
$S$ is approached.

As in the proof of Proposition \ref{PT.1a}, let $(t, x)$ be coordinates
on $M_{\rm ext}$ as given by
Proposition
\ref{PD.2}, normalized so that $A=1$ in all
the ends and so that $g_{\mu\nu}$ is asymptotically
conformal (with constant
conformal factor) to the standard
Minkowski metric.
Again, we let $\og$ be any smooth Lorentzian metric on $M$ which,
coincides with $g$ on $M_{\rm int}$, and such that
\[
\og = -dt^2 + \sum_{i=1}^3 {(dx^i)}^2
\]
for all $x$ such that $r(x) \ge R$, for some $R \ge R_i$, $i =
1,\ldots,I$, on all ends
$M_i$.  Let  ${\hat
X}$ be any smooth vector field defined on a neighborhood of
${\Sigma}$, which is transverse to ${\Sigma}$, and
which coincides with $X$ for $r(x) \ge R$ and which coincides with
the unit normal, $n^a$, to $\Sigma$ on a neighborhood of
$S$ in $\Sigma$. Again, on $M_{\rm ext}$ define
\[\rho = \Box_{ \og} \,t ,
\]
where $\Box_{ \og}$ is the d'Alembertian of the metric $\og$. Let
$\hat{\rho}_1 \in
C^\infty(M)$ be any function satisfying $\hat{\rho}_1 = \rho$ for $r \ge R$
and $\hat{\rho}_1 = \hat{\rho}$ in the intersection of ${\stackrel{o}{\cal
D}\!\!(\Sigma)}$ with some
open neighborhood of $S$. Let $\hat{\lambda}_1 \in
C^\infty(\Sigma \setminus S)$ be any function satisfying
$\hat{\lambda}_1 >0$,
$\hat{\lambda}_1 = 1$ for $r \ge R$
and $\hat{\lambda}_1 = \hat{\lambda}$ in in the intersection
of $\Sigma \setminus S$ with some
open neighborhood of $S$. Let $\hat{\sigma}_1 \in
C^\infty(\Sigma \setminus S)$ be any function satisfying
$\hat{\sigma}_1 = 0$ for $r \ge R$
and $\hat{\sigma}_1 = \hat{t}$ in in the intersection
of $\Sigma \setminus S$ with some
open neighborhood of $S$. Define $\hat{t}_1 \in
C^\infty({\stackrel{o}{\cal D}\!\!(\Sigma)})$ as the unique solution of the
Cauchy problem
\begin{eqnarray}
\Box_{ \og} \, \hat{t}_1 &= &\hat{\rho}_1 ,\nonumber\\
\hat{t}_1 \big|_{\Sigma} &= &\hat{\sigma}_1 ,\label{(newproblem)}\\
({\cal L}_{{\hat X}} \hat{t}_1)\big|_{\Sigma} &= &\hat{\lambda}_1,
\nonumber
\end{eqnarray}
Then $\hat{t}_1$ coincides with $\hat{t}$ in the intersection of
${\stackrel{o}{\cal D}\!\!(\Sigma)}$
with some
neighborhood of $S$.
On the neighbourhood ${\cal U}$ of $S$ defined above we have two
families of maps defined, $\phi_t[X]$ and $\phi'_t[X]$, and we claim
that these maps coincide, wherever this statement makes sense. This follows
from the fact that $\phi_t[X]$ maps geodesics starting at $S$ to
geodesics
starting at $S$, and preserves the geodesic distance along them.
Consequently, the corresponding
function $\hat{\tau}$ {in ${\stackrel{o}{\cal D}\!\!(\Sigma)}$ obtained
by the construction of Proposition \ref{PT.1a} will coincide in a
sufficiently small neighborhood of $S$ with the pullback
under $\psi^{-1}$ of the function $\tau'$ on $N$ constructed above. It then
follows
that the}
level surface $\hat{\tau} = 0$ can be smoothly extended to $S$ so as
to obtain a hypersurface with boundary satisfying all of the conditions
of the Lemma.
\hfill $\Box$

{\bf Proof of Proposition \ref{PT.1b}:} Let $\Sigma_0$ be as in Lemma
\ref{W.2b}, and let $\widetilde{\Sigma}$ be any smooth, spacelike
extension of $\Sigma_0$ through $S$. Let $\tilde{n}^a$ denote the vector
field -- defined in a neighborhood, ${\cal \widetilde O}$, of
$\widetilde{\Sigma}$ -- of tangents to the geodesics normal to
$\widetilde{\Sigma}$. In ${\cal \widetilde O} \cup M_{\rm ext}$,
define the
vector field $\widetilde{X}$ by
\[\widetilde{X}^a = \chi \tilde{n}^a + (1 - \chi) X^a
\]
where $\chi$ is any nonnegative, smooth function on $M$ such that
$\chi = 1$ in $M_{\rm int}$ and for $r \le R$ in $M_{\rm ext}$
and $\chi = 0$ for $r \ge 2R$, where
$R$ is defined as in the proof of Proposition \ref{PT.1a}. Then
$\widetilde{X}$ is transverse to $\widetilde{\Sigma}$, and by the same
arguments as in the proof of the first part of Theorem \ref{Q6b}, there
exists $\epsilon > 0$ such that the induced action
$\phi[\widetilde{X}]_{\tilde{t}}$
yields a spacelike foliation of an open neighborhood of
$\Sigma_0$ for $\tilde{t} \in (- 2\epsilon, 2\epsilon)$. By decreasing
$\epsilon$
if necessary, we can ensure that the hypersurfaces
$\widetilde{\Sigma}_{\pm \epsilon}$,
defined by $t = \pm \epsilon$, intersect ${\cal H}$ in cross-sections
of ${\cal H}_+$ and ${\cal H}_-$, respectively. By further decreasing
$\epsilon$ if necessary, we can ensure that
$\widetilde{\Sigma}_{\pm \epsilon} \cap M_{\rm int}$ lies
outside the causal
future of the portion of $\widetilde{\Sigma}$ with $r \ge {2}R$.

We claim, now, that the hypersurfaces
$\widetilde{\Sigma}_{\pm \epsilon}$ are transverse to $X$ in an
open neighborhood of
$D(\Sigma) = D(\Sigma_0)$.
Namely, in $M_{\rm ext}$, this is obvious because $X$ is timelike.
In $M_{\rm int}$, the vector field $\tilde n^a$ is normal to
$\widetilde{\Sigma}_{\pm \epsilon}$. However, $X_a \tilde n^a$ is
the inner product of a Killing field with a geodesic tangent, and
thus is constant along
the geodesics tangent to $\tilde n^a$. For any point
$q \in \widetilde{\Sigma}_{\pm \epsilon} \cap M_{\rm int}$
the normal geodesic through $q$ is a timelike curve and, hence,
must intersect
$\widetilde{\Sigma}$ in $\Sigma_0 \setminus S$.
Since $X$ is transverse to
$\Sigma_0 \setminus S$, we have $X_a \tilde n^a <0$ everywhere on
$\Sigma_0$, and, hence, $X_a \tilde  n^a < 0$
everywhere on $\widetilde{\Sigma}_{\pm \epsilon} \cap D(\Sigma_0)$.
This implies that $X$ is transverse to $\widetilde{\Sigma}_{\pm \epsilon}$
in an open neighborhood of $D(\Sigma)$.

It follows immediately that the action of $\phi[X]_{\tilde{t} - \epsilon}$
on $\widetilde{\Sigma}_{\epsilon}$ for $\tilde{t} > \epsilon$ yields a
foliation. As a consequence of Corollary \ref{W1} and
Lemma \ref{AQ4}, this foliation covers an open region containing
$[I^+(\widetilde{\Sigma}_{\epsilon})] \cap [{\stackrel{o}{\cal D}\!\!(\Sigma)}
\cup {\cal H}]$.
On the other hand, the transversality of $X$ to
$\widetilde{\Sigma}_{\epsilon}$ and the tangency of $X$ to ${\cal H}$
imply that no point in
$[J^-(\widetilde{\Sigma}_{\epsilon})] \cap [{\stackrel{o}{\cal D}\!\!(\Sigma)}
\cup {\cal H}]$ is covered
by this foliation. Similarly, the action of $\phi[X]_{\tilde{t} + \epsilon}$
on $\widetilde{\Sigma}_{-\epsilon}$ for $\tilde{t} < -\epsilon$ yields a
foliation covering an open region containing
$[I^-(\widetilde{\Sigma}_{-\epsilon})] \cap [{\stackrel{o}{\cal D}\!\!(\Sigma)}
\cup {\cal H}]$
but containing no point in
$[J^+(\widetilde{\Sigma}_{-\epsilon})] \cap [{\stackrel{o}{\cal D}\!\!(\Sigma)}
\cup {\cal H}]$.
Consequently, by merging these foliations with the foliation defined
by $\widetilde{X}$ for $\tilde{t} \in [-\epsilon, \epsilon]$, we obtain
a foliation covering an open region containing ${\stackrel{o}{\cal
D}\!\!(\Sigma)} \cup {\cal H}$.
The time function, $\tilde{t}$, of this foliation is easily verified to
satisfy all the properties of this Proposition with one exception:
although $\tilde{t}$ is continuous, it is only piecewise smooth in
that it need not be smooth on $\widetilde{\Sigma}_{\pm \epsilon}$
{when $r > 2R$.}
However, since this possible
non--smoothness
of $\tilde{t}$ within $D(\tilde\Sigma)$
occurs only on a set which intersects  $D(\Sigma)$ compactly,
we may smooth it
{\em e.g.}, by convolution with localized
Friedrich mollifiers) to obtain a time function $\tilde{\tau}$ satisfying
all the conditions of the Proposition.

\hfill$\Box$

Before proving Theorems \ref{TM.1} and \ref{TM.2} we prove the
following Lemma.

\begin{Lemma}
\label{LT.0}
Let $(M, g_{ab})$ be a space-time of class (a) or (b), let $t$ be a
time function defined
in a globally hyperbolic neighbourhood ${\cal O}$ of $\Sigma$,  such
that
\begin{enumerate}
\item
$\Sigma$ is a Cauchy surface for ${\cal O}$, and
\item there exist constants $\theta \in [0, 1)$,
$T \in
\R$ such that for all $i = 1,\ldots,I$,
\[
{\cal O} \cap M_i \subset \Omega_{\theta, {R}_i,T} = \{r
\ge {\widetilde
R}, |t| < \theta (r-R_i) + T\}\ ,
\]
\item
and $t\big|_{{\cal
O} \cap M_{\rm ext}}$ coincides with the Killing time based on
$\Sigma\cap M_{\rm ext}$, as defined in
Proposition
\ref{PD.2}, for $r \ge R$ for some $R$.
\end{enumerate}
 Let\footnote{ The reader should be warned about a slight clash of
notation  here, due to the fact that we have run out of reasonable
symbols for denoting hypersurfaces: $\Sigma_1$ in this Lemma is {\em
not} the asymptotic end of $\Sigma$, as defined in Definition
\ref{D1}.} $\Sigma_1 \subset {\cal
O}$ be a 
smooth, embedded, acausal
hypersurface which is a graph of a function $u_1$ over $\Sigma$:
\[
\Sigma_1 = \{p \in {\cal O} : t(p) = u_1(q), q \in \Sigma\} ,
\]
If $(M, g_{ab})$ is of class (b), suppose moreover that $u_1\big|_S =
0$.  Then
$\Sigma_1$ is a Cauchy surface for ${\stackrel{o}{\cal D}\!\!(\Sigma)}$.
\end{Lemma}

{\bf Proof}:
We can identify ${\cal O}$ with $\{(t,q): t_-(q)<t<t_+(q),q\in\Sigma\}$,
for some functions $t_\pm\in\R\cup\{\pm\infty\}$, by dragging
$q\in\Sigma$ along the integral curves of $\nabla t$. Let $\lambda$ be
a piecewise differentiable inextendible causal curve in $({\cal
O},g_{ab})$, so $\lambda:(s_-,s_+)\ni s \ra (s,\gamma(s))\in {\cal
O}$ for some piecewise differentiable curve
$\gamma:(s_-,s_+) \ra \Sigma$. Since $\Sigma$ is a Cauchy
surface for ${\cal O}$ we have $\lambda\cap \Sigma\ne\emptyset$,
and hence
$0\in(s_-,s_+)$.
For any
$\theta \in (0, 1)$ one
can choose ${\cal R}(\theta)$ large enough such that for $p \in M_i$,
$i = 1,\ldots,i$
and $r(p) \ge {\cal R}(\theta)$ the slopes of the light cones, in the
coordinates of
Proposition \ref{PD.2}, lie between $\frac{1+\theta}{2}$ and
$\frac{3-\theta}{2}$, which easily implies that there exists $C$ such
that   $\gamma(s_-,s_+)\subset\Sigma_{0,C}$. Consider a sequence
$s_i \ra s_-$.
Compactness of $\Sigma_{0,C}$ implies that there exists
$q_-\in\Sigma_{0,C}$ and a subsequence,
still denoted by $s_i$, such that $\gamma(s_i) \ra q_-$.
Suppose that $s_-\ne t_-(q_-)$. Then the timelike curve $(
t_-(q_-),s_-)\ni s \ra (s,q_-)$ extends $\lambda$ which is not
possible; thus $s_-= t_-(q_-)$. By compactness of  $\Sigma_{0,C}$ there
exists $\epsilon>0$ such that $(u-t_-)\big|_{\Sigma_{0,C}}\ge
\epsilon$. We thus have $\lim_{s_i \ra s_-}u(\gamma(s_i))=
u(q_-)\ge\epsilon +t_-(q_-)=\epsilon +s_-$. Set $f(s)=s-u(\gamma(s))$;
continuity of $u(\gamma(s))$ implies that there exists $s_1\in
(s_-,s_+)$ such that $f\big|_{(s_-,s_1)}<0$. One shows similarly that
there exists $s_2\in
(s_-,s_+)$ such that  $f\big|_{(s_2,s_+)}>0$, by continuity of $f$
there exists $s_0\in
(s_-,s_+)$ such that $f(s_0)=0$, so
$(s_0,\gamma(s_0))\in\Sigma_1\cap\lambda\ne\emptyset$, which
is what had to be
established.
\hfill $\Box$

{\bf Proof of Theorem \ref{TM.1}}:  Let $\tau$ be the time function as
given by
Proposition \ref{PT.1a};  we want to show that Bartnik's interior
condition
(\cite{Ba1} [eq. (5.4)] or \cite{BCOM} [eq. (4.4)]) holds.  Set
\[
M_\delta = \bigcup_{t \in (-\infty,\infty)} \phi[X]_t(\Sigma_{0,\delta})
\ ,
\]
where $\Sigma_{0,\delta}\equiv \Sigma_{0,\delta}(\tau)$,
with $\Sigma_{0,\delta}(\tau)$  defined by (\ref{hypdelta}). Let $q\ \in
\pt M_\delta$, $p \in M_\delta \setminus I(q)$,
{where $I(q)$ denotes the set of points causally related to $q$.}
By {Corollary
\ref{W1}} there
exists $t_0$ such that $\phi[X]_{t_0}(q) \in \Sigma_0(\tau)$.
  We have
\begin{eqnarray*}
|\tau(p) - \tau(q)| = |\tau(\phi[X]_{t_0}(p))| &\le &\sup_{p' \in
M_\delta \setminus
I(\phi[X]_{t_0}(q))} |\tau(p')|\\
&\le &\sup_{q'\in\pt\Sigma_{0,\delta}, p'\in M_\delta \setminus I(q')}
|\tau(p')| .
\end{eqnarray*}
For $q' \in \pt \Sigma_{0,\delta}$ let ${\cal N} = \pt (M_\delta
\setminus I(q'))$;  since
${\cal N}$ is null, uniform boundedness of light--cone slopes in
${ M}_{\rm ext}$
implies the existence of a  constant $C=C(\delta)$ such that
we have
\[
|\tau_{|{\cal N}}| \le C\,.
\]
It follows that
$M_\delta \setminus I(q') \subset{\tilde{\cal D}}
\left(\bigcup_{|t|\le C}
\Sigma_{t,\delta}\right)$ which
by Corollary \ref{compactness1} is a compact set, and therefore there exists a
constant $C_1$ such that
\[
|\tau(p) - \tau(q)| \le C_1 ,
\]
which had to be established. Existence of a maximal slice $\hat\Sigma_0$
follows by the arguments of
Proposition
4.3 and Theorem 4.1 of \cite{BCOM}.  $\hat\Sigma_0$ is an asymptotically
flat graph
by construction and thus a Cauchy surface by Lemma \ref{LT.0}.  The
hypersurfaces
$\hat\Sigma_t = \phi[X]_t(\hat\Sigma_0)$ cover ${\stackrel{o}{\cal
D}\!\!(\Sigma)}$ by {Corollary
\ref{W1}}.   If
moreover the
timelike convergence condition holds, the family $\{\hat\Sigma_t\}$ forms
a foliation
of ${\stackrel{o}{\cal D}\!\!(\Sigma)}$
({\em cf.\ e.g.} the arguments of the proof of Theorem 5.4
of
\cite{BCOM},
and \cite{Ba2} [Section 5]) which completes the proof of
Theorem
\ref{TM.1}. \hfill $\Box$

{\bf Proof of Theorem \ref{TM.2}}:  Let ${\tilde \tau}$ be the time
function given by Proposition
\ref{PT.1b}. By Theorem 3.1 of \cite{BCOM}
$\Sigma_0({\tilde \tau})$  can be
deformed outside of a compact set $K \subset \Sigma_0({\tilde \tau})$,
$\pt K \supset
S$, to an asymptotically flat hypersurface $\tS$ the mean extrinsic
curvature of
which is compactly supported.
By construction, $\tS$ is an asymptotically flat ${\tilde
\tau}$--graph of a function
${\tilde u}$, with ${\tilde u}\big|_K = 0$, so that $\tS$ is a Cauchy
surface for ${\stackrel{o}{\cal D}\!\!(\Sigma)}$ by
Lemma \ref{LT.0}.  Enlarging $K$ if necessary the function ${\tilde
\tau}_1 = {\tilde
\tau} - {\tilde u}$ is a time function on a neighbourhood $\U$ as
given by Proposition
\ref{PT.1b}, and ${\tilde \tau}_{1|M_{\rm ext}}$ is the Killing time
based on $\tS$, as
described in Proposition \ref{PD.2}.  Let $\rho$ be such that the 
hypersurfaces $\tS_t$ are \locally
maximal outside of $\tS_{t,\rho}$, set $\tS_{\rm ext} = \tS
\setminus\tS_{0,\rho}$,
define
\[
\pt {\widetilde M}_{\sigma} = \bigcup_{t \in [-{\sigma}, {\sigma}]}
\pt \tS_{t,\rho} .
\]
Because the light cones in ${\widetilde M}_{\rm ext} = \bigcup_{t\in
(-\infty,\infty)}
\tS_{t,{\rm ext}}$ have uniformly bounded slopes it follows that there
exists a constant
$K_1$ such that if $\Sigma$ is a spacelike hypersurface then we have
the implication
\begin{equation}
\label{(MD.8)}
\Sigma \cap \pt {\widetilde M}_1 \ne \emptyset \Longrightarrow \Sigma
\cap \pt {\widetilde
M}_{\rm ext} \subset \pt {\widetilde M}_{K_1} .
\end{equation}
By Theorem 4.2 of \cite{Ba1} and by 
Theorem \ref{Q6b} for $i \in \N$
there exists a
\locally  maximal surface $\Sigma_i$ such that $\pt \Sigma_i = \pt
\tS_{0, i}$.

For a set $\Omega$ define
\begin{eqnarray*}
t_+(\Omega) &= & \sup \{{\tilde \tau}_1(p) : p \in \Omega \cap \pt
M_{\rm
ext}\}\\
t_-(\Omega) &= & \inf \{{\tilde \tau}_1(p) : p \in \Omega \cap \pt
M_{\rm ext}\}
{}.
\end{eqnarray*}
We have
\[
t_\pm (\phi[X]_t(\Omega)) = t_\pm(\Omega) + t ,
\]
thus there exists a constant $t_i$ such that
\begin{equation}
\label{(MD.9)}
t_-(\phi[X]_{t_i}(\Sigma_i)) = 1 .
\end{equation}
Set $\phi[X]_i\equiv\phi[X]_{t_i}$. (\ref{(MD.9)}) and (\ref{(MD.8)}) imply
\[
\phi[X]_i(\Sigma_i) \cap \pt \widetilde M_{\rm ext} \subset \pt {\widetilde
M}_{K_1}\,,
\]
and spacelikeness of $\Sigma_i$ gives
\[
\phi[X]_i(\Sigma_i) \cap \widetilde M_{\rm int} \subset K =
{\tilde{\cal D}}\left(\bigcup_{t\in[-K_1,K_1]}
\phi[X]_t(\tS_{\rm
int})
\right)
\ ,
\]
where $\widetilde M_{\rm int}=M\setminus \widetilde M_{\rm ext}$.
$K$
is compact by 
Corollary \ref{compactness2}, thus
\[
|{\tilde \tau}_1|\Big|_{\phi[X]_i(\Sigma_i)\cap \widetilde M_{\rm int}}
\le C = \sup_{p \in K}
|{\tilde \tau}_1(p)| .
\]
Suppose first that $t_i\le 0$. It follows that ${\tilde
\tau}_1\big|_{\pt \phi[X]_i(\Sigma_i)} \le 0$, and  we conclude
from the estimate
(3.14) of \cite{Ba1} [Theorem 3.1] that there exists a constant $C$
such that the ``tilt
function" $\nu_i=\nu(\phi[X]_i(\Sigma_i))$ ({\em cf.} \cite{Ba1} for
details) satisfies
\[
|\nu_i|\Big|_{\phi[X]_i(\Sigma_i) \cap \pt {\widetilde M}_{\rm ext}} \le C
{}.
\]
Thus the argument of the proof of Theorem 5.3 of \cite{Ba1} applies to
give
\be
\label{5.35.1}
t_i \ge -C_1
\ee
for some constant $C_1$.  Suppose now that $t_i \ge K_1+1$.  It follows
from
(\ref{(MD.8)})--(\ref{(MD.9)}) that there exists $0 > s_i \ge -K_1-1$
such that
\begin{equation}
\label{(MD.10)}
t_+(\phi[X]_{t_i+s_i}(\Sigma_i)) = -1 ,
\end{equation}
and since $t_i + s_i \ge 0$ one can again conclude as above, from
Theorems 3.1 and 5.3 of \cite{Ba1}, that
\be
\label{5.37}
t_i + s_i \le C_2 \Longrightarrow t_i \le C_2 + K_1+1 \ .
\ee
(\ref{5.37}) together with (\ref{5.35.1}) gives
\[
|t_i| \le C_3
\]
for some constant $C_3$, and
from what has been said
it follows that
\[
|{\tilde \tau}_1|\Big|_{\Sigma_i} \le C_4 .
\]
{}From the estimate (3.10) of \cite{Ba1} we conclude that
\[
|\nu(\Sigma_i)| \le C_5
\]
and a standard method, together with barrier considerations using the
barriers $u_\pm
= \pm C r^{-\alpha}$ if $\alpha \in (0,1)$, $u_\pm = \pm C r^{-1} \log
r$ if $\alpha =
1$ ({\em cf.\ e.g.} Lemma 2.2 of \cite{BCOM}) shows that there exists
a subsequence
converging to an asymptotically flat \locally  maximal surface
$\hat\Sigma_0$.  The
hypersurfaces $\hat\Sigma_t$ are obtained by setting $\hat\Sigma_t =
\phi[X]_t(\hat\Sigma_0)$.  By {Corollary
\ref{W1}}
for $p \in {\stackrel{o}{\cal D}\!\!(\Sigma)}$ every orbit $\phi[X]_t(p)$
intersects
$\hat\Sigma_0$, so
that the hypersurfaces $\{\hat\Sigma_t\}_{t\in\R}$ cover ${\stackrel{o}{\cal
D}\!\!(\Sigma)}$.  If the
timelike convergence
condition holds, then  the hypersurfaces $\hat\Sigma_t\setminus S$
foliate ${\stackrel{o}{\cal D}\!\!(\Sigma)}$
and the
proof is completed. \hfill$\Box$

\appendix

\section{A generalization --- \stationary\ space--times.}
\label{rotating}

It is an interesting  feature of the causal structure theory presented in
Section \ref{structure} that several results follow just from
the fact that the orbits of the isometry group are {\em
time--oriented} ``near spacelike infinity'', so that timelikeness
of the Killing vector is never used. This leads to the following
curious generalization of our results to space--times which are
asymptotically
``\stationary'', in the following sense:

\begin{Definition}
\label{AD1}
A spacetime $(M, g_{ab})$ possessing an acausal slice $\Sigma$
expressed in the form $\Sigma = \Sigma_1 \cup \Sigma'$ will be called
$(k, \alpha)$--asymptotically \stationary\ with respect to the ``end"
$\Sigma_1$ if 
conditions \ref{D1.1} and \ref{D1.2} of Definition \ref{D1} hold, moreover:
\begin{enumerate}
\setcounter{enumi}{2}

\item
\label{AD1.3}
On $M$ there exists a Killing vector field $X$ the orbits of which are
complete. 
Furthermore, in the coordinate system of point \ref{D1.1} of
Definition \ref{D1} we have as $r
\rai$,
\be
X=(A + O(r^{-\alpha}))\frac{\pt}{\pt t}+
(\omega^i_jx^j+
O(r^{-\alpha})) \frac{\pt}{\pt x^i}
\label{A(D.4)}
\ee
for some constant $A\,(\ne 0)$,
and some constant coefficients matrix $\omega_i^j$ satisfying
$\omega_i^j=-\omega_j^i$. We moreover assume that the
vector field
$$Y=X-\hat X$$
is {\em uniformly} timelike on $\Sigma_{1}$ ($g_{\mu\nu}Y^\mu
Y^\nu<-\epsilon$ for some constant $\epsilon>0$), where
$$
\hat X^\mu\pt_\mu=\omega_i^jx^i\pt_j\ .
$$
If $\omega_i^j\ne0$ we shall moreover
require that the metric satisfies
\be
\label{A(D.4.-)}
\forall\quad 0\le |\beta|+i
\le k\qquad
\pt_x^{\beta}{\cal L}_{\hat X}^{i}
(g_{\mu\nu}-\eta_{\mu\nu})=O(r^{-\alpha-|\beta|-i})\ ,
\ee
\be
\forall \quad
|\beta|+i
= k,\ 
|x-y|\le r(x)/2
\qquad 
\label{A(D.4.-0)}
\Big| 
\pt_x^{\beta}{\cal L}_{\hat X}^{i}
g_{\mu\nu}(0,x^k) - \pt_x^{\beta}{\cal L}_{\hat X}^{i}
g_{\mu\nu}(0,y^k)
\Big|\le Cr^{-\alpha-k-\lambda}|x-y|^\lambda\ ,
\ee
\be
\forall \quad
|\beta|+i
= k,\ 
0\le|s|\le 2\pi A|\omega|^{-1} \qquad
\label{A(D.4.-1)}
\Big|\pt_x^{\beta}{\cal L}_{\hat X}^{i}
\Big[\Big(R(s)-{\rm id}\Big)^* g_{\mu\nu}\Big](0,x^i)
\Big|\le Cr^{-\alpha-k-\lambda}
|s|^\lambda\ ,
\ee
where $\beta$ 
is a multi--index with
{\em space} indices only, 
$R^j_i(s)=\exp\{s\omega_i^j\}$, $(R(s)-{\rm id}\Big)^*$ is the
pull--back by the map $R^j_i(s)x^i-x^j$ and  ${\cal L}_{\hat X}$ is the
Lie derivative.  We shall also assume, for $\omega_i^j\ne0$, that the
vector field $Y$ 
satisfies

\be
\label{A(D.4.+)}
\forall\quad
0\le |\beta|+i
\le k\qquad
\pt_x^{\beta}{\cal L}_{\hat X}^{i}
Y_{\mu}
=O(r^{-\alpha-|\beta|-i})\ ,
\ee
\be
\forall \quad
|\beta|+i
= k,\ 
|x-y|\le r(x)/2
\qquad 
\label{A(D.4.+0)}
\Big| 
\pt_x^{\beta}{\cal L}_{\hat X}^{i}
Y_{\mu}(0,x^k) - \pt_x^{\beta}{\cal L}_{\hat X}^{i}Y_{\mu}(0,y^k)
\Big|\le Cr^{-\alpha-k-\lambda}|x-y|^\lambda\ ,
\ee
\be
\forall \quad
|\beta|+i
= k,\ 
0\le |s|\le 2\pi A|\omega|^{-1} \qquad
\label{A(D.4.+1)}
\Big|\pt_x^{\beta}{\cal L}_{\hat X}^{i}
\Big[\Big(R(s)-{\rm id}\Big)^*Y_{\mu}
\Big](0,x^i)
\Big|\le Cr^{-\alpha-k-\lambda}
|s|^\lambda\ .
\ee
\end{enumerate}
\end{Definition}
{\bf Remarks:}
\begin{enumerate}
\item
If $|\omega|=0$, then Definition \ref{AD1} clearly reduces to
Definition \ref{D1}.

\item
If $\hat X=\omega^i_jx^j\pt_i$ is also a Killing vector field,
such that $[X,\hat X]=0$,
as occurs {\em e.g.} in the case of the Kerr space--time, then
(\ref{A(D.4.-)})--(\ref{A(D.4.+1)}) will clearly hold if
(\ref{(D.1)})--(\ref{(D.3)}) are fulfilled.
\item
If (\ref{A(D.4.-)}) and (\ref{A(D.4.+)}) hold with $1 \leq i
\leq k+1$, then (\ref{A(D.4.-0)})--(\ref{A(D.4.-1)}) and
(\ref{A(D.4.+0)})--(\ref{A(D.4.+1)}) will hold as well (with any $0
\leq \lambda \leq 1$).
\end{enumerate}

\begin{Proposition}[Killing time based on $\Sigma\cap M_1
$]
\label{APD.2}
Let $(M, g_{ab})$ with $\Sigma = \Sigma_1 \cup \Sigma'$ be
asymptotically stationary with respect to $\Sigma_1$. Then there exists a
global coordinate system on $M_1$ such that $M_1 \approx \R \times
(\R^3 \setminus B(R_1))$, $\Sigma_1\subset\{t=0\}$, and
\[
g_{00} < -C^{-1} , \qquad A\frac{\pt g_{\mu\nu}}{\pt t}
+\omega^i_jx^j\frac{\pt g_{\mu\nu}}{\pt x^i}= 0
\qquad\qquad\qquad\qquad
\]
\be
\qquad\qquad\qquad\qquad\qquad\left(\Longrightarrow  \quad
\cases{ g_{\mu\nu}(t+s,x^i)=g_{\mu\nu}(t,x^i),\ s\in\R &  if
$|\omega|=
0 ,$\cr
g_{\mu\nu}(t+2\pi A|\omega|^{-1},x^i)=g_{\mu\nu}(t,x^i), &
otherwise, 
\cr}
\right)
\label{A(D.7)}
\ee
\begin{eqnarray}
\forall Z^i\in \R^3\quad g_{ij}Z^iZ^j&\ge&
C^{-1}\sum_{i=1}^3(Z^i)^2\label{A(D.7.0)} \\
|\pt_{\sigma_1} \ldots \pt_{\sigma_i} (g_{\mu\nu}- \eta_{\mu\nu})|
&\leq &C
r^{-\alpha-i} \qquad 0 \leq i \leq k \label{A(D.9)}
\end{eqnarray}
for some constant $C$, with $|\omega| = \sum_{i<j} \sqrt
{(\omega^i_j)^2}$;  moreover, $\pt_{\sigma_1} \ldots \pt_{\sigma_k}
g_{\mu\nu}$ satisfy an obvious weighted H\"older condition.
\end{Proposition}

{\bf Proof:} 
\label{Killing}
Let  $x^\mu$ be the coordinates of point \ref{D1.1} of
definition \ref{D1}, let $\phi[X]^\mu(s,y^i)$ be the unique solution of
the problem
\be
\label{ODE.1}
\cases{\frac{d\phi[X]^\mu}{ds}(s,y^i)=X^\mu\Big(\phi[X]^\mu(s,y^i)\Big)\ ,
\cr
\phi[X]^0(0,y^i)=0, \quad \phi[X]^j(0,x^i)=x^j\ .}
\ee
Let the new coordinates
$\Big(y^\mu(x^\alpha)\Big)=\Big(As(x^\alpha),y^i(x^\alpha)\Big)$ be
implicitly defined by the equations
\[
x^o=\phi[X]^0(s,y^i)\ ,
\]
\be
\label{ODE.2}
x^i=
\phi[X]^j(s,y^k)\ .
\ee
Since the derivatives of $X$ satisfy uniform decay conditions up to
order $k$, it follows from (\ref{ODE.1}) by standard ODE theory that
the derivatives of $\phi[X]^\mu$ will satisfy uniform decay conditions up
to order $k$, then by  (\ref{ODE.2}) the derivatives of
$\pt x^\mu/\pt y^\nu$ will satisfy uniform decay conditions up
to order $k-1$; it turns out, however, that decay conditions for the
derivatives of the metric 
hold 
at order $k$ as well, which
can be seen as follows: differentiating equation (\ref{ODE.2}) with
respect to $x^\mu$ one obtains:
\[
1=\frac{\pt x^0}{\pt x^0}=
X^0\frac{\pt s}{\pt t}+ \frac{\pt\phi[X]^0}{\pt y^i}
\frac{\pt y^i}{\pt t}\ ,
\]
\[
0=\frac{\pt x^0}{\pt x^i}=
X^0\frac{\pt s}{\pt x^i}+ \frac{\pt\phi[X]^0}{\pt y^j}
\frac{\pt y^j}{\pt  x^i}\ ,
\]
%
\[
0=\frac{\pt x^i}{\pt t}=
X^j\frac{\pt s}{\pt t} + \frac{\pt\phi[X]^j}{\pt y^k}
\frac{\pt y^k} {\pt t}\ ,
\]
\be
\delta^i_j=
X^\ell\frac{\pt s}{\pt x^j} + \frac{\pt\phi[X]^\ell}{\pt y^k}
\frac{\pt y^k} {\pt  x^j}\ .
\label{ODE.3}
\ee
{}From (\ref{ODE.1}) and (\ref{ODE.3}) at $s=0$ 
it follows that
\[
\frac{\pt s}{\pt t}\Big|_{s=0}=(X^0)^{-1}\Big|_{s=0}, \quad
\frac{\pt s}{\pt  x^i}\Big|_{s=0}=0\ ,
\]
\be
\label{ODE.4}
\frac{\pt y^i}{\pt
t}\Big|_{s=0}=-(X^0)^{-1}X^i\Big|_{s=0}
,
\quad
\frac{\pt y^i}{\pt x^j}\Big|_{s=0}=\delta^i_j\ .
\ee
To avoid ambiguities let us write the metric in the form $g=g_{x^\mu
x^\nu}dx^\mu dx^\nu = g_{y^\mu
y^\nu}dy^\mu dy^\nu $; we have $X=X^{x^0}\pt/\pt
t+X^{x^i}\pt/\pt{x^i}=
X^{y^0}\pt/\pt y^0+X^{y^i}\pt/\pt{y^i} =A\pt/\pt
y^0 $, 
and from (\ref{ODE.4}) one obtains
\be
\label{ODE.5}
g_{y^i
y^j}(y^0,y^i)=g_{x^i
x^j}(x^0=0,x^i=y^i)\ ,
\ee
\be
\label{ODE.6}
g_{y^0
y^i}(y^0,y^i)=A^{-1}[g_{x^i
x^\ell}X^{x^\ell}+g_{x^i
x^0}X^{x^0}](x^0=0,x^i=y^i)\ ,
\ee
\be
g_{y^0
y^0}(y^0,y^i)=A^{-2}[g_{x^0
x^0}(X^{x^0})^2+g_{x^k
x^\ell}X^{x^k}X^{x^\ell}
+2g_{x^k
x^0}X^{x^0}X^{x^k}](x^0=0,x^i=y^i)
\ ,
\label{ODE.7}
\ee
and for $\omega=0$ the asymptotic bounds readily follow from
(\ref{(D.1)})--(\ref{(D.4)}) and (\ref{(D.6)}). If $\omega\ne 0$ the
coordinates $\{y^\mu\}$ are, however, not asymptotically flat. This is
easily cured as follows:
let $R_i^j(s)\equiv\exp\{s\omega_i^j\}$; thus $R_i^j(s)$ is a rotation
by angle $|\omega|s$ around the appropriately oriented axis $\hat e^i$
defined by
$\omega_i^j\hat e^i=0$, where
$|\omega|=\sqrt{\sum_{i<j}(\omega_i^j)^2}$, and
\[
\frac{dR_i^j(s)}{ds}=R_i^k(s)\omega_k^j=\omega_i^kR^j_k(s)\ .
\]
Let new coordinates $\{z^\mu\}$ be defined by
\[
z^0=y^0\ , \qquad z^i=R^i_k(s)y^k \ ; 
\]
one easily finds
\[
g_{z^i
z^j}(z^0,z^i)=g_{x^k
x^\ell}(x^0=0,x^i=R^i_j(-A^{-1}z^0)z^j)R^k_i(-A^{-1}z^0)R^\ell_j(-A^{-1}z^0)\ ,
\]
\[
g_{z^0
z^i}(z^0,z^i)=A^{-1}[g_{x^k
x^\ell}(X^{x^\ell}-\omega^\ell_jx^j)+g_{x^k
x^0}X^{x^0}](x^0=0,x^i=R^i_j(-A^{-1}z^0)z^j)R^k_i(-A^{-1}z^0)\ ,
\]
\[
g_{z^0
z^0}(z^0,z^i)=A^{-2}[g_{x^0
x^0}(X^{x^0})^2+g_{x^k
x^\ell}(X^{x^k}
-\omega^k_jx^j)(X^{x^\ell}-\omega^\ell_mx^m)
\qquad\qquad\qquad\qquad
\]
\[
\qquad\qquad\qquad\qquad
+2g_{x^k
x^0}X^{x^0}(X^{x^k}-\omega^k_jx^j)](x^0=0,x^i=R^i_j(-A^{-1}z^0)z^j)
\ .
\]
We have, {\em e.g.},
\[
\pt_{z_0}g_{z^iz^j}=O(r^{-\alpha-1})\quad \Longleftrightarrow
\quad {\cal L}_{\hat X} g_{x^kx^\ell}=O(r^{-\alpha-1})\ ,
\]
with $\hat X=\omega^i_jx^j\pt _i$, and (\ref{A(D.7)})--(\ref{A(D.9)})
follow in a similar manner from  (\ref{A(D.4.-)})--(\ref{A(D.4.+1)}).
\hfill$\Box$

{\bf Remark:} Equations (\ref{ODE.5})--(\ref{ODE.7}) provide a very
effective way of constructing stationary metrics of the form
(\ref{example}) with some ``desirable'' pathological properties (not
necessarily, however, satisfying some reasonable field equations and/or energy
inequalities): fix a
three dimensional manifold $\Sigma$, with a metric $g_{x^i
x^j}dx^idx^j$, a scalar field $g_{x^0
x^0}$, a one form  $g_{x^i
x^0}dx^i$, a vector field $X^i\partial/\partial x^i$, and finally
a scalar field $X^0>0$; make sure that the matrix $g_{x^\mu
x^\nu}$ has Lorentzian signature. Then the manifold $\R\times \Sigma$ with the
metric $g_{y^\mu
y^\nu}dy^\mu dy^\nu$, with $g_{y^\mu
y^\nu}$ given by (\ref{ODE.5})--(\ref{ODE.7}), is a Lorentzian
manifold with Killing vector  $X^\mu\partial/\partial
x^\mu=\partial/\partial y^0$.
As an illustration, consider $\Sigma=\R^3$ with $g_{x^\mu
x^\nu}(x^i)\,dx^\mu dx^\nu=-(dx^0)^2+\sum(dx^i)^2$,
$X^\mu(x^i)\,\partial/\partial x^\mu=\phi(x^i)\partial/\partial
x^0+\partial/\partial x^1$, where $\phi$ is any smooth radially symmetric
function satisfying
\[
\phi\ \cases{=2\ , & $r\le 10^3\ ,$
\cr
\in [1+10^{-400},2]\ , & $10^3\le r\le 10^4\ ,$
\cr
=1+r^{-100}\ ,& $r\ge 10^4\ .$
}
\]
Then $X^\mu\partial/\partial x^\mu$ is timelike everywhere,
asymptotically approaches the null vector $\partial/\partial
x^0+\partial/\partial x^1$, and we have $g_{y^\mu
y^\nu}{dy^\mu
dy^\nu}=(1-\phi^2)(dy^0)^2 +2 dy^0dy^1+\sum (dy^i)^2$. Introducing
$z^0=y^0$, $z^1=y^1+y^0$, $z^A=y^A$, $A=1,2$, one has $g_{z^\mu
z^\nu}{dz^\mu
dz^\nu}=-\tilde\phi^2\,(dz^0)^2 +\sum (dz^i)^2$, with
$\tilde\phi(z^i)=\phi(z^1-z^0,z^2,z^3)$, thus  all the conditions of
Definition \ref{D1} are satisfied (with  any $2\le k\le 98$, $\alpha=99-k$
and $\lambda=1$) except for uniform timelikeness of $X$ and for
equation (\ref{(D.4)}).
Moreover it is clear that {\em no} coordinates  as in  Proposition \ref{PD.2}
exist for this metric.

We have the following

\begin{Theorem}
\label{metatheorem}
All the results of this paper hold with {\em
stationary} replaced by {\em \stationary}.
\end{Theorem}

The reader who wishes to check the validity of this result will
notice that {essentially} all the proofs have been worded in a way which
generalizes
immediately to the \stationary\ case.
Let us mention the following. First,
it is easily seen from Proposition \ref{APD.2} that for any $p=(t,x)\in
M_{\rm ext}$ the orbits $\gamma_p(t)$ are future or past oriented,
according to the sign of $A$ (the sequence $t_i$ can be taken to be equal
to $2\pi iA|\omega|^{-1}, i\in {\bf Z}$).

A property of the orbits of $\phi[X]$ which is often used
is the following:
let $p,q\in M_i$, then there exists $T$ such that,
changing the time orientation if necessary 
for all $t\le T$ we
have $\phi[X]_t(p)\in I^-(q)$. This can be seen as follows:
 joining points with curves of the form $(t+Bs,r+s)$ and
$(t+Bs,\gamma(s))$, where $t,r$ refer to the coordinates of
Proposition \ref{APD.2}, $B$ is a constant large enough, and
$\gamma(s)$ is either a constant curve or, say, a geodesic arc in the
standard round metric on the spheres $r=const$, one finds that there
exists a constant $C=C(p,q)$
 such that all points
$p^\prime=(t(p^\prime),x(p^\prime))$
with
$t(p^\prime)\ge\max(t(p),t(q))+C$,
$r_0=\min(r(p),r(q))\le
r(p^\prime)\le \max(r(p),r(q))$ satisfy $p^\prime\in I^+(p)$, and
$p^\prime\in I^+(q)$; similarly if
$t(p^\prime)\le\min(t(p),t(q))-C$
and
$r_0=\min(r(p),r(q))\le r(p^\prime)\le \max(r(p),r(q))$, then
$p^\prime\in I^-(p)$, and $p^\prime\in I^-(q)$. In the Killing
coordinates on $M_i$ we have $\phi[X]_s(t,x)=(t+As,R(s)x)$,
where $R(s)$ is a rotation (around an appropriate axis, {\em cf.\ }the
proof of Proposition \ref{APD.2}) by an angle $|\omega| s$, and setting
$T= -A^{-1}(\max(|t(p)|,|t(q)|)+C)$
the result follows.

\section{An Alternate Proof of Existence of a Maximal Foliation
for Spacetimes of Class (b) Satisfying the Strong Energy Condition}
\label{gluing}
In this Appendix we shall outline a
somewhat simpler though certainly less elegant proof of the
Theorems of existence of maximal slices for space--times of class (b)
when one assumes that the timelike convergence condition (\ref{(TCC)})
holds:

\begin{Theorem}
\label{TA.3}
Let $(M, g)$ be a space-time of class (b), suppose that
(\ref{(M.1)}) holds and that the
timelike convergence condition (\ref{(TCC)})
is satisfied.  Then ${\stackrel{o}{\cal D}\!\!(\Sigma)}$ can be covered by a
family of maximal (spacelike)
asymptotically flat hypersurfaces $\Sigma_s$, $s\in\R$,  with boundary
$\pt \Sigma_s = \pt \Sigma = S$, such that
\[
\phi[X]_t(\Sigma_s) = \Sigma_{s+t} , \quad \mbox{and} \quad \Sigma_s
\approx \Sigma .
\] 
Moreover $\Sigma_s\setminus S$ are Cauchy
surfaces for $\ds$, and the hypersurfaces $\Sigma_s\setminus S$
foliate ${\stackrel{o}{\cal D}\!\!(\Sigma)}$.
\end{Theorem}
{\bf Proof:} The idea of the proof is similar to that of the proof of
Theorem \ref{TM.2}, the essential difference being how the appropriate
time function is constructed. By \cite{BCOM} one can deform $\Sigma$
outside of a compact set
so that, still denoting the deformed hypersurface by $\Sigma$,
$\Sigma \setminus\Sigma_{0,\rho}$ is \locally  maximal. By
Theorem 4.1 of \cite{Ba2} there exists a maximal surface $\check
\Sigma\approx \Sigma_{0,\rho}$ such that $\pt \check \Sigma = S\cup \pt
\Sigma_{0,\rho}$. By the gluing Lemma \ref{LM1}
 which is proved below the
hypersurface $\check \Sigma \cup (\Sigma \setminus\Sigma_{0,\rho})$
can be smoothed out to a smooth
hypersurface $\hat
\Sigma$; by construction there exists $\epsilon$ such
$\hat\Sigma_{0,\rho-\epsilon}$ and
$\hat\Sigma\setminus\hat\Sigma_{0,\rho+\epsilon} $ are \locally
maximal. We can choose $\epsilon$ small enough so that
$\pt\Sigma_{0,\rho\pm\epsilon}$ are smooth coordinate spheres in $M_{\rm ext}$.
Because the slopes of the light cones in $M_{\rm ext}$ are uniformly
bounded, there exists a constant $K_1$ such that for any spacelike
surface $\tilde \Sigma$ we have the implication
\[
\tilde \Sigma\cap \Tau_1\ne\emptyset\ \Longrightarrow \ \tilde
\Sigma\cap \Tau_\infty \subset \Tau_{K_1}\ ,
\]
where
\[
\Tau_s\equiv \cup_{t\in(-s,s)}\phi[X]_t(\hat\Sigma_{0,\rho+\epsilon} \setminus
\hat\Sigma_{0,\rho-\epsilon})\ .
\]
Let $\Sigma_i$ be a maximal surface such that $\pt \Sigma_i=S\cup
\pt\hat\Sigma_{0,i}$. Let $s$ be such that
$\phi[X]_s(\Sigma_i)\cap\Tau_1\ne\emptyset$. We then have
$\phi[X]_s(\Sigma_i)\cap\Tau
_\infty \subset \Tau_{K_1}$. Set
\[
M_{\rm int} \equiv M
\setminus[\cup_{t\in\R}(\hat\Sigma_t\setminus\hat\Sigma_{t,\rho-\epsilon})]\ .
\]
It follows that $\phi[X]_s(\Sigma_i)\cap M_{\rm int}
$ is a maximal
surface with boundary included in
$S\cup_{t\in[-K_1,K_1]}\pt\hat\Sigma_{t,\rho-\epsilon} $. Because
$\phi[X]_{\pm K_1}(\hat\Sigma_{0,\rho-\epsilon})$ are maximal, the
comparison principle implies that  $\phi[X]_s(\Sigma_i)\cap M_{\rm int}
$ is a subset of the compact set
$\cup_{t\in[-K_1,K_1]}\hat\Sigma_{t,\rho-\epsilon} $. This observation
allows one to carry out the remaining arguments of the proof as in
Theorem \ref{TM.2}, by using any time function which coincides with
the Killing time in $M_{\rm ext}$, and
covers a neighbourhood of
$\cup_{t\in[-K_1-1,K_1+1]}\hat\Sigma_{t,\rho+\epsilon} $. (Such a time
function can be easily constructed spanning maximal surfaces on some
timelike hypersurface containing $\pt\Sigma$ and on
$\cup_{t\in[-K_1-1,K_1+1]}\pt\hat\Sigma_{t,\rho+\epsilon} $, and
appropriately
``rounding off'' any nondifferentiability if necessary.)
\hfill $\Box$

Let us finally prove the ``gluing Lemma'':

\begin{Lemma}
\label{LM1}
Let $\Sigma$ be a $C^{k,\alpha}$, $k \ge 1$, $0 < \alpha \le 1$,
spacelike embedded
hypersurface in a Lorentzian space-time $(M, g_{ab})$ with a
$C^{k-1,\alpha}$ metric,
suppose that $\tS$ is a $C^{k,\alpha}$ spacelike embedded hypersurface
in $M$
such that $\pt \tS
\subset \Sigma$ is a compact $C^{k,\alpha}$ submanifold of $\Sigma$,
and assume that
$\Sigma$ is the disjoint union of $\Sigma_{\rm int}$, $\pt \tS$ and
$\Sigma_{\rm ext}$, with $\partial \Sigma_{\rm int}=\pt \tS=
\pt\Sigma_{\rm ext}$.
Define the (Lipschitz) piecewise $C^{k,\alpha}$ hypersurface $\hS$
by:
\be
\label{5.22.1}
\hS = \tS \cup \Sigma_{\rm ext} .
\ee
Let $d(p)$ be the distance on $\hS$ from $p$ to $\pt \tS$.  For every
$\epsilon > 0$ there
exists a spacelike embedded $C^{k,\alpha}$ hypersurface $\hS_\epsilon$
--- ``a smoothed out deformation of $\hat \Sigma$'' --- such
that
\[
p \in \hS , \quad  p \not\in \hS_\epsilon \Longrightarrow d(p) <
\epsilon\ ; \qquad \quad {\rm edge}\, \hat\Sigma_\epsilon= {\rm edge}\,
\hat\Sigma\ .
\]
(In other words, $\hS$ coincides with $\hS_\epsilon$ except for points
closer than
$\epsilon$ to $\pt \tS$.)
\end{Lemma}

{\bf Proof:}
Let $t$ be any time function defined in a neighbourhood
${\cal O}$ of $\pt
\tS$ such that $\Sigma \cap {\cal O} = \{t = 0\}$.  [Such a function
can {\em  e.g.} be constructed as follows: let ${\cal V}
\subset \Sigma$ be a
conditionally compact neighbourhood of $\pt \tS$, let ${\tilde
g_{ab}}$ be any smooth
Lorentzian metric in a neighbourhood ${\cal W} \subset M$ of $\bar{\cal
V}$ such that 
$\Sigma \cap {\cal W}$ is spacelike with respect to the metric
$\tilde g_{ab}$, and such that in every tangent space the solid light
cones of ${\tilde
g_{ab}}$ are proper subsets of 
the solid light cones of $g_{ab}$. Let $t \in
C^\infty({\widetilde{\cal D}}({\cal V}))$
be a solution of the problem
\begin{eqnarray*}
&&{\tilde g}^{\mu\nu} {\tilde \nabla}_\mu {\tilde \nabla}_\nu
t = 0 ,\\
&&t\big|_{{\cal V}\cap\Sigma} = 0 , \qquad {\tilde n^\mu \pt_\mu
t}\big|_{{\cal
V}\cap\Sigma} = 1\,,
\end{eqnarray*}
where ${\widetilde \nabla}_\mu$ is the covariant derivative of the
metric ${\tilde  g_{ab}}$,
${\tilde n}^\mu$ is a vector field transverse to ${\cal
W}\cap\Sigma$ and ${\widetilde{\cal D}}({\cal V})$ is the domain of
dependence of ${\cal V}$ in the spacetime $({\cal W},\tilde g_{ab})$.
Embeddedness of $\Sigma$ implies that there exists an open
neighbourhood ${\cal O}$
of $\pt \tS$ in ${\cal W}$ such that $t$ is a time function for the
metric $g_{ab}$ on ${\cal O}$, and we have 
$\Sigma \cap {\cal O} = \{t = 0\}$.]
Let $r$ be any $C^\infty$ function on $\U =
\Sigma \cap {\cal  O}$ such that $|dr| > 0$, $r\big|_{\pt \tS} = 1$.
Let $v$ be any
coordinates on $\pt \tS$,
one can extend $v$ to $\U$ by Lie dragging along the integral curves
of $\nabla r$;  $r$
and $v$ can be extended to ${\cal O}$ by Lie dragging along $\nabla
t$.  By
compactness of $\pt \tS$ there exists $0 < \delta \le \frac{1}{2}$
such that we have
\[
(t, r, v) \in K_\delta =[-\delta, \delta] \times [1-\delta, 1+\delta]
\times \pt \tS \subset {\cal O} \ .
\]
Since ${\rm edge}\tilde \Sigma=\pt \Sigma\subset\Sigma$, decreasing
$\delta$ if necessary one can find a $C^{k,\alpha}$
function ${\tilde u}$ such
that $\tS \cap K_\delta$ is a graph:
\[
\tS \cap K_\delta = \{t = {\tilde u}(r, v),\; r \in [1-\delta, 1],\; v
\in \pt \tS\} .
\]
It follows that $\hS$, as defined by (\ref{5.22.1}), is a graph of the
following $C^{0,1}$ function $u$:
\begin{equation}
\label{(L1.0)}
u(r, v) = \left\{
\begin{array}{ll}
{\tilde u}(r, v) , & r \in [1-\delta, 1]\\
0 & r \in [1, 1+\delta] .
\end{array} \right.
\end{equation}
By Whitney's extension Lemma ({\em cf.\ e.g.} \cite{GT}) there exists
a function $\apu
\in C^{k,\alpha}([1-\delta, 1+ \delta] \times \pt \tS)$ such that
$\apu\big|_{[1-
\delta,1]\times\pt\tS} = {\tilde u}$.  Let $\apS$ be the graph of
$\apu$, decreasing
$\delta$ if necessary we may assume that $\apS$ is a spacelike
hypersurface in
$K_\delta$.  Let $m$ be the unit (spacelike) vector field proportional
to $\frac{\pt}{\pt
r}$, define $n$ to be the
future oriented
timelike, unit vector field, orthogonal to $\frac{\pt}{\pt r}$, of the
form $  a \frac{\pt}{\pt t} + b \frac{\pt}{\pt r}$; we
thus have
\begin{eqnarray}
\frac{\pt}{\pt t} &= &A(t, r, v) n + B(t, r, v) m \qquad (|B| <
A)\label{(L1.1)}\\
\frac{\pt}{\pt r} &= & C(t, r, v) m ,\\
n^2 &= &-m^2 = -1 , \qquad n \cdot m = 0 .
\end{eqnarray}
By compactness of $K_\delta$ it follows that there exists $\delta_1 >
0$ such that
\begin{equation}
\label{(L1.2)}
C > \delta_1 , \qquad A > \delta_1
\end{equation}
The vector field $Y =Y ^\mu \pt_\mu = \frac{\pt \apu}{\pt r}
\frac{\pt} {\pt t} +
\frac{\pt} {\pt r} = A \frac{\pt \apu}{\pt r} n + \left(C + B
\frac{\pt \apu}{\pt r}\right)  m$ is tangent to $\apS$, so that
compactness of $K_\delta$ and
spacelikeness of $\apS$
imply that there exists $\delta_2 > 0$ such that
\[
C(\apu(r,v), r, v) \ge \left[A(\apu(r, v),  r, v)   - {\rm sgn}
\left(\frac{\pt \apu}{\pt r}(r, v) \right) B (\apu(r, v),
r, v) \right]\Big| \frac{\pt \apu}{\pt r} (r, v) \Big| + \delta_2 ,
\]
where ${\rm sgn}(\cdot)$ denotes the sign of $\cdot$.  This together
with continuity of
$A$,
$B$ and $C$ shows that, decreasing $\delta$ if necessary, there exists
$\delta_3 > 0$
such that for all $(t,r,v)\in K_\delta$ one has
\begin{equation}
\label{(L1.3)}
C(t,r,v) \ge \left[A(t,r,v) - {\rm sgn}\left(\frac{\pt \apu}{\pt r}
(r,v)\right) B(t,r,v)\right]
\left[\Big|\frac{\pt \apu}{\pt r} (r,v)\Big| + \delta_3\right] .
\end{equation}
Let $\epsilon$ be as described in the statement of this Lemma, there
exists a constant $K
> 0$ such that for $|r-1| \le K\epsilon$ one has $d(r, v) \le
\epsilon$, where $d(p)$ is the
geodesic distance on $\tS \cup \Sigma_{\rm ext}$ from $p=(r,v)$ to $\pt
\tS$.
Without loss of generality
we may assume $K\epsilon \le \delta/4$.  Let $\phi[X]_\epsilon \in
C^\infty(\R)$ be any
function satisfying ${\rm supp}\, \phi[X]_\epsilon \subset
\left(1 - \frac{K\epsilon}{2} , 1 + \frac{K\epsilon}{2}\right)$, $0
\le \phi[X]_\epsilon \le
1$, $\phi[X]_\epsilon(x) = 1$ for $x \in [1 - K\epsilon/4, 1 +
K\epsilon/4]$.  Let $\psi \in
C^\infty(\R)$ satisfy ${\rm supp}\, \psi \in [-1, 1]$, $0 \le \psi \le
1$, $\int_{-
\infty}^\infty \psi(x) dx = 1$.  For $0 < \nu < K\epsilon/2$ set
$\psi_\nu (x)= \psi
\left(\frac{x}{\nu}\right)$, define
\begin{equation}
\label{(L1.4)}
u_{\nu,\epsilon}(r,v) =  \int_{-\infty}^\infty \psi_\nu(r-s)\,
\phi[X]_\epsilon(s)\, u(s, v)\, ds + (1
- \phi[X]_\epsilon(r))\, u(r, v) .
\end{equation}
It follows from (\ref{(L1.4)}) that
\begin{enumerate}
\item $u_{\nu,\epsilon} \le C^{k,\alpha}([1-\delta, 1+\delta] \times
\pt \tS)$,

\item
$u_{\nu,\epsilon}$ converges to $u$ in $C^{0,1}([1-\delta,
1+\delta] \times \pt
\tS)$ as $\nu \longrightarrow 0$.

\item For $r \not\in [1 - K\epsilon, 1+K\epsilon]$ we have
$u_{\nu,\epsilon}(r, v) =
u(r, v)$;  it follows that the hypersurfaces $\Sigma_{\nu,\epsilon}$
defined as those
coinciding with $\tS$ or $\Sigma_{\rm ext}$ outside of $K_\delta$, and
defined as the
graph of $u_{\nu,\epsilon}$ in $K_\delta$, are $C^{k,\alpha}$
hypersurfaces.

\item For fixed $r \in [1-\delta, 1+\delta]$ the functions
$u_{\nu,\epsilon}(r,\cdot) \in
C^{k,\alpha}(\pt \tS)$ converge as $\nu \longrightarrow 0$ to
$u(r,\cdot)$ in
$C^{k,\alpha}(\pt \tS)$;  in particular there exists $\nu_0 > 0$ such
that for $\nu \le
\nu_0$ the vector fields $\frac{\pt u_{\nu,\epsilon}}{\pt v^A}
\frac{\pt}{\pt t} +
\frac{\pt}{\pt v^A}$ are spacelike.
\end{enumerate}

To show spacelikeness of $\Sigma_{\nu,\epsilon}$ for $\nu$ small
enough it thus
remains to show that  the vector fields $Y_{\nu,\epsilon} = \frac{\pt
u_{\nu,\epsilon}}{\pt v} \frac{\pt}{\pt t} + \frac{\pt}{\pt v}$ are
spacelike.  This is
equivalent to the inequality
\begin{equation}
\label{(L1.6)}
C(u_{\nu,\epsilon}(r,v), r, v) > \left[A(u_{\nu,\epsilon}(r, v), r, v) -
{\rm sgn}\left(\frac{\pt u_{\nu,\epsilon}}{\pt r} (r, v)\right)
B(u_{\nu,\epsilon} (r, v), r,
v) \right] 
\Big|\frac{\pt u_{\nu,\epsilon}}{\pt r} (r, v)\Big| .
\end{equation}
It follows from (\ref{(L1.3)}), (\ref{(L1.6)}) and compactness of
$K_\delta$ that
there exists $\delta_4
> 0$ such that if
\[
\beta(r,v)\equiv \Big|\frac{\pt u_{\nu,\epsilon}}{\pt r} (r, v)\Big|
\le 2 \delta_4 \quad
\Longrightarrow  \quad
Y_{\nu,\epsilon} \quad \mbox{is spacelike} .
\]
Set $\Xi = \{v \in \pt \tS : \beta(1,v)\ge \delta_4\}$.  There exists
$\delta_5
> 0$ such that for
$(r,v) \in [1-\delta_5, 1+\delta_5] \times \Xi$ we have ${\rm sgn}
\left(\frac{\pt u_{\nu,\epsilon}}{\pt r} (r, v)\right) =
{\rm sgn} \left(\frac{\pt u_{\nu,\epsilon}}{\pt r} (1, v)\right)$.
Let $\delta_6 =
\min(\delta_5, K\epsilon/8)$.  For $(r,v) \in \Omega \equiv
([1-\delta, 1-\delta_6] \cup
[1+\delta_6, 1+\delta]) \times \pt \tS$ the functions
$u_{\nu,\epsilon}$ converge in
$C^{k,\alpha} \subset C^1$ to $u$, it follows that decreasing $\nu_0$
if necessary
spacelikeness of $Y_{\nu,\epsilon}$ for $\nu \le \nu_0 \le
K\epsilon/8$ will hold on
$\Omega$.  On $[1-\delta_6, 1+\delta_6] \times \pt \tS$ we have
(recall that $u = 0$ for
$r \ge 1$)
\begin{eqnarray}
\frac{\pt u_{\nu,\epsilon}}{\pt r} &= &I_{\nu,\epsilon} +
II_{\nu,\epsilon}
,\label{(L1.5)}\\
I_{\nu,\epsilon} &= &\int_{-\infty}^\infty \psi_\nu(r - s)\; \frac{\pt
\apu}{\pt r} (s, v) ds
,\label{(L1.5.1)}\\
II_{\nu,\epsilon} &= &-\int_{1}^\infty \psi_\nu(r - s)\;
\frac{\pt \apu}{\pt r} (s, v)
ds\label{(L1.5.2)}
\end{eqnarray}
(since $r \le K\epsilon/8$ and $\nu \le K\epsilon/8$ we have
$\phi[X]_\epsilon(r) = 1$,
$\psi_\nu(r - s) \phi[X]_\epsilon(s) = \psi_\nu(r - s))$.  For $v \not\in
\Xi$ spacelikeness of
$Y_{\nu,\epsilon}$ has already been established;  for $\nu \in \Xi$ the
sign of $\frac{\pt
\apu}{\pt r}$ is constant along the curves $v = {\rm const}$, so that
\[
\Big|\frac{\pt u_{\nu,\epsilon}}{\pt r}\Big| \le |I_{\nu,\epsilon}|
\]
and since $I_{\nu,\epsilon}$ converges to $\frac{\pt
\apu_{\nu,\epsilon}}{\pt r}$ the
inequality (\ref{(L1.6)}) 
follows from (\ref{(L1.3)}). \hfill$\Box$

\end{document}